\documentclass[usegraphicx,usenatbib]{mn2e}
\usepackage{amssymb}
\usepackage{graphicx}
\usepackage{longtable}

\def\Teff{$T_{\rm eff}$}
\def\logg{$\log\,g$}
\def\loggf{$\log\,gf$}

\def\Vt{V${\rm t}$}
\newcommand{\kms}{km~s$^{-1}$}
\newcommand{\R}{log$R^{'}_{\rm HK}$}
\newcommand{\vsini}{$v$sin$i$}

\newcommand {\apgt} {\ {\raise-.5ex\hbox{$\buildrel>\over\sim$}}\ }
\newcommand {\aplt} {\ {\raise-.5ex\hbox{$\buildrel<\over\sim$}}\ }

\title[stars with massive planets]
{Chemical composition of stars with
massive planets
\thanks{Based on observations collected at OHP observatory, France}
}
\author[T.~Mishenina  et al.]
{T.~Mishenina$^{1}$,
  N.~Basak$^{1}$,
        V.~Adibekyan$^{2}$,
        C.~Soubiran$^{3}$,
        V.~Kovtyukh$^{1}$ \\
        \\
$^{1}$Astronomical Observatory, Odessa National University,
       Shevchenko Park, 65014, Odessa, Ukraine, \\
 $^{2}$ Instituto de Astrof\'isica e Ci\^encias do Espa\c{c}o, Universidade do Porto, CAUP,
Rua das Estrelas, 4150-762 Porto, Portugal \\
$^{3}$  Laboratoire d'astrophysique de Bordeaux, Univ. Bordeaux, CNRS, B18N, all\'ee Geoffroy Saint-Hilaire, 33615 Pessac, France \\
                                }

\begin{document}

\date{Accepted, Received, in original form \today}
\pagerange{\pageref{firstpage}--\pageref{lastpage}}
\pubyear{2015}

\maketitle

\label{firstpage}

\begin{abstract}
Stellar parameters of 25 planet-hosting stars and abundances of
Li, C, O, Na, Mg, Al, S, Si, Ca, Sc, Ti, V, Cr, Mn, Fe, Ni,
Zn, Y, Zr, Ba, Ce, Pr, Nd, Sm and Eu, were studied based on
homogeneous high resolution spectra and uniform techniques.
The iron abundance [Fe/H] and key elements (Li, C, O, Mg, Si)
indicative of the planet formation, as well as the dependencies of
[El/Fe] on $T_{cond}$, were analyzed. The iron abundances determined
in our sample stars with detected massive planets range within
--0.3 $ <$ [Fe/H] $ <$ 0.4. The behaviour of [C/Fe], [O/Fe],
[Mg/Fe] and [Si/Fe] relative to [Fe/H] is consistent with the
Galactic Chemical Evolution trends. The  mean values of C/O and
[C/O] are $<$C/O$>$  = 0.48  $\pm$0.07 and $<$[C/O]$>$ = --0.07 $\pm$0.07,
which are slightly lower than solar ones. The Mg/Si ratios range
from 0.83 to 0.95 for four stars in our sample and from 1.0 to 1.86
for the remaining 21 stars.
Various slopes of [El/Fe] vs. $T_{cond}$ were found.
The dependencies of the planetary mass on metallicity,
the lithium abundance, the C/O and Mg/Si ratios, and also on the
[El/Fe]--$T_{cond}$ slopes were considered.
\end{abstract}

\

\section{Introduction}

Over the past few decades, there has been an ongoing search for opportunities to study extrasolar planets based on characteristics of their hosting stars. The chemical composition features of parent stars considered  as possible indicators of the presence of planetary systems, while elemental relationships can be essential for studying the formation of planets of different masses. Initially, the iron abundance [Fe/H]  \citep[][and references therein]{gonzalez:97, santos:01, udry:07}, and the lithium abundance in stellar atmospheres \citep[e.g.][]{gonzalez:00, gonzalez:08, israelian:04}
came to the attention of researchers. Stars with massive Jupiter-like planets exhibited a [Fe/H] excess \citep[e.g.][]{santos:01, fischer:05, sousa:08}, however such an excess was not obvious in stars hosting less massive planets, like Neptune- and Earth-sized ones \citep[e.g.][]{udry:07, sousa:08}; later, extrasolar planets were also found around metal-deficient stars \citep{sousa:08, adibekyan:12a, adibekyan:12b}. A detailed review with the results on the relationship between stellar metallicity and planet occurrence was recently presented by \cite{adibekyan:19}.

As for lithium, in particular  $^{7}$Li, its lower abundance in stars with planetary systems, as compared to those without discovered planets, has been interpreted in favour of the presence of planets in these stars \citep[e.g.][]{gonzalez:00, israelian:04, gonzalez:10, israelian:09, delgado:15, figueira:14, mishenina:16}. Meanwhile, an overabundance of the $^{6}$Li isotope could suggest the presence of massive planets \citep[e.g.][]{israelian:01, montalban:02}.
For solar analogues (with \Teff~ $\sim$ 5600-5900 K), a significant difference in lithium abundance between stars with and without planets was detected by \cite{delgado:14}; the authors noted that it was due not only to the differences in \Teff, [Fe/H] or age, but also to the possible presence of planets that caused additional rotationally-induced mixing in the external layers of planet-hosting stars. Yet,  the observed lithium depletion, represented by a robust correlation between lithium abundance and the age of solar-like stars, \citep[see, e.g.][]{baumann:10, carlos:16}, may reflect the evolution of the star rather than the presence of a planet.

Studying the CNO abundances in stars with and without planetary systems has not revealed any differences in the levels of these elements between the two groups of stars \citep[e.g.][]{delgado:10, dasilva:11, dasilva:15, nissen:14}, while the elemental abundance trends with metallicities favored the Galactic chemical enrichment. \cite{suarez:16, suarez:17} demonstrated  that the linear relationship between [N/Fe] and [Fe/H] was associated with the Galactic chemical evolution, as the planet-hosting stars tended to be richer in metals. They also found two different trends for [C/Fe] versus [Fe/H] $>$ 0 and [Fe/H] $<$ 0 and determined a flat distribution of the [C/Fe] ratio for all planetary masses.

The key role of carbon-to-oxygen (C/O) and magnesium-to-silicon (Mg/Si) ratios in planet-hosting stars, which can provide information about the protoplanetary discs wherein the respective planets are formed, has been revealed in, e.g., \cite{kuchner:05}, \cite{bond:10}. Theoretical calculations displayed that C/O and Mg/Si are the most important elemental ratios in determining the mineralogy of planets. The C/O ratio determines the distribution of Si among carbide and oxide species, while Mg/Si gives information about the silicate mineralogy \citep{bond:10}. 
However, currently available data on the abundances of these elements, in particular the C/O ratios, are quite contradictory and ambiguous, and there have been noticeable discrepancies in the estimates of the C and O abundances in various studies. For instance, \cite{petigura:11} reported that planet-bearing systems were enriched in carbon with ratio $<$C/O$>$ =1.00. Meanwhile, other studies have not confirmed that \citep[e.g.][]{nissen:13, nissen:14} and shown that the stars with planets do not exhibit C/O ratios exceeding 0.8. According to \cite{teske:14}, the mean value of the C/O ratio in their sample of transiting exoplanet-host stars was 0.54, being equal to the solar value (C/O$_\odot$ = 0.54) and lower than the earlier measured $<$C/O$>$ value of about 0.65-0.75 in host stars.
\cite{pavlenko:19} reported that metal-rich dwarfs with planets were overabundant in carbon relative to Sun with an average difference $<$[C/O]$>$ = 0.05 $\pm$0.05.

\cite{adibekyan:15} found that the magnesium-to-silicon ratio [Mg/Si] in stars with low-mass planets was higher than in stars without planets. Despite the fact that the [Mg/Si] ratio depends significantly on metallicity through the Galactic chemical evolution, the difference in the [Mg/Si] ratio between low-mass planet hosts and non-hosts remains even after removing the Galactic evolution trend \citep{adibekyan:15}. \cite{adibekyan:17} took into account a plausible departure from Local Thermodynamic Equilibrium (NLTE) during the formation of the Mg lines and found out that the correction for the NLTE effects resulted in a small difference in [Mg/Si] ratios and that high [Mg/Si] values in the stars hosting super-Earth or Neptune-like planets were likely to be associated with their formation.
The correlation between the C/O and Mg/Si ratios, that is critical for determining the mineralogy of planetary companions, has been studied in \cite{suarez:18}. The authors obtained that 86\% of stars among all stars with high-mass companions have the carbon-to-oxygen ratio in the range of 0.4 $<$ C/O $<$ 0.8, the remaining 14\% having a C/O below 0.4; and the Mg/Si ratio is ranged within 1 $<$ Mg/Si $<$ 2 for all the stars with low-mass planets  and for 85\% of stars with high-mass companions (the other 15\% have Mg/Si $<$ 1). Planet hosts with low-mass companions yielded C/O and Mg/Si ratios similar to the solar ones, whereas stars with high-mass companions exhibited lower C/O ratios. The existing contradictions in the data obtained inspire the need for new investigations of the abundances of elements to study the relationship of the abundances with the formation of planetary systems.

\cite{udry:07} pointed out that the relationship between volatile and refractory elements may be indicative of the presence of planets.
\cite{melendez:09} showed that the Sun exhibited a depletion of refractory elements relative to volatiles as compared to the solar twins. Such a peculiarity could be related to the formation of planetary systems like our own, in particular to the existence of terrestrial planets \citep{melendez:09}. Moreover, later, it was shown that similar relationships could reflect a wide diversity of exoplanetary systems observed nowadays, as well as a variety of scenarios which could occur within the circumstellar discs  \citep{spina:16}. Alternatively, it was also suggested that similar trends could also be associated with the correlation of elemental abundances with the age and place of birth in the Galaxy \citep[e.g.][]{gonzalez:13, adibekyan:14, adibekyan:15, nissen:15}, and hence not related to the presence of planets.

This paper aims to study the parameters and chemical composition of stars with detected massive planets in order to establish and verify the relationship between the presence of Jupiter-like planets and chemical composition of the respective parent stars. It provides an excellent opportunity to check how the peculiarities of the chemical composition of  the planet host stars  characterize the presence of planets and their properties, such as, masses. To this end, the lithium and the key elements for the formation of planetary systems (C, O, Mg, Si and their ratios), as well as the dependence of the contents of volatile and refractory elements on the condensation temperature $T_{cond}$ will be considered as a function of planetary mass.

The paper is organized as follow.
Observations and selection of stars are described in \S \ref{sec: observ select}. The  main stellar parameters are defined in \S \ref{sec: param}.
The abundance determinations  and an error analysis are presented in \S  \ref{chem: comp}.
The analysis of the relationship between the presence of Jupiter-like  planets and the chemical composition of the parent stars is reported in \S \ref{sec: result}.

\section{Sample stars and observations }
\label{sec: observ select}

A sample of planet-hosting  stars with detected massive planets consisting of 24 stars with the planetary masses, $M_{pl}$, from 0.78 to 19.02 $M_{J}$ and one star with Neptun-size mass planet, that is 0.05 $M_{J}$ was selected from several papers related to a series of the SOPHIE Exoplanet Consortium programmes \citep{bouchy:09}. We relied on the following sources: \cite{boisse:12, moutou:14, courcol:15, hebrard:16, diaz:16, santos:16, rey:17, borgniet:19}. The stellar and planetary characteristics of the sample  planet hosts are presented in Table \ref{parplan}.

We used the stellar spectra from the OHP archive \citep{moultaka:04} obtained
with the echelle spectrograph SOPHIE \citep{perruchot:11} at the resolving
power R = 75,000 restricted to the wavelength range 4,400 -- 6,800 \AA\ and
signal-to-noise ratio (S/N) higher than 80. The spectral
processing, including normalisation of individual spectra to the
local continuum, identification of spectral lines of different
chemical elements, the line depth and equivalent width (EW)
measurements, etc., was performed employing the spectrum with
the highest S/N ratio for each star using the DECH30 software
package developed by G.A. Galazutdinov
\footnote{http://www.gazinur.com/DECH-software.html}.

\begin{table*}
\caption{Stellar characteristics  and planetary masses of the planet hosts sample.}
\label{parplan}
\begin{tabular}{lrrrrrrrrrlcl}
\hline
\hline
Star&     \Teff & \logg  & [Fe/H] & \Vt &\Teff & \logg  & [Fe/H] & \vsini & \R & planet&$M_{pl}$& ref  \\
     &       (K) &        &        & (\kms)  &(K)    &        &    &  (\kms)     &    &       &  ($M_{J}$)   &      \\
\hline
\multicolumn{1}{c}{}&\multicolumn{3}{c}{(our data)}&\multicolumn{5}{c}{(from literature)}&\multicolumn{4}{c}{}\\
\hline
HD12484&        5831&   4.40&    0.05& 1.1&  5920&   4.65&   0.05&   8.2&    --4.43&  HD12484b&      2.98&   \cite{hebrard:16}       \\
HD13908&        6150&   4.00      &0.05& 1.0&         6255&   4.11&   0.01&   4.2&    --4.90&   HD13908b&      0.86&   \cite{moutou:14}        \\
                        &      &                 &   &    &        &     &     &    &     &      HD13908c&      5.10     &    \\
HD16175&        5977&   4.30&0.37 &1.0       &         6022& 4.21&   0.37&   5.1&    --4.77&HD16175b&        4.77&   \cite{diaz:16}  \\
HD17674&        5904&   4.28&   --0.15& 0.8 &5904&     4.34&--0.16&     2.3&    --4.91&HD17674b&        0.87&   \cite{rey:17}    \\
HD24040&        5800&   4.25&   0.23& 0.8&  --&     --&     --&     --&     --&     HD24040b&      4.0      &       \cite{boisse:12}        \\
HD29021&        5550&   4.30&--0.18& 1.1&     5560&   4.44&   -0.24&2.7&      --5.00&     HD29021b&      2.4&    \cite{rey:17}   \\
HD35759&        5917&   4.10&--0.04& 0.9&      6060&   4.24&   0.04&   3.5&    --5.36&HD35759b&        3.76&   \cite{hebrard:16}       \\
HD89307&        5975&   4.50&--0.05&  0.7&    --&     --&     --&     --&     --&     HD89307b&      2.0&      \cite{boisse:12}                \\
HD113337&       6750&   4.30& 0.14&  1.3 &    6669&   4.21&   0.09&   6.0&      --      &HD113337b&     3.00&      \cite{borgniet:19}      \\
HD141399& 5515& 4.35&   0.47&  0.7& 5600&   4.28&   0.35&   2.9&    --5.26&HD141399b&       0.45&   \cite{hebrard:16}       \\
        &   &      &    &    &     &     &    &   &              &       HD141399c&      1.33&   \\
        &   &      &    &    &     &     &    &   &              &       HD141399d&      1.18&    \\
        &   &      &    &    &     &     &    &   &              &       HD141399e&      0.66&    \\
HD143105&       6280&   4.30&    0.01& 1.7&  6380&   4.37&   0.15&   9.1&--5.00& HD143105b&      1.21&   \cite{hebrard:16}       \\
HD150706&       5950&   4.50&    0.05& 0.8&   5961&4.50&--0.01&3.7&--4.47       &HD150706b&     2.71&   \cite{boisse:12}        \\
HD154345&       5503&   4.40&    --0.05& 0.6& --&     --&     --&     --&     --&     HD154345b&      1.0&              \cite{boisse:12}  \\
HD159243&       6085&   4.55&   0.14& 0.8&   6123&   4.55&   0.05&   3.8&    --4.65&  HD159243b&      1.13&           \cite{moutou:14}        \\
&   &      &    &    &     &     &    &    &             &               HD159243c&      1.9&   \\
HD164595&       5725&   4.40&--0.01&0.6&      5790&   4.44&   --0.04&  2.1&--4.86&      HD164595b&  0.05&       \cite{courcol:15}       \\
HD191806&       5807&   4.30&0.36& 0.7&      6010&   4.45&   0.30&    3.3&    --4.85&  HD191806b&      8.52&           \cite{diaz:16}  \\
HD214823&       5980&   3.90&0.13& 1.2&      6215&   4.05&   0.17&   5.7&    --4.79&  HD191806c&      19.4&           \cite{diaz:16}  \\
HD219828&       5850&   4.10&    0.18& 1.0&   5891&   4.08&   0.19&   2.9&--5.12&      HD219828b&      15.1&   \cite{santos:16}        \\
        &   &      &    &    &     &     &    &     &            &       HD219828c&0.07&   \\
HD220842&       5938&   4.15&   --0.10& 1.0&   5960&   4.24&   --0.17&  3.4&--5.17&      HD220842b&      3.18&\cite{hebrard:16}          \\
HD221585&       5562&   4.10&    0.28& 0.8&  5620&   4.05&   0.29&   3.7&    --4.86&  HD221585b&      1.61&   \cite{diaz:16}  \\
HD222155&       5665&   3.95&   --0.05&0.6&  5765&   4.10&--0.11&     3.2&--5.06&      HD222155b&      1.90&    \cite{boisse:12}        \\
HIP65407&       5354&   4.50&    0.35& 0.5&  5460&   4.47&   0.25&   2.8&    --4.60&   HIP65407b&      0.43&  \cite{hebrard:16}       \\
&   &      &    &    &     &     &    &     &            &       HIP65407c&      0.78&  \\
HIP91258&       5515&   4.65&   0.30& 0.9&   5519&   4.53&   0.23&   3.5&    --4.65&  HIP91258b&      1.07&   \cite{moutou:14}        \\
HIP109384&5179& 4.40&--0.24& 0.8&     5180&   4.43&   --0.26&  2.7&    --5.02&  HIP109384b&     1.56&           \cite{hebrard:16}       \\
HIP109600&      5525&   4.65&    --0.07&0.7& 5530&   4.45&   --0.12&  2.7&--5.07&      HIP109600b&     2.68&   \cite{hebrard:16}       \\
\hline
\end{tabular}
\end{table*}

\section{Stellar parameters }
\label{sec: param}

The initial effective temperatures, \Teff, were derived through
a method based on line-depth ratios
for spectral-line pairs with different low-level excitation potentials
taking advantage of the calibrations developed by \cite{kovtyukh:03}.
Then, the final effective temperatures
were determined  by imposing the condition of independence of the Fe I
abundance on the excitation potential. These latter temperatures are
given in Table 1 and represent the values adopted for the abundance
analysis.

The surface gravities \logg\  were obtained through the
Fe {\sc i}/Fe {\sc ii} ionisation balance, which means that the iron
abundances derived from the lines of neutral and ionised iron were
forced to be equal. In our case, the difference between these values
does not exceed 0.03 dex. The microturbulent velocity \Vt\ was determined
by removing any correlation between the Fe {\sc i} abundances and EWs of the lines.
Using such iterative procedure, we determined the parameters by changing
them sequentially, requiring that the slope of the Fe {\sc i} abundances with
respect to the excitation potential (for \Teff) or EW (for \Vt)
was almost zero and that the difference between the mean iron abundances
obtained from the Fe {\sc i} and Fe {\sc ii} lines did not exceed 0.03 dex (for \logg).
This procedure made it possible to determine \Teff~ with an accuracy (1 $\sigma$)
exceeding 60 K and both surface gravities and microturbulent velocity with
an accuracy of about 0.1 dex.
We employed the iron abundance obtained from the Fe {\sc i} lines as the
metallicity [Fe/H] of the studied stars.
The neutral iron lines were used due to their large
number in the spectra with known reliable atomic parameters.

\begin{table*}
\caption{Mean difference and standard deviation between atmospheric parameters derived in this study and those obtained by other authors ($n$  is the number of stars in common).}
\label{tabcomp2}
\begin{tabular}{rrrrl}
\hline
\hline
$\Delta$\Teff, K & $\Delta$\logg & $\Delta$[Fe/H]&  n      & source\\
\hline
--22 $\pm$ 36  & --0.02$\pm$0.12 & --0.04 $\pm$0.04  & 9    &  \protect\cite{brewer:16} \\
--10 $\pm$  28 & --0.04$\pm$0.08& --0.02 $\pm$0.10 & 3    & \protect\cite{takeda:05}, \protect\cite{takeda:07} \\
25 $\pm$  35   &  0.04$\pm$0.06 & --0.05 $\pm$0.07&  7   &  \protect\cite{ramirez:12} \\
10 $\pm$  84   & --0.01$\pm$0.08 & --0.05 $\pm$0.12&  11   &  \protect\cite{petigura:11} \\
--48 $\pm$ 55 & --0.11$\pm$0.12 & --0.06 $\pm$0.03&  6   &  \protect\cite{gonzalez:10} \\
29 $\pm$  44   &  0.05$\pm$0.08 & --0.04 $\pm$0.03 &  7   &  \protect\cite{santos:13} \\
38 $\pm$  64   &  0.00$\pm$0.08 & --0.06 $\pm$0.06&  7   &  \protect\cite{sousa:15} \\
--9 $\pm$  44   & --0.03$\pm$0.10& --0.08 $\pm$0.07&  7   &  \protect\cite{aguilera:18} \\
\hline
\end{tabular}
\end{table*}

A comparison of our parameter values with the data reported in the papers used
for sampling the target stars is shown in Table \ref{parplan}. 
The difference between atmospheric  parameters derived in this study
and those earlier obtained by other authors (for all stars in common)
is shown in Table \ref{tabcomp2}. As results from the comparison of our \Teff, \logg, and [Fe/H] with
those determined by other authors, provided that the star HD 12484 is
excluded \citep{petigura:11}, the standard deviation of the difference
does not exceed $\pm$70 K, $\pm$0.12, and $\pm$0.10, respectively.
The mean difference and average standard deviation are 1.6$\pm$49 K,
--0.02 $\pm$0.09, and --0.05 $\pm$0.07, respectively.
We also compared the parameter definitions reported by other
authors for each investigated star.
The differences between our parameter determinations and average
values reported by other authors, 
but without measurements
from the SOPHIE exoplanet consortium, are shown in
Table \ref{compind} for each star of our sample.
Such a comparison was made for 21 stars (as there
were no parameter determinations reported by other
authors for the remaining four stars). The differences
and standard deviations resulting from the comparison
of our data with the average values of \Teff, \logg,
and [Fe/H] reported by other authors are --9.9$\pm$38,
0.01$\pm$0.10, 0.06$\pm$0.06, respectively.
The comparison of our data and those from literature 
is very good, as illustrated in
Table \ref{tabcomp2} and Table \ref{compind}.

\begin{table*}
\caption{Comparison of stellar parameters obtained in this study and
by other authors for some individual stars of our sample
("n" is the number of measurements of \Teff, \logg, and [Fe/H]).}
\label{compind}
\begin{tabular}{lcrrrrrrrrrrrr}
\hline
\hline
HD/HIP& \Teff, K& \logg  & [Fe/H] & $<$\Teff$>$, K& $\sigma$, $\pm$ & $<$\logg$>$ & $\sigma$, $\pm$& $<$[Fe/H]$>$ & $\sigma$, $\pm$& n&$\delta$ \Teff, K  &$\delta$ \logg  & $\delta$ [Fe/H]\\
\multicolumn{1}{c}{}&\multicolumn{3}{c}{(our data)}&\multicolumn{6}{c}{(mean from literature)}&\multicolumn{4}{c}{}\\
\hline
12484   &       5831    &       4.40    &       0.05    &       5870    &       47      &       4.47    &       0.08    &       0.12    &       0.02    &       5,5,5   &      --39     &      --0.07   &      --0.07   \\
13908   &       6150    &       4.00    &       0.05    &       6199    &       112     &       4.01    &       --      &      --0.08   &       --      &       3,1,1   &      --49     &      --0.01   &       0.13    \\
16175   &       5977    &       4.30    &       0.37    &       5976    &       52      &       4.16    &       0.10    &       0.30    &       0.04    &       11,9,9  &       1       &       0.14    &       0.07    \\
17674   &       5904    &       4.28    &      --0.15   &       5893    &       22      &       4.09    &       0.09    &      --0.18   &       0.05    &       8,4,5   &       11      &       0.19    &       0.03    \\
24040   &       5800    &       4.25    &       0.23    &       5800    &       61      &       4.24    &       0.10    &       0.17    &       0.03    &       16,14,13&       0       &       0.01    &       0.06    \\
29021   &       5550    &       4.30    &      --0.18   &       5594    &       46      &       4.36    &       0.05    &       -0.22   &       0.07    &       5,3,3   &      --44     &      --0.06   &       0.04    \\
35759   &       5917    &       4.10    &      --0.04   &       5998    &       --      &       --      &       --      &       --      &       --      &       1       &      --81     &       --      &       --      \\
89307   &       5975    &       4.50    &      --0.05   &       5944    &       46      &       4.39    &       0.13    &      --0.15   &       0.05    &       16,15,14&       31      &       0.11    &       0.10    \\
113337  &       6750    &       4.30    &       0.14    &       6696    &       90      &       4.40    &       0.30    &       0.17    &       --      &       3,2,1   &       54      &      --0.10   &      --0.03   \\
141399  &       5515    &       4.35    &       0.47    &       5570    &       44      &       4.21    &       0.04    &       0.34    &       0.02    &       2,2,2   &      --55     &       0.14    &       0.13    \\
143105  &       6280    &       4.30    &       0.01    &       6262    &       42      &       --      &       --      &       --      &       --      &       2       &       18      &       --      &       --      \\
150706  &       5950    &       4.50    &       0.05    &       5930    &       34      &       4.48    &       0.06    &      --0.04   &       0.04    &       15,13,13&       20      &       0.02    &       0.09    \\
154345  &       5503    &       4.40    &      --0.05   &       5458    &       61      &       4.44    &       0.11    &      --0.13   &       0.05    &       21,19,17&       45      &      --0.04   &       0.08    \\
159243  &       6085    &       4.55    &       0.14    &       6078    &       10      &       4.44    &       --      &       --      &       --      &       2,1,0   &       7       &       0.11    &       --      \\
164595  &       5725    &       4.40    &      --0.01   &       5726    &       27      &       4.40    &       0.06    &      --0.08   &       0.03    &       18,16,16&       --1     &       0.00    &       0.07    \\
191806  &       5807    &       4.30    &       0.36    &       -       &       --      &       --      &       --      &       --      &       --      &       --      &       --      &       --      &       --      \\
214823  &       5980    &       3.90    &       0.13    &       6008    &       148     &       3.90    &       0.04    &       0.09    &       0.07    &       4,2,2   &      --28     &       0.00    &       0.04    \\
219828  &       5850    &       4.10    &       0.18    &       5856    &       52      &       4.13    &       0.06    &       0.15    &       0.03    &       14,13,12&       --6     &      --0.03   &       0.03    \\
220842  &       5938    &       4.15    &      --0.10   &       5909    &       50      &       4.19    &       0.03    &      --0.24   &       0.04    &       5,3,3   &       29      &      --0.04   &       0.14    \\
221585  &       5562    &       4.10    &       0.28    &       5563    &       44      &       4.00    &       0.06    &       0.29    &       0.05    &       7,7,7   &       --1     &       0.10    &      --0.01   \\
222155  &       5665    &       3.95    &      --0.05   &       5735    &       77      &       3.99    &       0.07    &      --0.16   &       0.05    &       9,8,7   &       --70    &     --0.04    &       0.11    \\
HIP65407    &   5354    &       4.50    &       0.35    &       --      &       --      &       --      &       --      &       --      &       --      &       --      &       --      &       --      &       --      \\
HIP91258    &   5515    &       4.65    &       0.30    &       --      &       --      &       --      &       --      &       --      &       --      &       --      &       --      &       --      &       --      \\
HIP109384   &   5179    &       4.40    &      --0.24   &       5224    &       71      &       4.59    &       0.04    &      --0.28   &       0.07    &       2,2,2   &       --45    &      --0.19   &       0.04    \\
HIP109600   &   5525    &       4.65    &      --0.07   &       --      &       --      &       --      &       --      &       --      &       --      &       --      &       --      &       --      &               \\
\hline
\end{tabular}
\end{table*}

\section{Chemical composition}
\label{chem: comp}

\begin{figure}
\begin{tabular}{cccc}
\includegraphics[width=7cm]{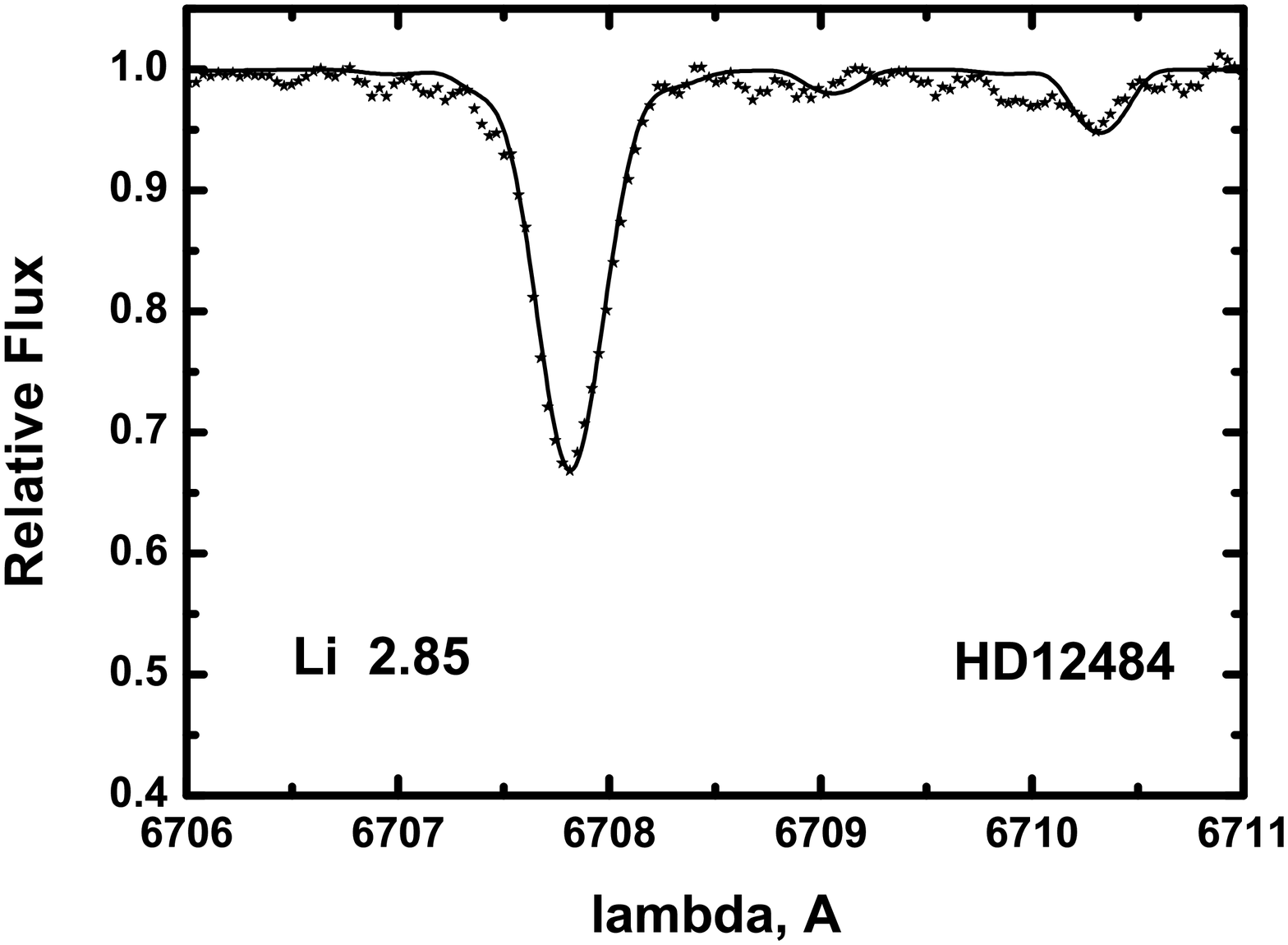}\\
\includegraphics[width=7cm]{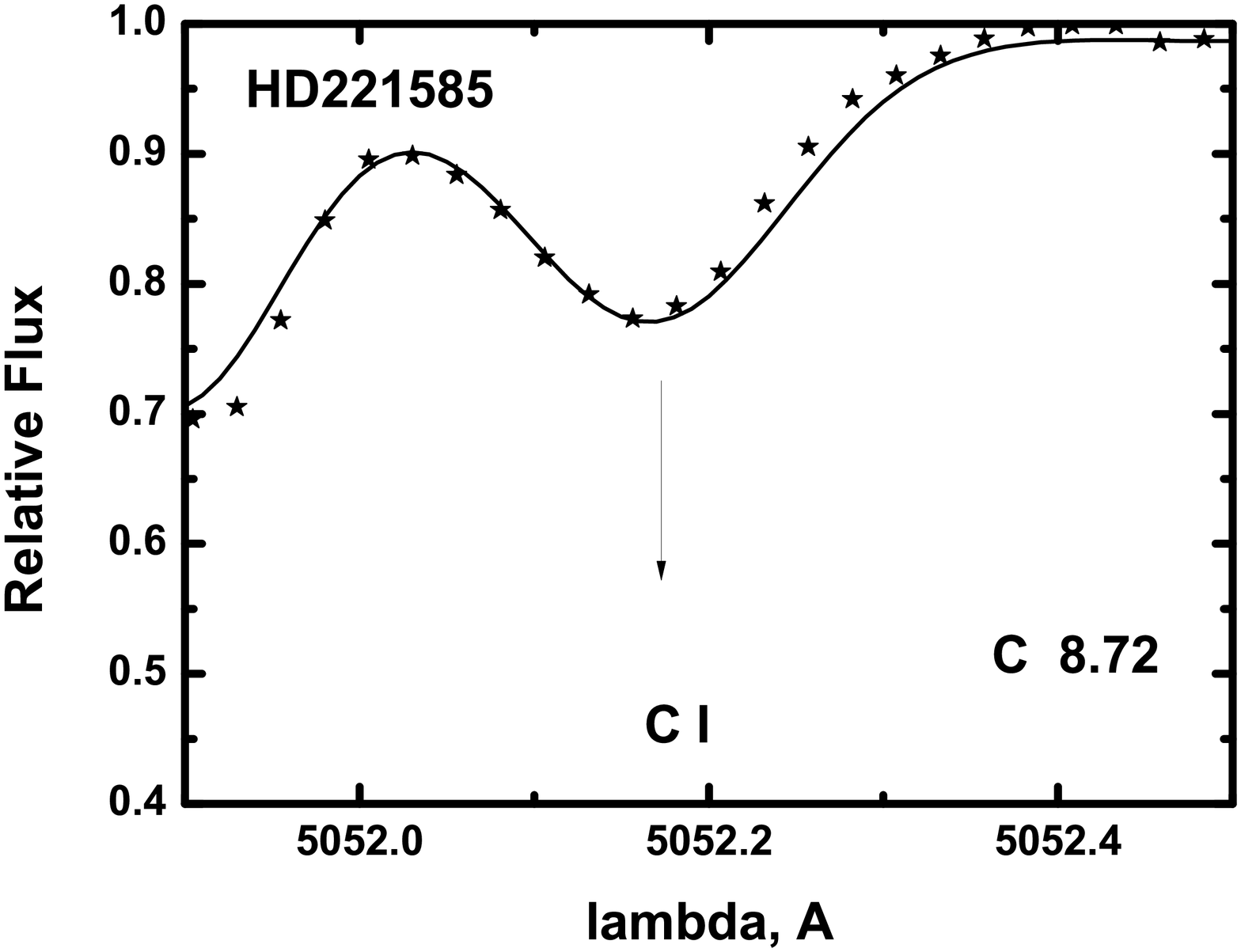}\\
\includegraphics[width=7cm]{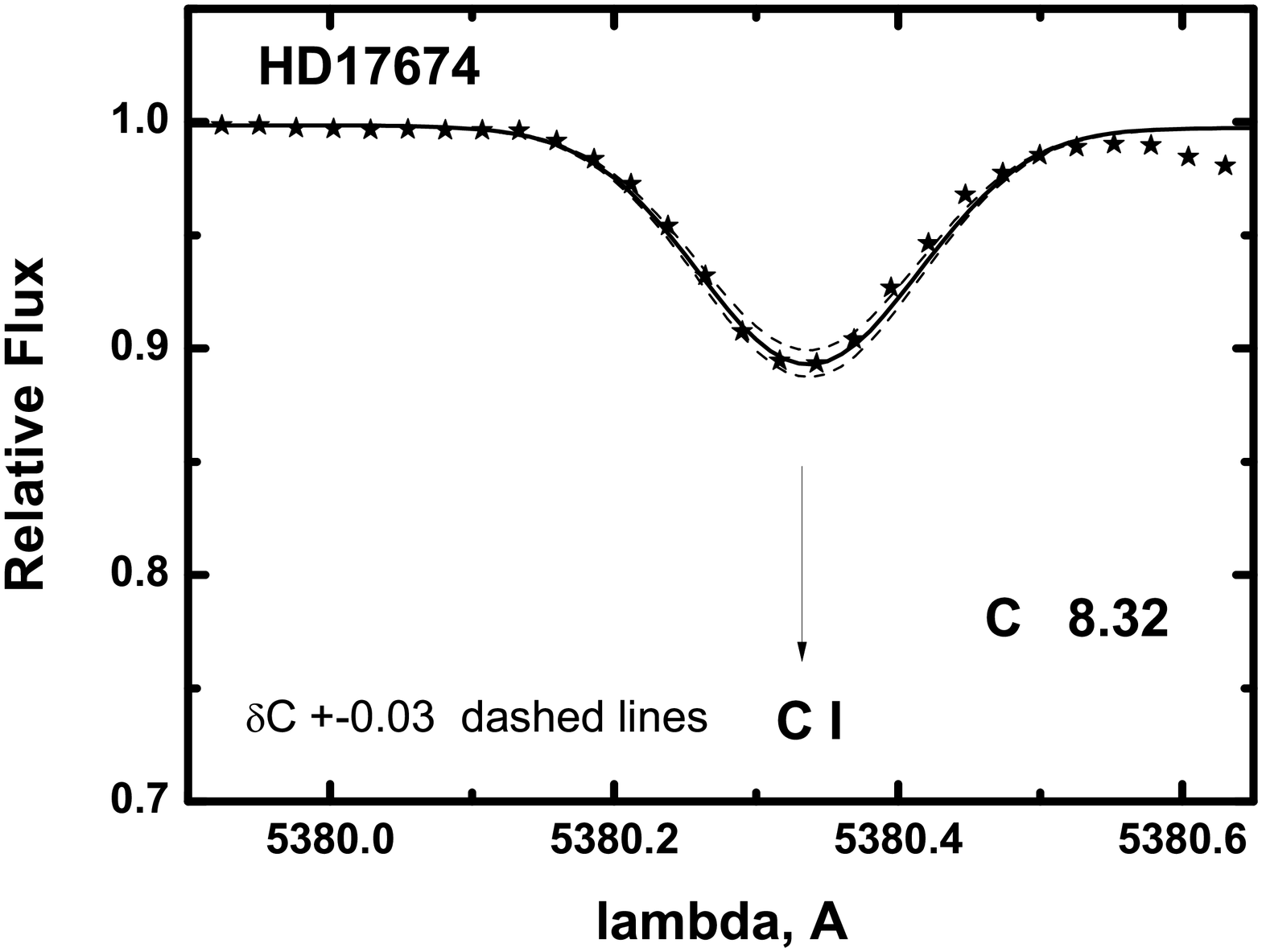}\\
\includegraphics[width=7cm]{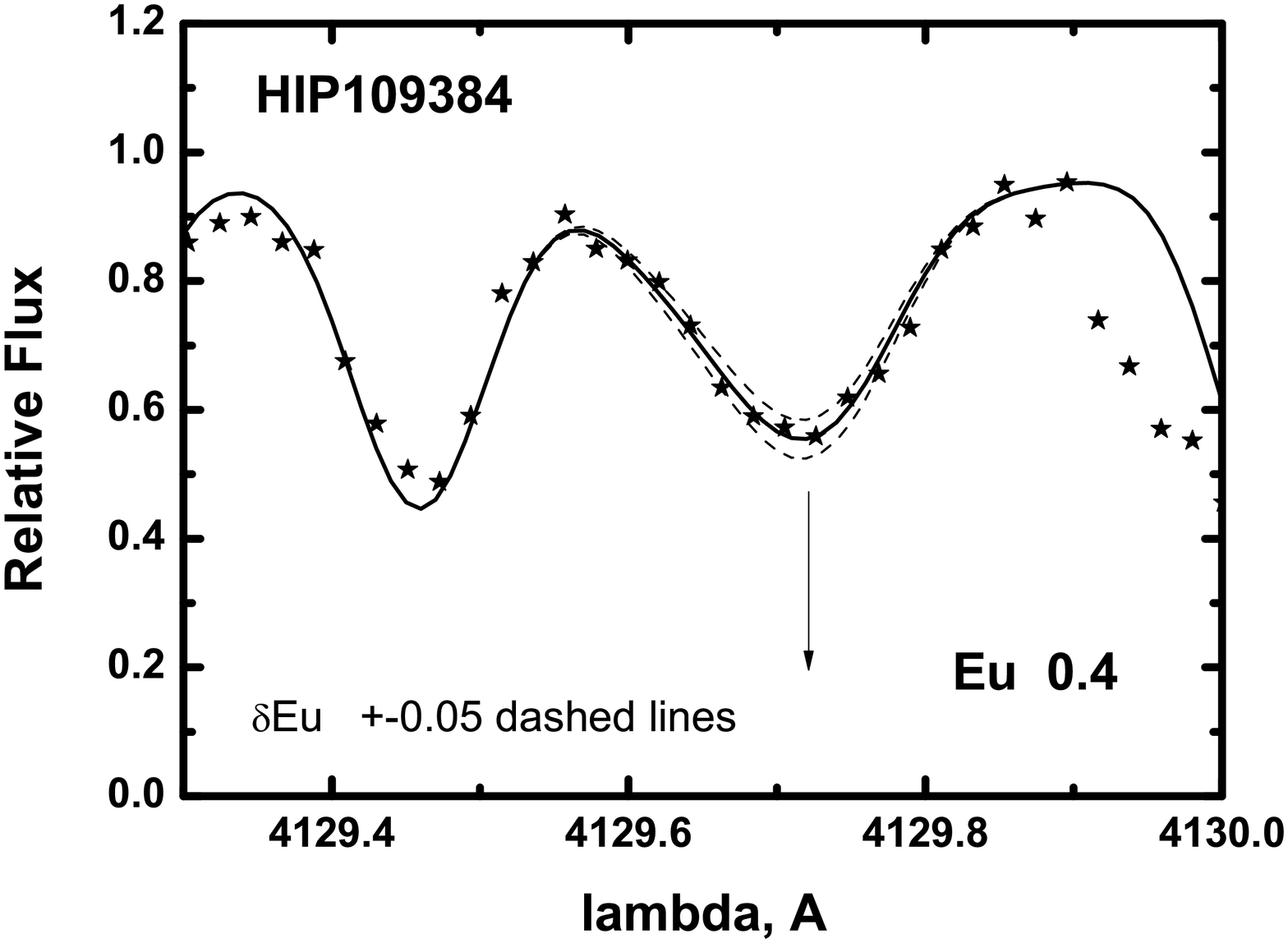}\\
\end{tabular}
\caption{Observed (asterisks) and calculated (solid line) spectra in the region
of Li I 6707 \AA~ line for HD 12484 (top panel), CI 5052 \AA~ line for HD 221585 and C I 5380 \AA~ line  
for HD17674 (middle panels), Eu II 4129 \AA~ line for HIP109384 (bottom panel).}
\label{c_synth}
\end{figure}

The elemental abundances were determined under the LTE condition
employing the grid of model atmospheres \citep{castelli:04}. The each
stellar model was obtained by interpolating on the grid in accordance
with the required combination of stellar parameters \Teff~ and \logg. 
The Kurucz WIDTH9 code was used to determine the LTE abundance  based
the equivalent  widths  of lines of Na, Mg, Al, S, Si, Ca, Ti, 
Cr, Fe, Ni, Zn, Y, Zr,  Ce, Pr, Nd and Sm elements. The list of used lines are presented 
in Table \ref{lines} in the Appendix (on line). We did not
use strong lines (with EWs $>$ 150 m\AA) due to noticeable effects of
damping. The latest modified spectral synthesis code STARSP \cite{tsymbal:96}
was employed  to calculate the line profiles 
for Li, C, O, Sc, V, Mn, Ba, and Eu. The determination of the abundances
of scandium, vanadium, manganese, barium and europium was carried out taking into account the
hyperfine structure (HFS). We used five lines (5432.5, 6013.5,
6016.7. 6021.7, 6021.8 \AA\AA) for manganese and  two lines
(4129.7 and 6645.1 \AA\AA) for europium with and the HFS data
from \cite{prochaska:00} and \cite{ivans:06}, respectively.
For scandium, vanadium, and barium we took the parameters of HFS from the last version of VALD database
\footnote{http://vald.astro.uu.se/~vald/php/vald.php}. We used five lines (5315.3, 5641.0, 5667.1, 
5684.2, 6245.6 \AA\AA) for scandium and four lines of vanadium (6081.4, 6090.2, 6242.8, 6243.1 \AA\AA).
For barium we explored three lines 5853.7, 6141.7, 6496.9 \AA\AA. The first two lines practically have an insignificant 
influence of the hyperfine structure; the 6496.9 \AA\ line requires taking into account the HFS \citep{korotin:15}. 
The damping constants for Ba lines are missing in VALD  
and we adopted the values from \cite{korotin:15}. 
All oscillator strengths were scaled by the solar isotopic ratios. 
Examples of some lines are shown in Fig. \ref{c_synth}.

To compute synthetic spectra, the following effects
were taken into consideration in addition to natural broadening of
atomic lines: 1) instrumental broadening through the resolving power (R)
of the spectrograph; 2) the projection of the rotation velocity \vsini\
obtained in the base papers and listed for each star in Table \ref{parplan};
and 3) large-scale atmospheric motions {$V_{\rm macro}$}, determined by
fitting the computed synthetic spectrum from several unblended iron lines,
factoring in \vsini.
Since for three stars, namely, HD 24040, HD 89307, and HD 154345 the \vsini\ values in the basic works
were missing, to calculate the synthetic
spectrum, we have computed their rotational velocities (\vsini) using a relation, calibrated by
\cite{queloz:98} and giving \vsini~ as a function of $\sigma_{RV}$,
the standard deviation of the ELODIE cross-correlation function
approximated by a Gaussian. We used the  average $\sigma_{RV}$
values from the ELODIE spectral archive of these stars.
 We obtained the following values: for HD 24040, \vsini = 2.3\kms,
for HD 89307, \vsini = 2.6\kms, and  for HD 154345, \vsini = 1.3\kms.
\cite{gonzalez:10} presented the \vsini\ for two of these stars, namely,
HD 89307, \vsini = 2.9\kms, and HD 154345, \vsini = 1.5\kms. All these values 
are consistent within the uncertainties and agree with those reported in
\cite{mishenina:12}. 

The determination of Li, C and O abundances based on the synthetic
spectra calculations will be discussed in more detail below.

The use of the LTE approximation in our analysis does
not introduce significant inaccuracies into the obtained values
of the abundances of elements, since the formation of the lines
used in this work is not subject to a noticeable deviation from
LTE in the considered range of stellar atmosphere parameters.
Thus, the NLTE corrections depend on \Teff\ and \logg, but their
values do not exceed 0.1 for the sodium lines
5688, 6154, 6160 \AA\AA\ \citep{korotin:14}; 0.05 for magnesium
lines of 4730, 5528, 5711, 6318, 6319.2, 6319.4 \AA\AA\ \citep{mishenina:04};
0.10-0.15 for aluminum 6696.0, 6698.6 \AA\AA\ lines \citep{andrievsky:08};
0.1 for barium 6141.71, 6496.90 \AA\AA\ lines, being insignificant for
5853.6 \AA~\citep{korotin:11}, the strong barium line at 4554 \AA~was
not included in our analysis; 0.05-0.10 for europium  4129.7 and
6645.1 \AA~lines \citep{mashonkina:00a}.
An examination of possible NLTE effects on the manganese lines based
on the abundance determinations from the lines of different multiplets
(which have also been used in this study) does not show any systematic
variations of abundances between these lines with the Mn abundances
being correct within an uncertainty of 0.1 dex \citep{mishenina:15}.
Besides, as shown, for instance, by \cite{adibekyan:17}, the NLTE
corrections to the Mg abundance determinations do not lead to any
significant changes in the results or conclusions drawn earlier
under LTE conditions.

 The elemental abundances determined relative to the solar ones with
the number of lines used for each element are listed in Table A2 in the
Appendix (on line) for all target stars.
The solar abundances were calculated by us earlier \citep{mishenina:17}
using the solar EWs, which were measured from the reflected spectra of the Moon and
asteroids obtained with the SOPHIE spectrograph. The oscillator
strengths,\loggf, of the lines are the same as the ones used in \cite{mishenina:17}
for the Sun, and are taken from the VALD \citep{kupka:99}.
A comparison of the adopted solar abundances with the solar elemental
abundances reported by \cite{asplund:09} was presented in
\cite{mishenina:17}. In this study, we use the same
list of lines as in \cite{mishenina:17} and the solar abundance
from \cite{asplund:09}.

\subsection{Li abundance}
\label{sec: abund}

The Li abundances log A(Li) (in unit of log A(H) = 12) in the investigated
stars were obtained by the $^{7}$Li~ line at 6707~\AA\, throught the spectral synthesis
method.  The line list around the Li I line is taken from  \cite{mishenina:97}. 
An example of spectral fitting for the star
HD~12484 is shown on the upper panel of Fig. \ref{c_synth}.
The study of departures from LTE in the case of the lithium line 6707~\AA\
was presented in \cite{lind:09}. For our target stars, the corrections
do not exceed $\pm$0.10 dex.

Our Li abundances and their comparison with those obtained by the 
authors are given in Table \ref{tabcompli}. For HD 16175 and 113337, the Li abundances 
have not been determined due to the distortion of the spectrum in the region of the lithium line. 

\begin{table*}
\caption{Our log A(Li) (column 1) compared with those obtained by other
authors:
2 - \protect\cite{aguilera:18}, 3 - \protect\cite{gonzalez:10}, 4 - \protect\cite{ramirez:12},
5 - \protect\cite{delgado:15}, 6 - \protect\cite{stonkute:20}.}
\label{tabcompli}
\begin{tabular}{rrrrrrrr}
\hline
HD&   HIP&   1 &        2 &     3 &     4&      5 & 6 \\
 \hline
12484&  9519&   2.93&   2.99&   --&     --&     --&     --  \\
13908&  10743&  1.38&   $<$1.35&        --&     --&     --&     --  \\
16175&  12191&  --&     2.68&   2.45&   --&     2.78&   --  \\
17674&  13291&  1.8&    --&     --&     --&     -- &    -- \\
24040&  17960&  1.1&    0.99&   --&     --&     --&     --  \\
29021&  21571&  1.0&    --&     --&     --&     -- &    -- \\
35759&  25883&  2.73&   --&     --&     --&     -- &    -- \\
89307&  50473&  2.25&   2.23&   2.1&    --&     2.31 &  -- \\
113337& 63584&  --&     --&     --&     --&     -- &    -- \\
141399& 77301&  0.7&    --&     --&     --&     -- &    -- \\
143105& 77838&  2.3&    --&     --&     --&     -- &    -- \\
150706& 80902&  2.6&    2.59&   2.46&   --&     --&     2.55  \\
154345& 83389&  $<$0.5& $<$-0.15&       $<$-0.29&       --&  --& $<$0.05  \\
159243& 85911&  2.6&    --&     --&     --&     -- &    -- \\
164595& 88194&  1.0&    0.78&   --&     0.75&   -- &    $<$1.05 \\
191806& 99306&  2.81&   --&     --&     --&     -- &    -- \\
214823& 111928& 1.5&    --&     --&     --&     -- &    -- \\
219828& 115100& 2.3&    2.28&   2.17&   --&     -- &    -- \\
220842& 115714& 1.16&   1.06&   --&     1.06&   -- &    -- \\
221585& 116221& 1.62&   --&     --&     --&     --  &   --\\
222155& 116616& 0.93&   1.05&   --&     0.58&   -- &    -- \\
--&     65407&  1.09&   --&     --&     --&     -- &    -- \\
-- &    91258&  0.0&    --&     --&     --&     -- &    --  \\
-- &    109384& 0.5&    0.34&   --&     0.34&   --&     --  \\
-- &    109600& 1.0&    --&     --&     --&     -- &    --  \\
\hline
\end{tabular}
\end{table*}

The results from different authors are in good agreement
with the exception of stars exhibiting low lithium abundances.
Such a discrepancy is mainly due to the difference in the quality of
the employed spectra, resolving power (R), signal-to-noise ratio (S/N)
and also to different stellar parameters.

\subsection{C and O abundances}

The carbon abundances were determined in LTE approximations and using the
following C I lines:  5052.17 \AA~(\loggf = --1.304),
5380.34 \AA\ (\loggf = --1.615),   6587.61 \AA\ (\loggf=--1.021).
Small NLTE corrections in the derived abundances lower
than 0.05 dex were applied for the lines listed above \citep{caffau:10}.
A comparison between the synthetic and observed spectra for
the CI lines at 5052\AA\ (HD~221585), and at 5380 \AA\ line
(HD~17674) is illustrated in Fig. \ref{c_synth} as an example.

\begin{figure}
\begin{tabular}{cccc}
\includegraphics[width=8cm]{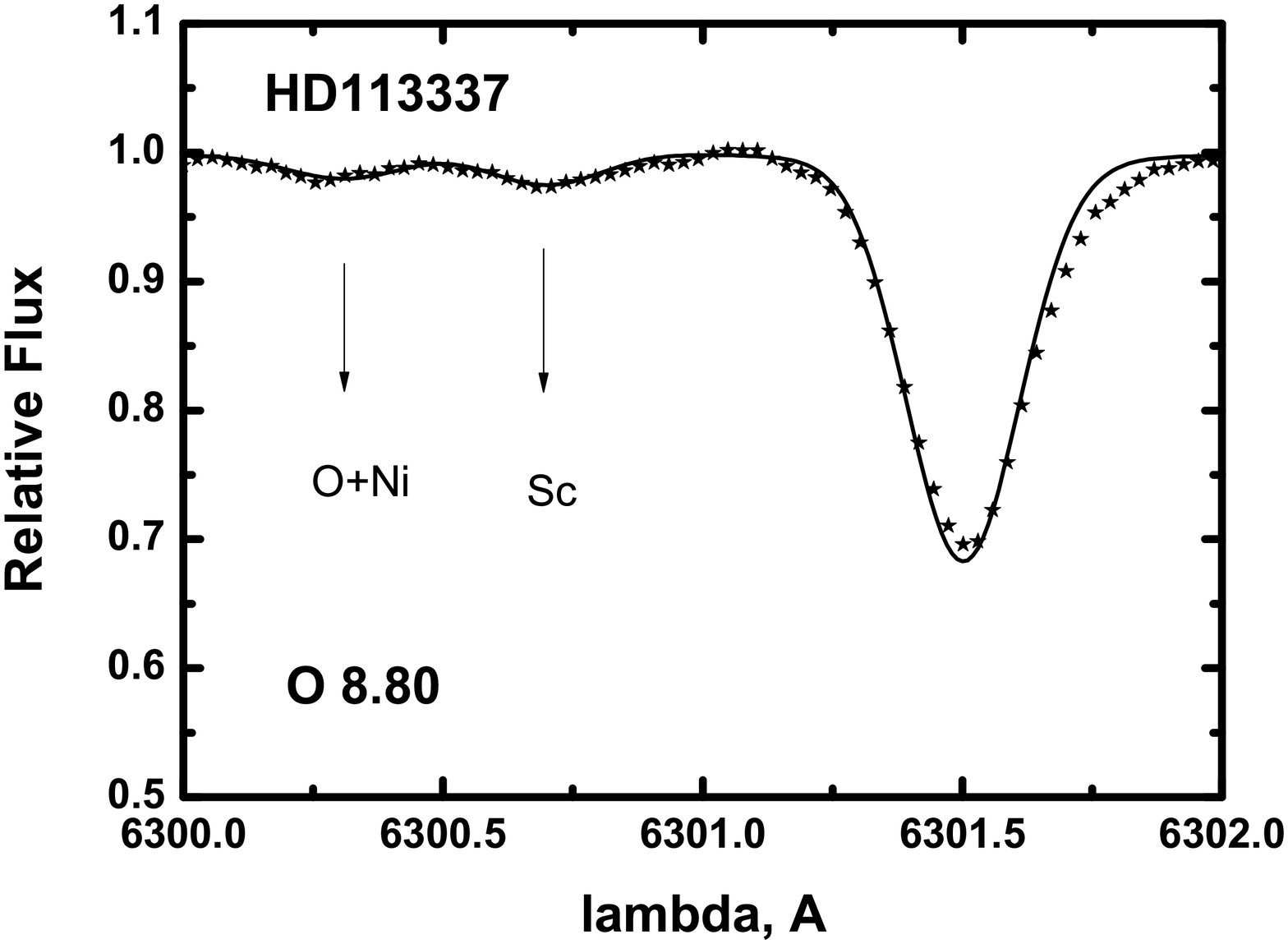}\\
\includegraphics[width=8cm]{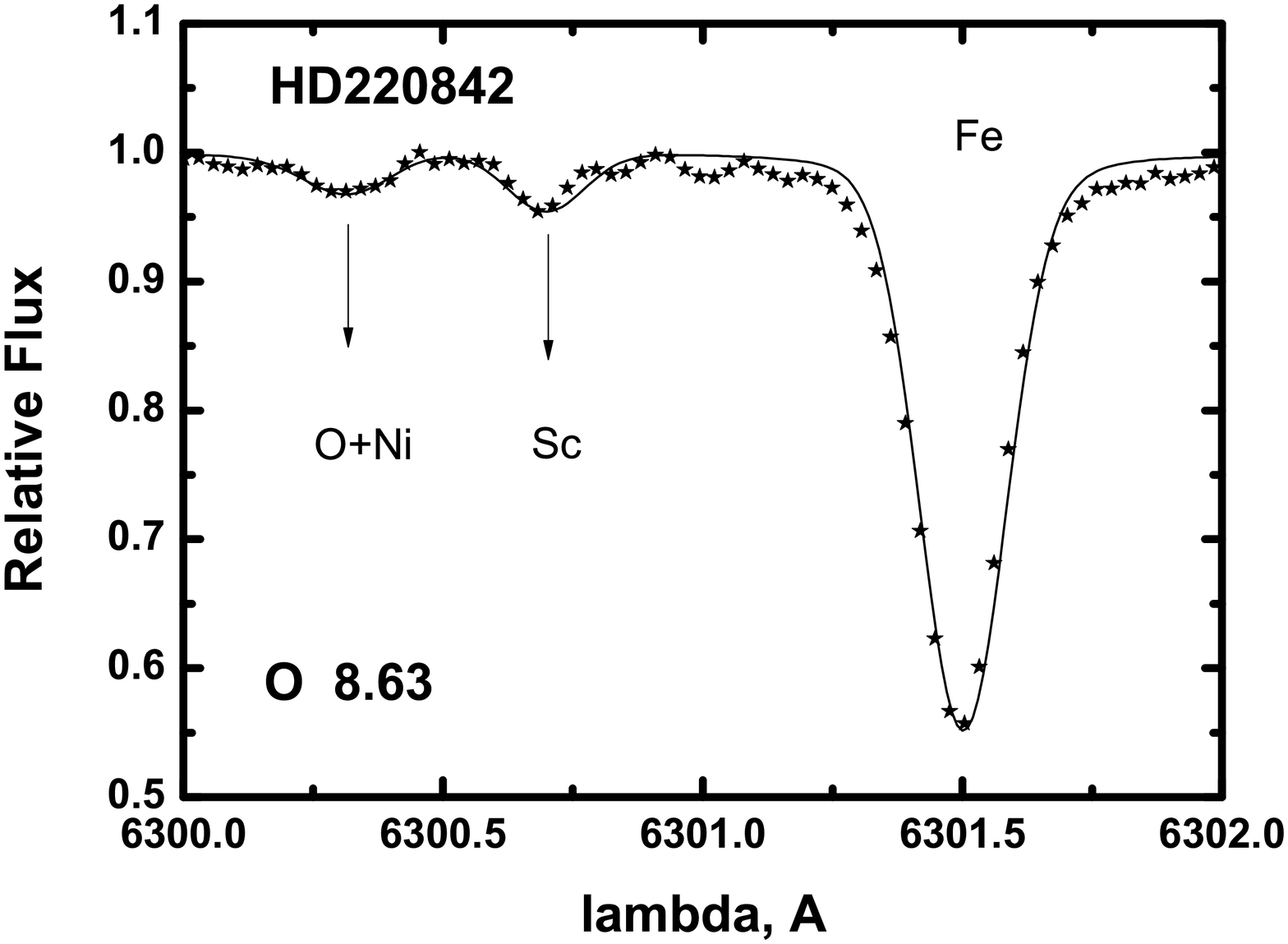}\\
\includegraphics[width=8cm]{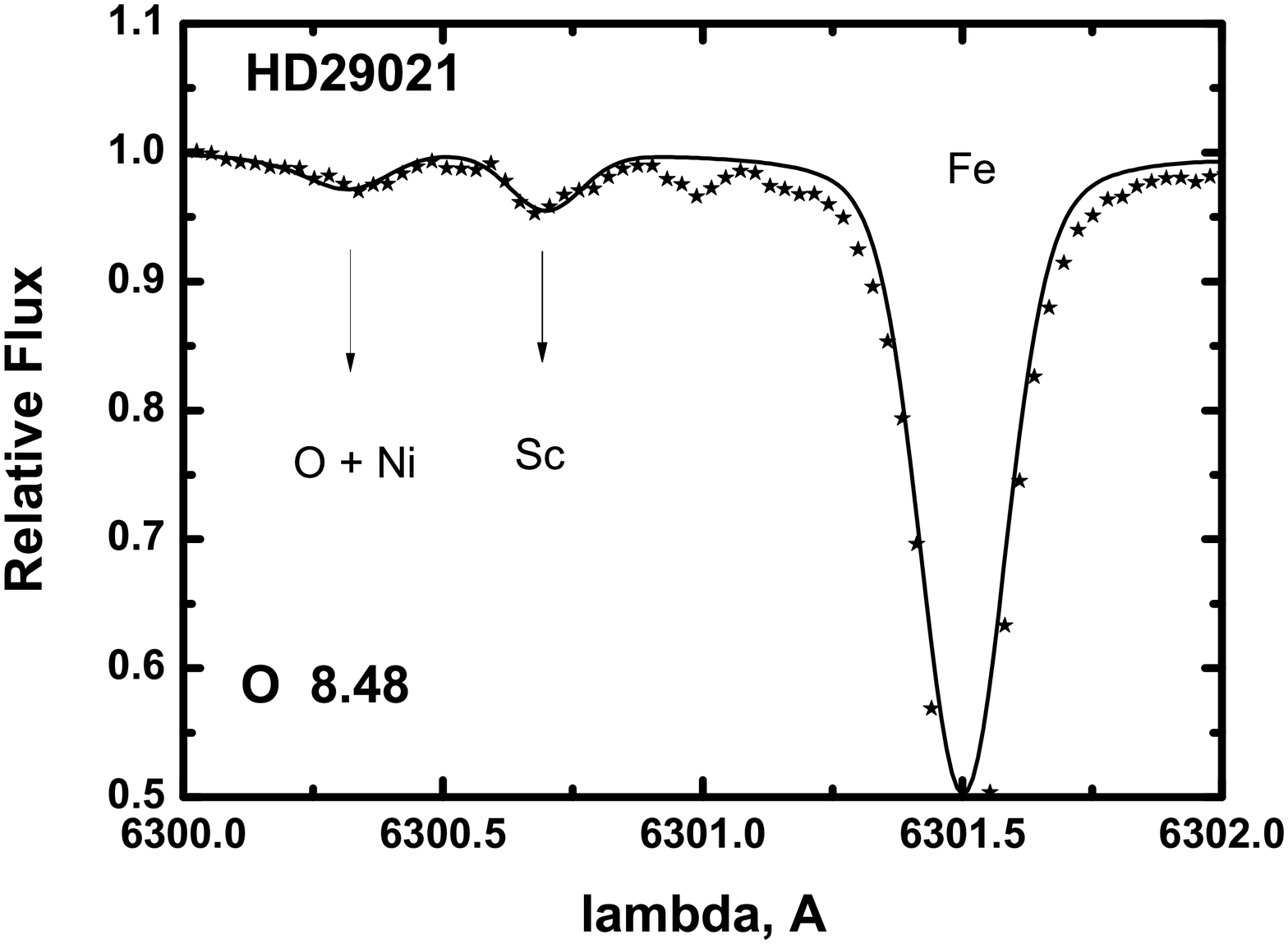}\\
\includegraphics[width=8cm]{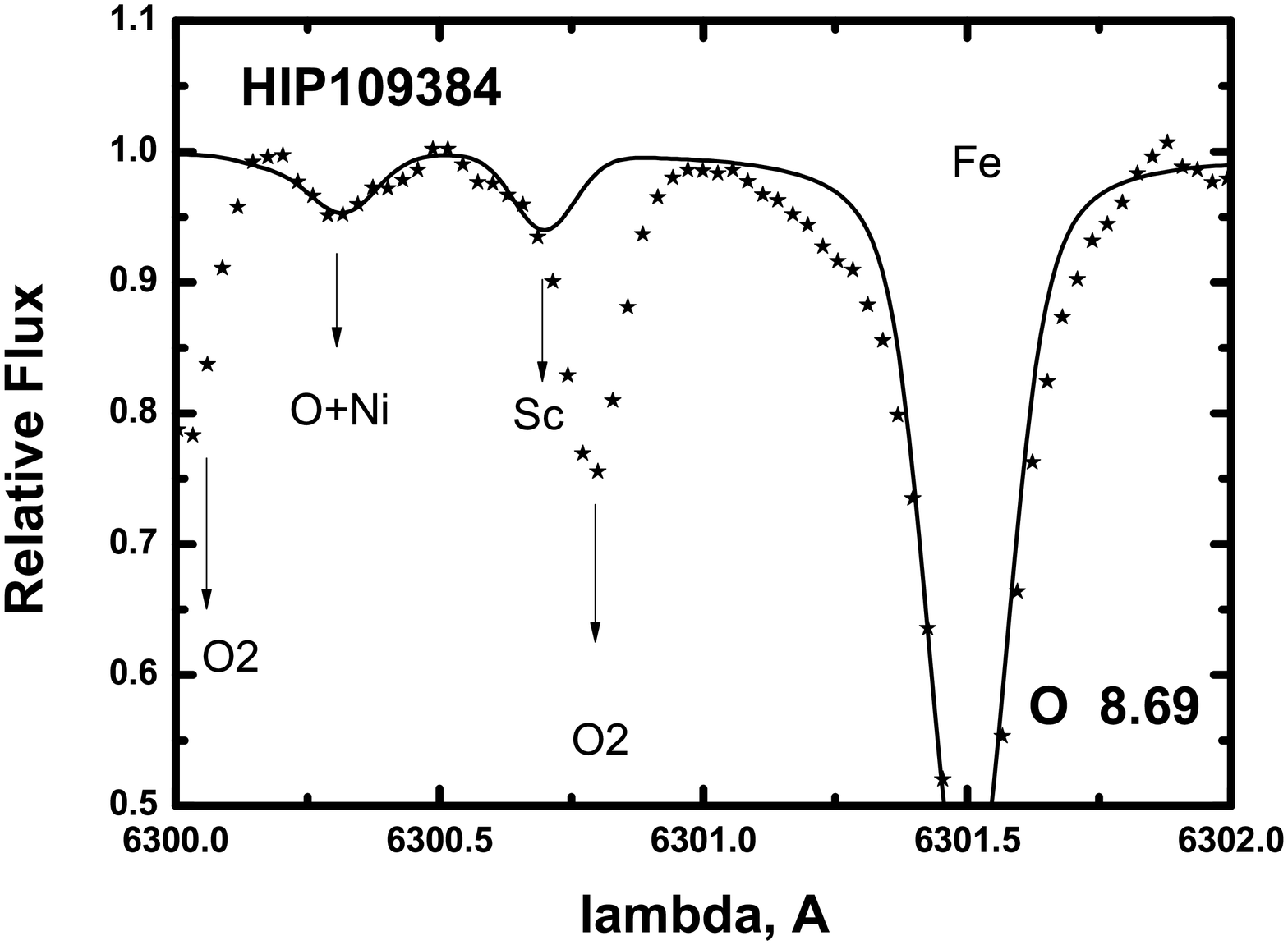}\\
\end{tabular}
\caption[]{Observed (asterisks) and calculated (solid line) spectra in the region
of [O I] 6300.3 \AA~ line for four stars with different stellar parameters.}
\label{o_synth}
\end{figure}

\begin{figure}
\begin{tabular}{cc}
\includegraphics[width=8.4cm]{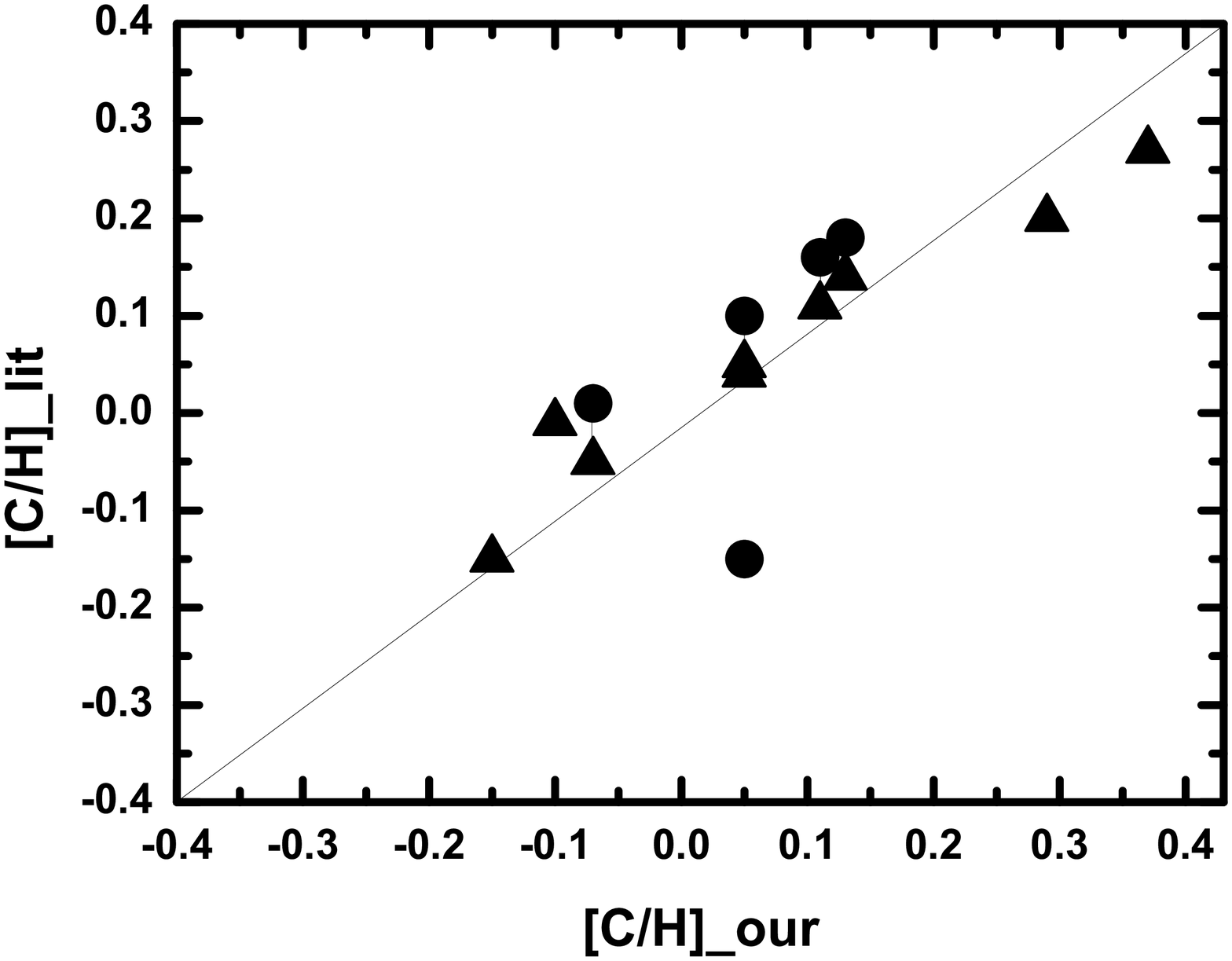}\\
\includegraphics[width=8.4cm]{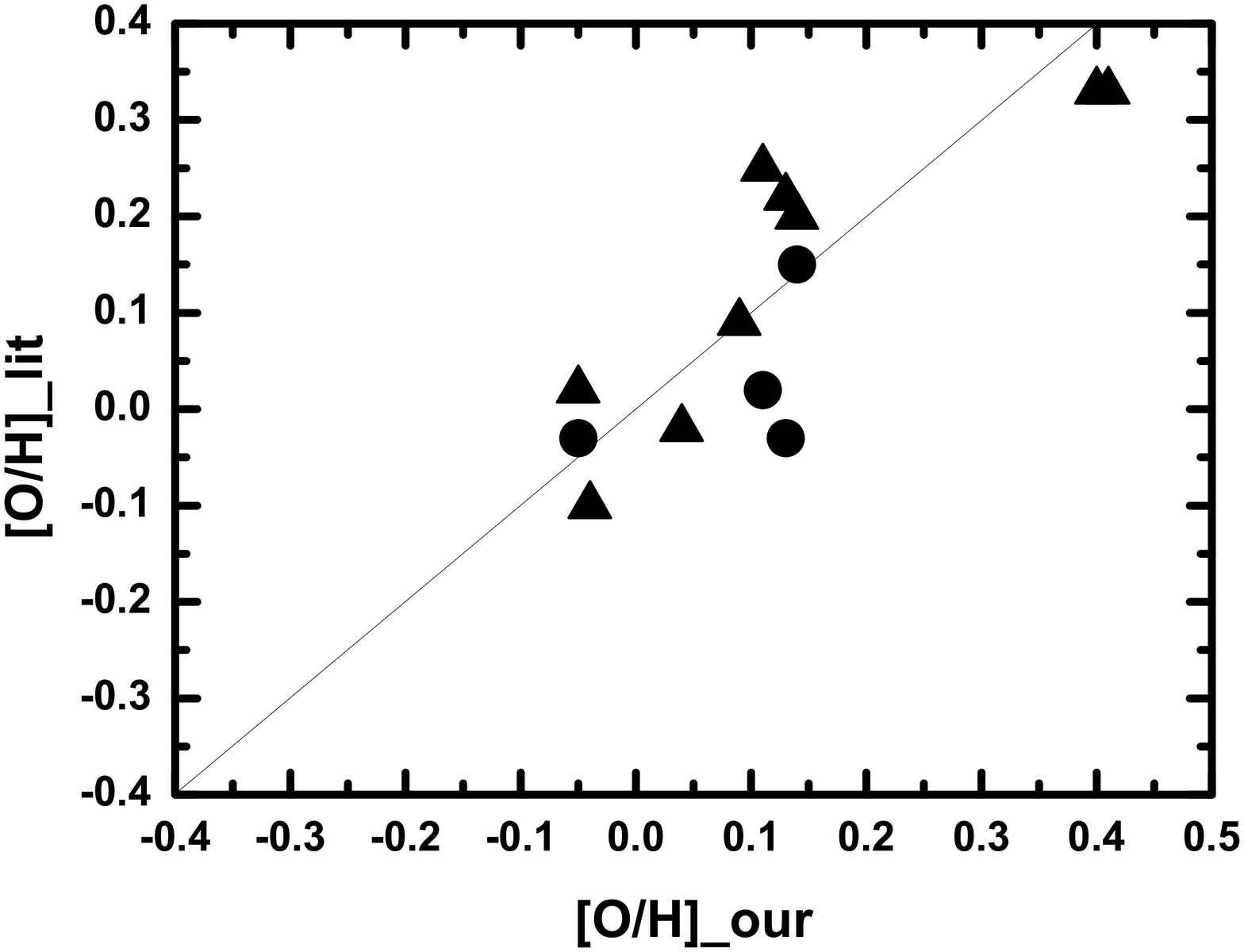}\\
\end{tabular}
\caption{Our [C/H] and [O/H] values compared with those reported
by \protect\cite{petigura:11} (circles) and \protect\cite{brewer:16} (triangles).}
\label{coh_comp}
\end{figure}

The LTE abundances of oxygen were determined using the list of lines from the
VALD database \citep[][version 2018]{kupka:99} by employing the [OI] line at
6300.30 \AA (\loggf = --9.818), which is unaffected by NLTE
\citep{mishenina:00}, taking into account the blending with Ni I
line at 6300.34 \AA, which is an isotopic splitting of $^{58}$Ni and
$^{60}$Ni, with the total oscillator strength for the two isotopes
\loggf= --2.11 \citep{johansson:03}.
As an example, in Fig. \ref{o_synth} we present the resulting
calculation of the oxygen line profile for four stars with different
parameters.
Our determinations of the C and O abundances and their comparison with the
data reported by other authors
are given in Table \ref{tabcompc} and plotted in Fig. \ref{coh_comp}.

\begin{table*}
\caption{Our C and O abundance determinations compared with those obtained by other authors:
 1 - \protect\cite{petigura:11}, 2 - \protect\cite{brewer:16}.}
\label{tabcompc}
\begin{tabular}{rrrrrrrrrrr}
\hline
  HD   &     HIP&   [O/H]&  [C/H]&  C/O& [O/H]&  [C/H]&  C/O&     [O/H]&       [C/H]&     C/O  \\
          &      &    our &   &    & 1   &   &   & 2  &   &  \\
\hline
12484&  9519&   0.09&   0.05&   0.50&    --  &--0.15& --&     0.09&   0.05&   0.50  \\
13908&  10743&  0.01&  --0.08&  0.45&   --&     --&     --&     --&     --&     --  \\
16175&  12191&  0.40&   0.29&   0.43&   --&     --&     -- &    0.33&   0.20&   0.41   \\
17674&  13291&  0.01&  --0.11&  0.48&   --&     --&     --&     --&     --&     --  \\
24040&  17960&  0.14&   0.13&   0.54&   0.15&   0.18&   0.59&   0.20&   0.14&   0.48 \\
29021&  21571& --0.21& --0.18&  0.59&   --&     --&     --&     --&     --&     --  \\
35759&  25883&  0.16&   0.04&   0.42&   --&     --&     --&     --&     --&     --  \\
89307&  50473&  0.08&  --0.16&  0.32&   --&     --&     --&     --&     --&     ---  \\
113337& 63584&  0.11&   0.06&   0.49&   --&     --&     --&     --&     --&     --  \\
141399& 77301&  0.41&   0.37&   0.40&    --&     --&     --&     0.33&   0.27&   0.48  \\
143105& 77838&  0.09&   0.10&   0.56&   --&     --&     --&     --&     --&     --  \\
150706& 80902&  0.04&  --0.10&  0.40&    --&     --&     --&    --0.02& --0.01&  0.56  \\
154345& 83389& --0.04& --0.15&  0.43&   --&     --&     --&    --0.10& --0.15&  0.49  \\
159243& 85911&  0.06&   0.04&   0.53&   --&     --&     --&     --&     --&     -- \\
164595& 88194& --0.05& --0.07&  0.53&  --0.03&  0.01&   0.60& 0.02&  --0.05&  0.47  \\
191806& 99306&  0.28&   0.25&   0.51&   --&     --&     --&     --&     --&     --  \\
214823& 111928& 0.11&   0.05&   0.48&   0.02&   0.10&   0.66& 0.25&   0.04&   0.34  \\
219828& 115100& 0.13&   0.11&   0.53& --0.03&  0.16&   0.87& 0.22&   0.11&   0.43 \\
220842& 115714&--0.06& --0.13&  0.47&   --&     --&     --&     --&     --&     --  \\
221585& 116221& 0.28&   0.29&   0.62&   --&     --&     --&     --&     --&     --  \\
222155& 116616& 0.07&  --0.11&  0.36&   --&     --&     --&     --&     --&     --  \\
--&     65407&  0.36&   0.27&   0.45&   --&     --&     --&     --&     --&     --  \\
--&     91258&  0.31&   0.27&   0.48&   --&     --&     --&     --&     --&     --\\
-- &    109384& 0.00&  --0.13&  0.41&   --&     --&     --&     --&     --&     --  \\
---&    109600&--0.06& --0.02&  0.60&           --&     --&     --&     --&     --&     --   \\
\hline
\end{tabular}
\end{table*}

Our carbon abundances agree well with those obtained by \cite{brewer:16};
however, the data reported by \cite{petigura:11}, slightly differ
from our values, which are lower.

\subsection{Errors in abundance determinations}

The statistical (internal) errors of abundance determinations  were calculated as the quadratic sum of errors arising from the uncertainty of the atmospheric parameters.
To estimate this we derived the elemental abundances for three stars with different parameters, namely, HD~113337 (\Teff = 6750 K, \logg = 4.30, \Vt = 1.3 \kms, [Fe/H] = 0.14),
HD~150706 (\Teff = 5950 K, \logg = 4.50, \Vt = 0.8 \kms, [Fe/H] = 0.05),
HIP~109384 (\Teff = 5179 K, \logg = 4.40, \Vt = 0.8 \kms, [Fe/H] = --0.24)
while changing \Teff\ of +70 K, \logg\ of +0.12, and \Vt\
of = 0.10 \kms.
These values are somewhat different from the errors of the parameters
indicated (stated) by us ($\pm$60 K, $\pm$0.10dex, $\pm$0.10\kms), but
corroborated with the results of other authors obtained for individual
stars and for the stars in common in different works and they may reflect
systematic (external) errors that are typically due to use of different model atmospheres, 
codes, etc.
The effects of uncertainties in atmospheric parameters on the accuracy of
elemental abundance determinations for these stars is given in
Table \ref{errors}.
As it can be seen from Table \ref{errors} the total abundance
determination error for all elements does not exceed 0.10 dex.
A comparison with the results of other authors has shown that
external or systematic errors caused by different approaches
and methods in the determination of stellar parameters, are comparable
to our error values and should not significantly affect the results
we obtained.

\begin{table*}
\caption{Compilation of random errors due to uncertainties in atmospheric parameters.
 HD~11337 (\Teff = 6750 K, \logg = 4.30, \Vt = 1.3 \kms, [Fe/H] = 0.14);
 HD~150706 (5950/4.50/0.8/ 0.05);  HIP~109384 (5179/4.40/0.8/ --0.24).}
\label{errors}
\begin{tabular}{lrrrrrrrrrrrr}
\hline
\multicolumn{1}{c}{}&\multicolumn{4}{c}{HD~113337}&\multicolumn{4}{c}{HD~150706}&\multicolumn{4}{c}{HIP~109384}\\
\hline
  El  & $\delta$ \Teff+&$\delta$\logg+&$\delta$\Vt+&tot+& $\delta$ \Teff+  & $\delta$ \logg+ & $\delta$ \Vt+ &tot+&$\delta$ \Teff+&$\delta$\logg+ & $\delta$ \Vt+ & tot+\\
\hline
Li {\sc i} &  0.05&  0.01&  0.01&0.05&  0.06&--0.01&0.00  &0.06&  0.08&--0.01&0.00  &0.08\\
C  {\sc i} &--0.02&  0.03&  0.00&0.04&--0.04&  0.06&0.00  &0.07&--0.06& 0.03 &0.00  &0.07\\
O  {\sc i} &  0.03&  0.04&  0.00&0.05&  0.01&  0.04&0.00  &0.04&--0.02& 0.05 &0.00  &0.05\\
Na {\sc i} &  0.03&--0.01&--0.01&0.03&  0.03&--0.02&--0.01&0.04&  0.05&--0.02&0.00  &0.05\\
Mg {\sc i} &  0.02&  0.01&--0.01&0.03&  0.02&--0.01&--0.01&0.03&  0.02&--0.02&0.00  &0.03\\
Al {\sc i} &  0.03&  0.00&  0.00&0.03&  0.04&--0.01&--0.01&0.05&  0.03&--0.01&--0.01&0.03\\
Si {\sc i} &  0.02&  0.00&--0.02&0.03&  0.01&--0.01&--0.01&0.02&--0.01&--0.01&--0.01&0.02\\
S  {\sc i} &--0.01&  0.02&  0.00&0.02&--0.04&  0.05&0.00  &0.06&--0.04& 0.03 &0.00  &0.03\\
Ca {\sc i} &  0.05&--0.01&--0.02&0.05&  0.04&--0.03&-0.02 &0.05&  0.06&--0.03&--0.02 &0.07\\
Sc {\sc ii}&  0.02&  0.04&--0.01&0.05&  0.00&  0.05&--0.01&0.05&--0.01& 0.04 &--0.01&0.04\\
Ti {\sc i} &  0.05&  0.01&  0.00&0.05&  0.05&--0.02&--0.01&0.05&  0.08&--0.01&--0.02&0.08\\
Ti {\sc ii}&  0.01&  0.03&--0.01&0.03&  0.00&  0.05&--0.01&0.05&--0.01&  0.05&--0.01&0.05\\
V  {\sc i} &  0.05&  0.00&  0.00&0.05&  0.07&--0.01&0.00  &0.07&  0.09&  0.00&--0.01&0.09\\
Cr {\sc i} &  0.03&  0.00&--0.01&0.03&  0.04&--0.02&--0.01&0.05&  0.06&  0.00& 0.00 &0.06\\
Mn {\sc i} &  0.03&--0.01&--0.02&0.04&  0.04&--0.02&--0.02&0.05&  0.04&--0.02&--0.03&0.05\\
Fe {\sc i} &  0.04&  0.00&--0.01&0.04&  0.05&--0.01&--0.01&0.05&  0.03&  0.00&--0.01&0.03\\
Fe {\sc ii}&  0.00&  0.03&--0.02&0.04&--0.02&  0.05&--0.01&0.05&--0.05& 0.04 &--0.01&0.06\\
Ni  {\sc i}&  0.04&  0.00&  0.00&0.04&  0.04&--0.01& 0.00 &0.04&  0.01& 0.02 &--0.01&0.02\\
Zn {\sc i} &  0.04&  0.01&--0.02&0.05&  0.01&  0.01&--0.02&0.03&--0.02&  0.01&--0.02&0.03\\
Y  {\sc ii}&  0.02&  0.04&  0.00&0.04&  0.00&  0.04& 0.00 &0.04&  0.00& 0.05 &0.00  &0.05\\
Zr {\sc ii}&  0.02&  0.04&  0.00&0.04&  0.01&  0.05& 0.00 &0.05&  0.00& 0.05 &0.00  &0.05\\
Ba {\sc ii}&  0.05&  0.01&--0.05&0.07&  0.03&  0.01&--0.04&0.05&  0.02& 0.01 &--0.03&0.04\\
Ce {\sc ii}&  0.03&  0.04&  0.00&0.05&  0.01&  0.04& 0.00 &0.04&  0.00&  0.04&--0.01&0.04\\
Pr {\sc ii}&  0.04&  0.04&  0.00&0.06&  0.01&  0.04& 0.00 &0.04&  0.00&  0.04&--0.01&0.04\\
Nd {\sc ii}&  0.03&  0.04&  0.00&0.05&  0.02&  0.05& 0.00 &0.05&  0.01& 0.05 &--0.01&0.05\\
Sm {\sc ii}&  0.02&  0.04&  0.00&0.04&  0.01&  0.04&--0.01&0.04&  0.02& 0.05 &0.00  &0.05\\
Eu {\sc ii}&  0.03&  0.04&  0.00&0.05&  0.00&  0.04& 0.00 &0.04&--0.01& 0.04 &--0.01&0.04\\
\hline
\end{tabular}
\end{table*}

\section{Results and discussion }
\label{sec: result}

The abundance of iron ([Fe/H]) and lithium (log A(Li)), as well as the
carbon-to-oxygen (C/O) and magnesium-to-silicon (Mg/Si) ratios, are
important parameters in studying planet host stars in depth.
The ratio of volatiles to refractory elements may be an important key
to understand the formation of low-mass terrestial planets and giant-planet
cores.
A sample of stars with well-detected planetary systems, as that examined in this
study, may clarify the possible constraints for the connection
between these chemical parameters and the presence of planets.This
is discussed in more detail below.

\subsection{[Fe/H] versus \Teff}

To explain the metallicity excess in massive exoplanet host stars,
two hypotheses were proposed: the self-enrichment (or pollution) mechanism
and a primordial origin of metal overabundance. The primordial cloud is 
the most likely origin for the metal richness of giant planet host stars \citep{gonzalez:97}.
This hypothesis has been substantiated  by models of planet formation and
evolution, based on  exoplanet formation scenarios as such the core accretion
(CA) and tidal downsizing \cite[see in detail,][and references
therein]{adibekyan:19}. The first statistics about extrasolar
planets \citep{mortier:12} show  that the presence of giant
planets apparently depends on the stellar metallicity and mass.
For instance, such planets are more frequently observed around
metal-rich stars with an exponential increase in the planet occurrence
rate with metallicity; the occurrence rate (or frequency) for hot Jupiters
around metal-poor
stars is less than 1.0\% at one sigma and  frequency is higher for star
with [Fe/H]  $>$--0.7 than those with [Fe/H] $<$--0.7 \citep{mortier:12}.

\begin{figure}
\begin{tabular}{cc}
\includegraphics[width=8cm]{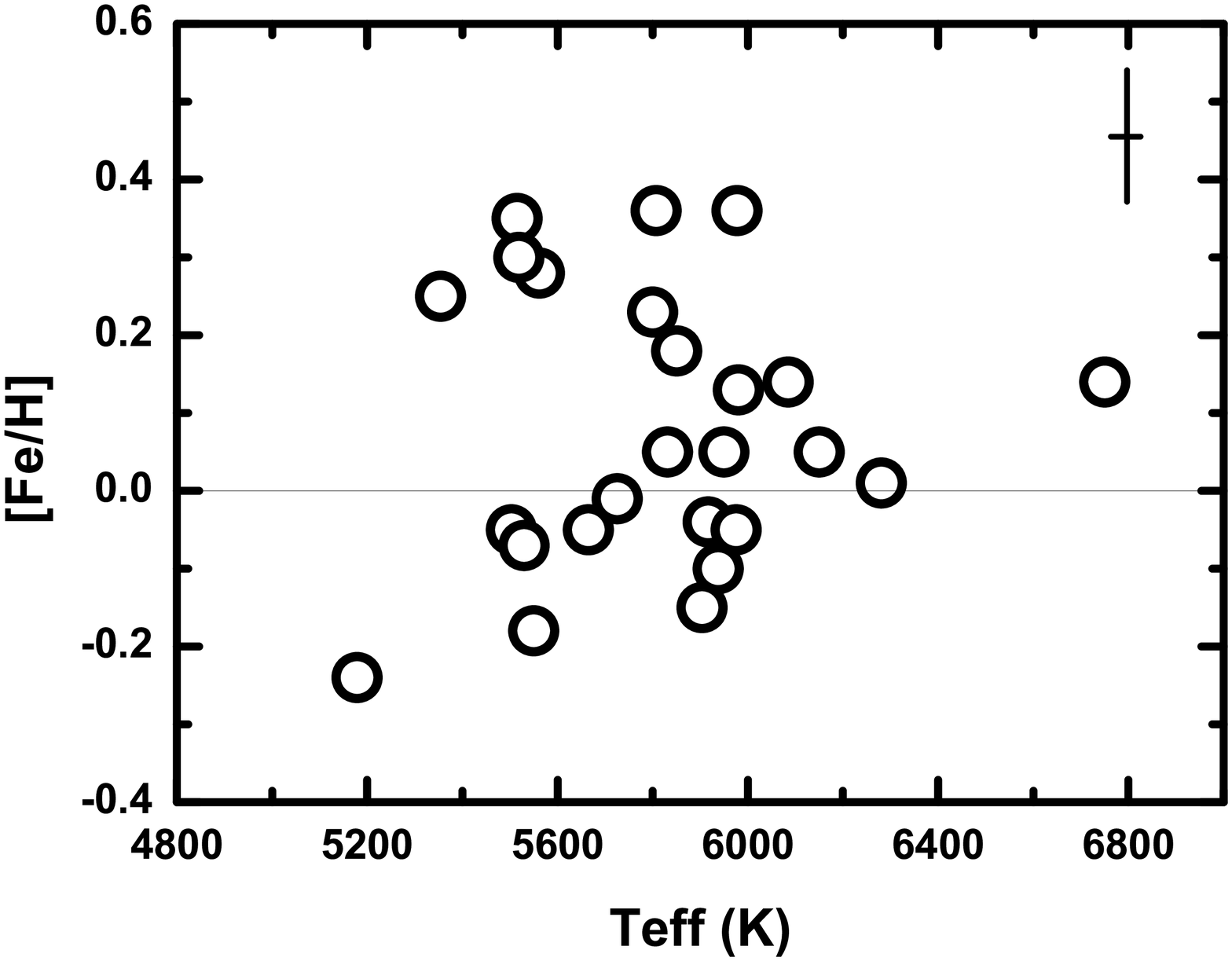}\\
\includegraphics[width=8cm]{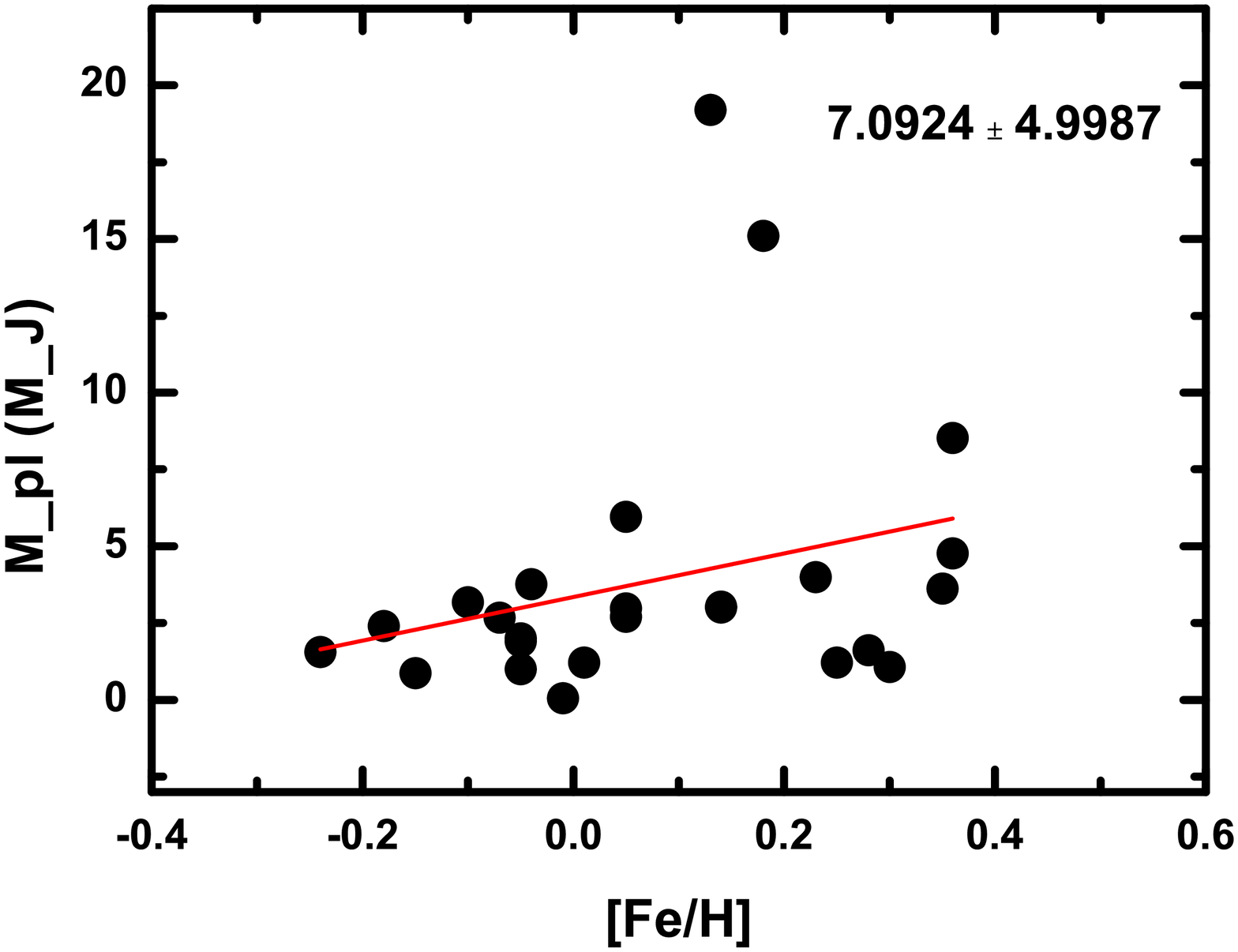}\\
\end{tabular}
\caption{[Fe/H] versus \Teff\ and  $M_{pl}$ versus [Fe/H] for our
target stars. The horizontal line corresponds to [Fe/H] = 0. 
The solid line in the lower panel represents a linear fit
to the data with the value of the slope indicated in the plot.}
\label{feh_t}
\end{figure}

The range of metallicities of the our sample is --0.3 $<$ [Fe/H] $<$ 0.4.
The correlation between [Fe/H] and the effective temperature \Teff\ for
our target stars is depicted in Fig. \ref{feh_t}.
As follows from the metallicity range of the studied stars 
massive planets in our
sample are located not only around stars with metallicities higher
than the solar one (in most cases), but also around stars with slight
metal deficiencies.
The bottom panel of Fig. \ref{feh_t} shows a positive correlation between planetary mass and metallicity.
The trend seem to be present (slope of 4.33$\pm$2.03) even if the two most massive planets with masses greater than 
13 $M_{J}$ (these objects can be Brown Dwarfs) are excluded.
As discussed in \cite{adibekyan:19}, there is a complex interdependence between planetary mass, stellar mass, and stellar metallicity, 
which makes difficult to identify the exact origin of the observed dependence of $M_{pl}$ on [Fe/H].

\subsection{Lithium abundance and exoplanets}

\begin{figure}
\begin{tabular}{ccc}
\includegraphics[width=7cm]{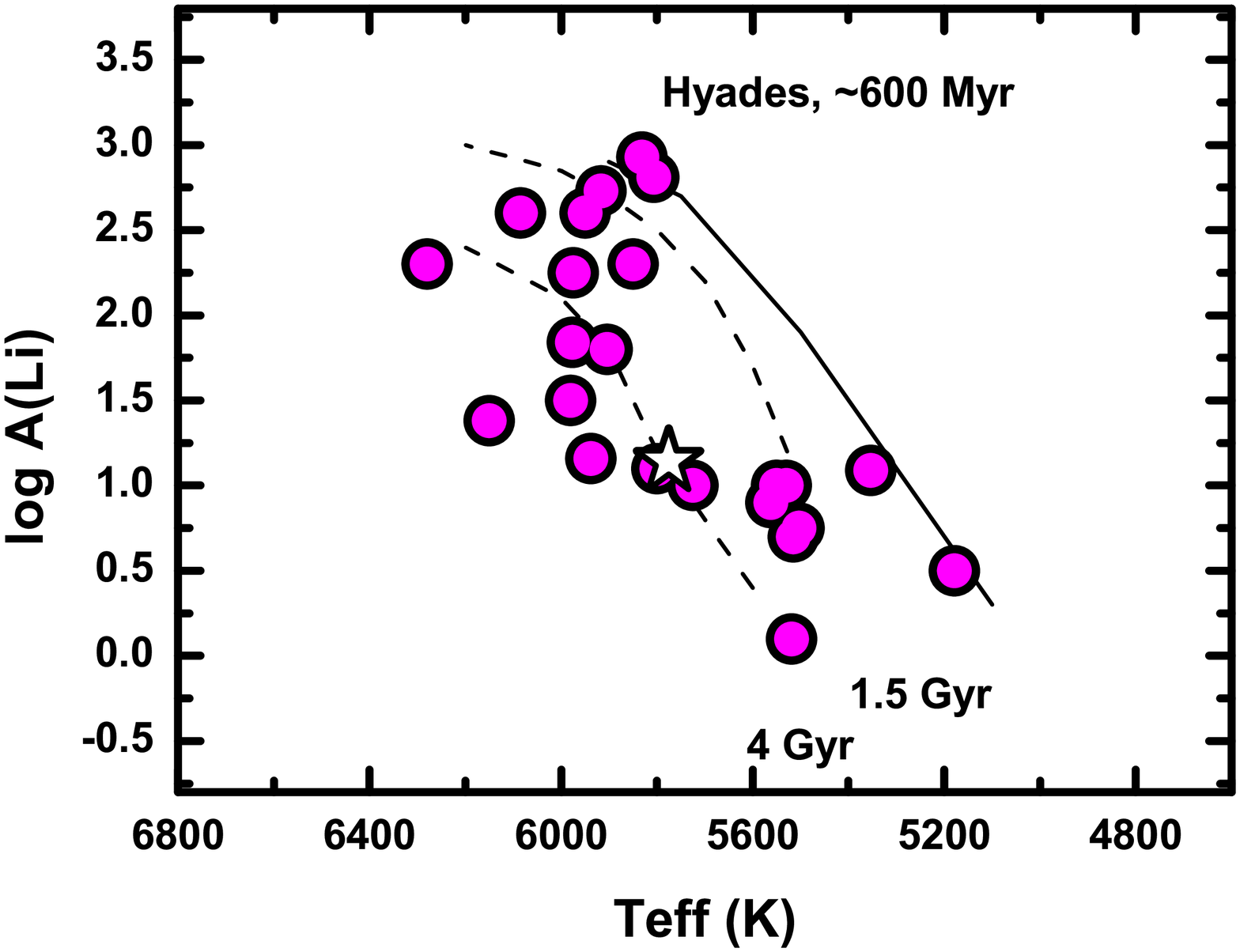}\\
\includegraphics[width=7cm]{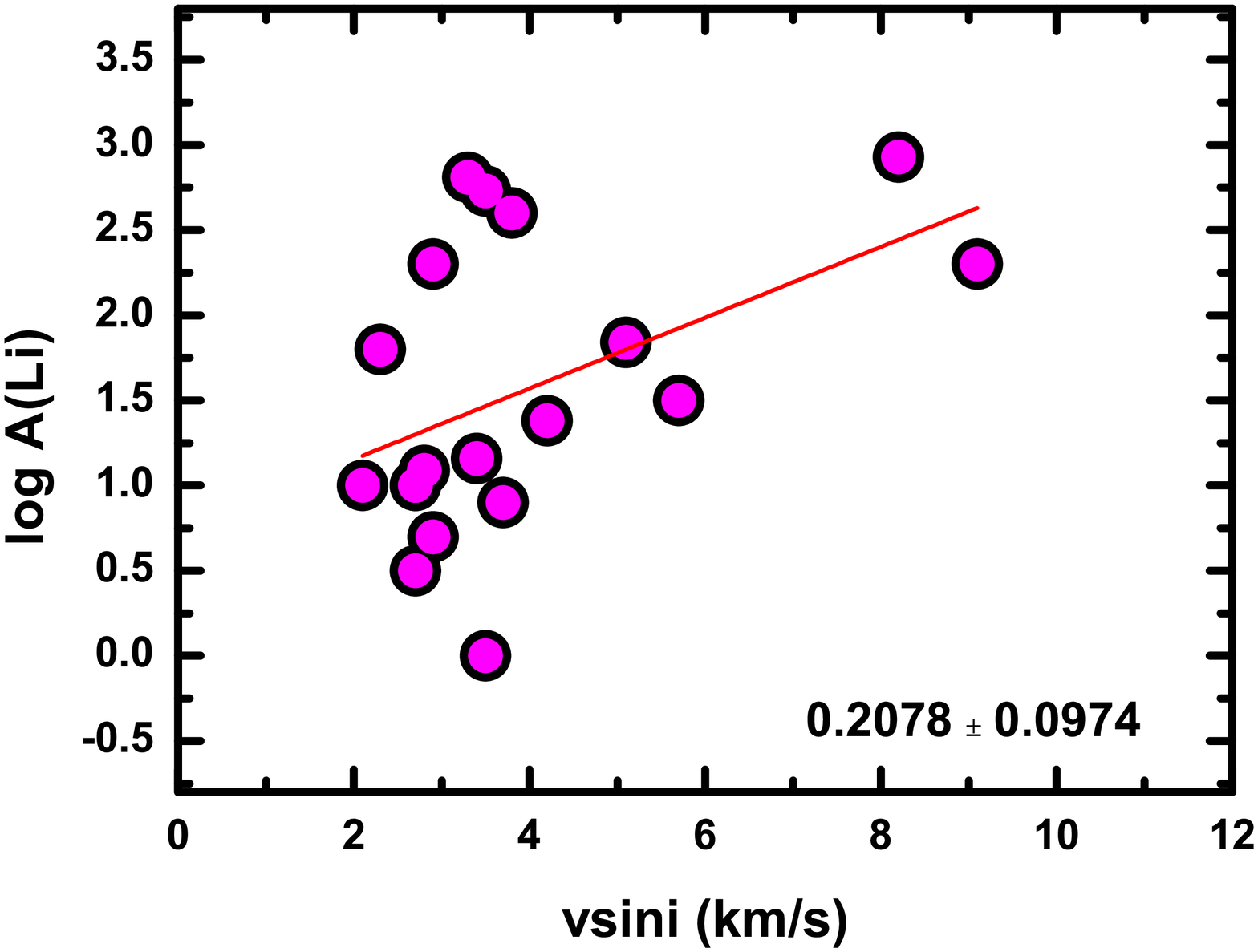}\\
\includegraphics[width=7cm]{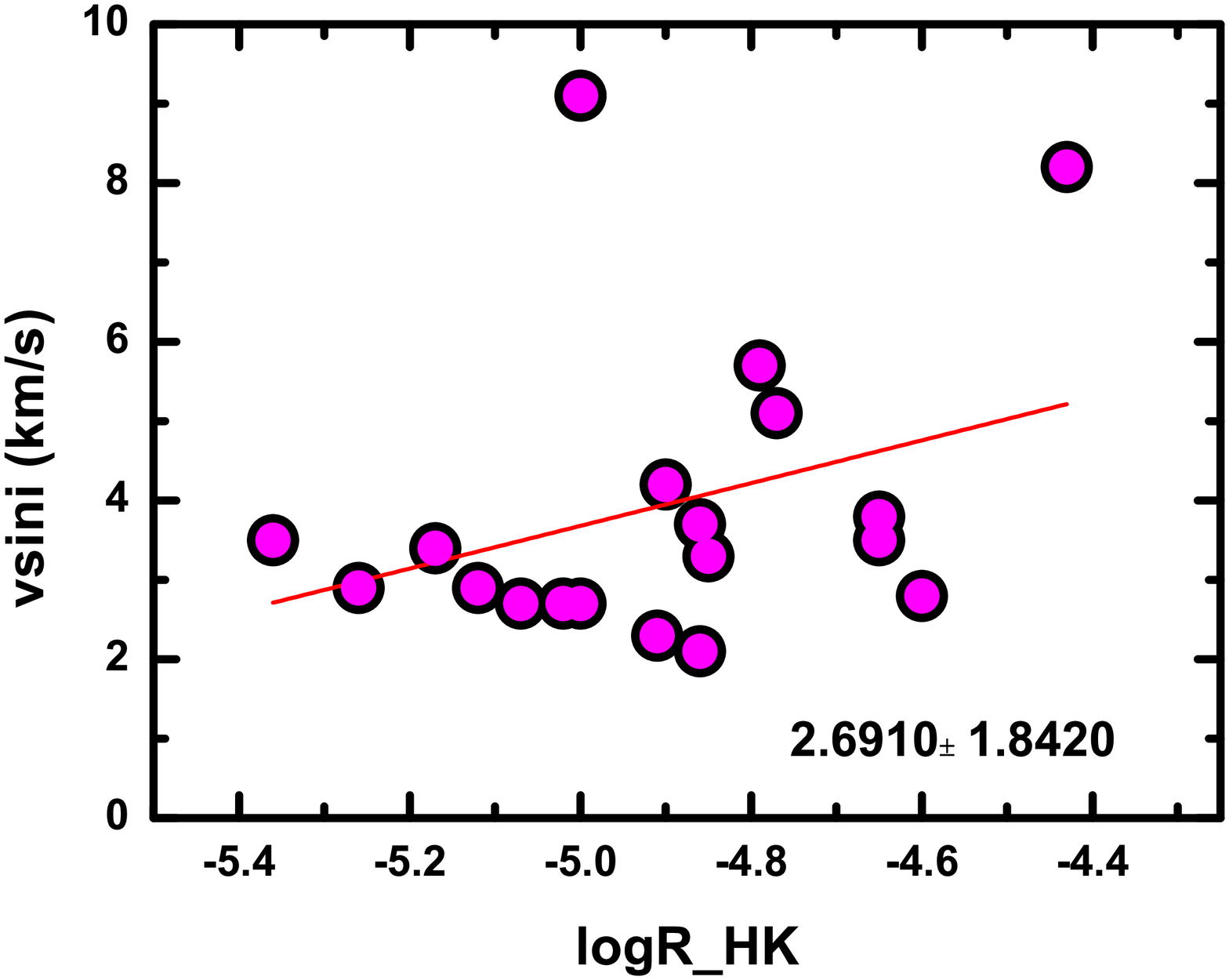}\\
\includegraphics[width=7cm]{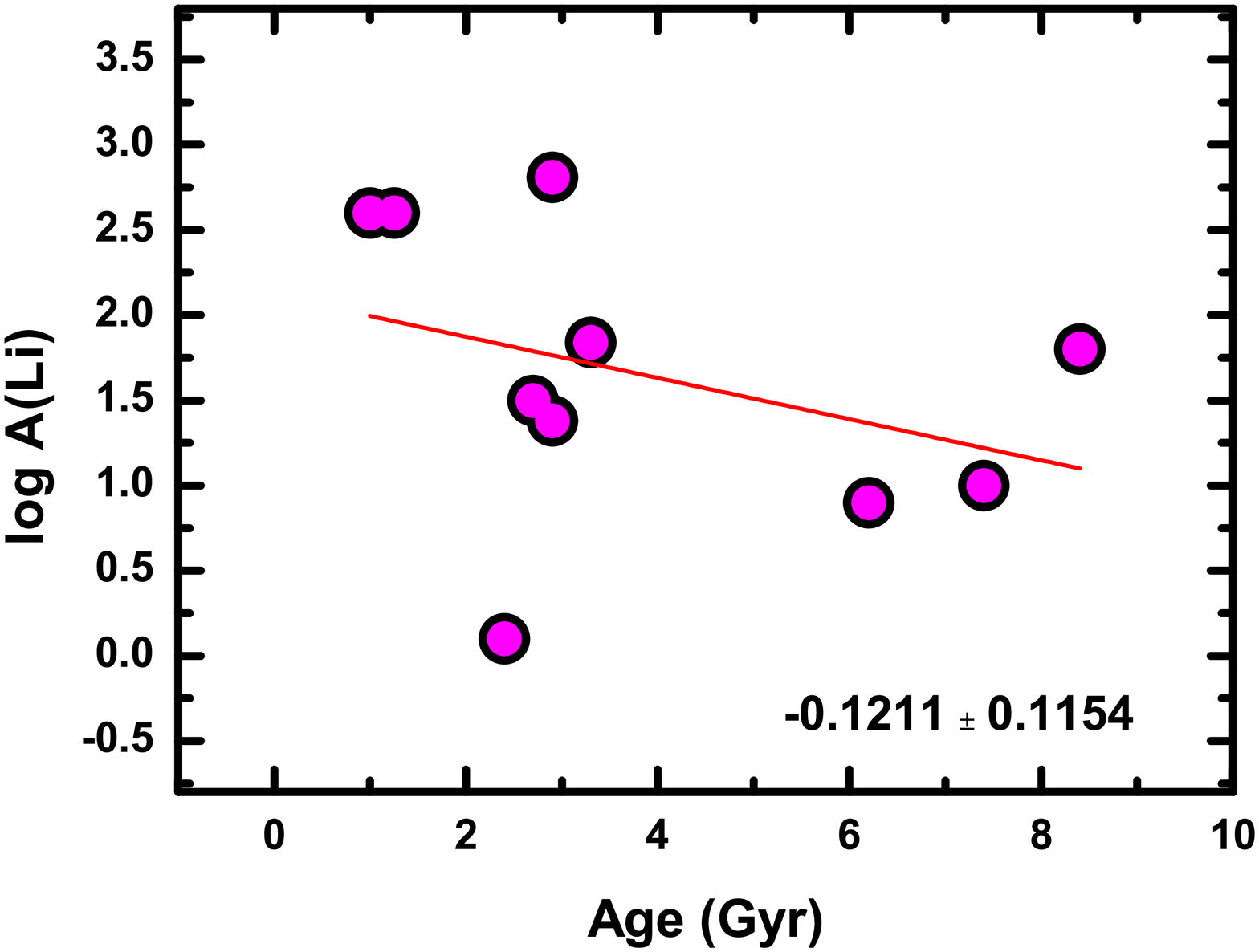}\\
\end{tabular}
\caption{Dependences of the Li abundance on stellar effective temperature (\Teff) and rotational velocity (\vsini) and stellar age for our target stars. Dependance of \vsini\ on chromospheric activity \R (third panel from the top). In the top panel, also plotted are the evolutionary tracks from \protect\cite{baraffe:17} for different ages (dashed lines) and the Li abundance envelope in the Hyades from \protect\cite{thorburn:93} (solid line). Position of the Sun is marked with an asterisk, the solar parameters are taken from \protect\cite{jofre:15}.}
\label{li_dep}
\end{figure}

The dependencies of the Li abundance on the slellar effective temperature \Teff,
rotational velocity \vsini\ and on age are plotted
in Fig.\ref{li_dep}. Evolutionary tracks for different
ages of \cite{baraffe:17} and the Li abundance envelope in the Hyades
\citep[see top panel,][]{thorburn:93} are overplotted.   \cite{baraffe:17}
studied lithium depletion in low-mass and solar-like stars as a function
of time, applying a new diffusion coefficient describing extra-mixing
occurring at the bottom of a convective envelope. 
We see that the lithium abundance behaviour reflects its depletion through stellar evolution.

The relationship between log A(Li) and rotational velocities \vsini\
is depicted in the middle panel in Fig.\ref{li_dep}. The surface Li abundance of a star
is highly sensitive to its rotational velocity during both main
and pre-main sequence evolution \cite[e.g.][]{beck:17, bouvier:16}.
\cite{beck:17} found that the measured A(Li) were in good agreement
with the A(Li) evolution predicted from a grid of models, which
included rotationally-induced internal mixing. 
The dependence obtained
in this study reflects the role of rotation in the lithium abundance
values, which could be associated with rotationally-induced internal
mixing and also to some extent with stellar activity (see, e.g., Fig.\ref{li_dep}, 
panel \vsini\ versus \R).

A very important
aspect of studying the lithium content in solar analogues is
related to the search for planetary systems and their occurrence.
The correlation between the Li abundance and planet occurrence
has been studied in a series of works, though there is still no
clarity in this respect. For instance, as reported by
\cite{gonzalez:00, gonzalez:08, israelian:04, israelian:09, delgado:14, figueira:14},
planet-hosting stars are likely to have lower lithium abundances
than stars without planetary  systems. 
Meanwhile, \cite{carlos:16} have discovered that the lithium abundance in solar-twin 
stars depends on stellar age while there was no indication
of any relationship between planet-hosting stars and strong lithium depletion.
Our determination of the Li abundance versus age is shown in Fig.\ref{li_dep} (bottom panel), 
using the age data from baseline papers. The plot shows a negative trend at about 1-$\sigma$ significance level.
Despite our sample has a wider \Teff\ range than solar twins, 
there seems to be a trend of Li versus age.
Given the presence of complex relationships between lithium content,
rotational velocities \vsini, stellar activity and the influence of
magnetic fields \cite[see, e.g.][]{mishenina:12, katsova:13}, it appears
that this dependence reflects the role of various physical processes that
affect the lithium abundance rather than results from the presence of
planetary systems.

\begin{figure}
\begin{tabular}{c}
\includegraphics[width=8cm]{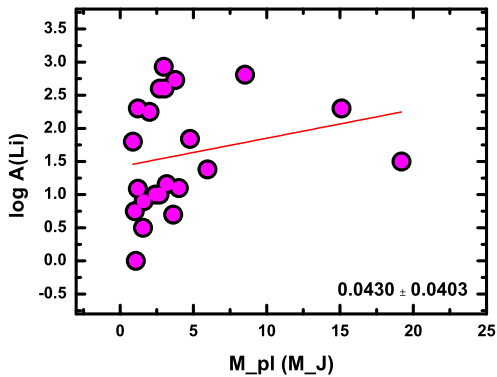}\\
\end{tabular}
\caption{Dependences of the Li abundance on planetary mass $M_{pl}$ for our target stars. }
\label{li_mpl}
\end{figure}

The relationship between the Li abundance and planetary mass 
can be seen in Fig.\ref{li_mpl}. The absence
of a large number of stars hosting planets with masses exceeding 5 $M_{J}$
in our sample does not allow us to draw reliable conclusions.

\subsection{C and O in exoplanet host stars}

\begin{figure}
\begin{tabular}{cccc}
\includegraphics[width=8.0cm]{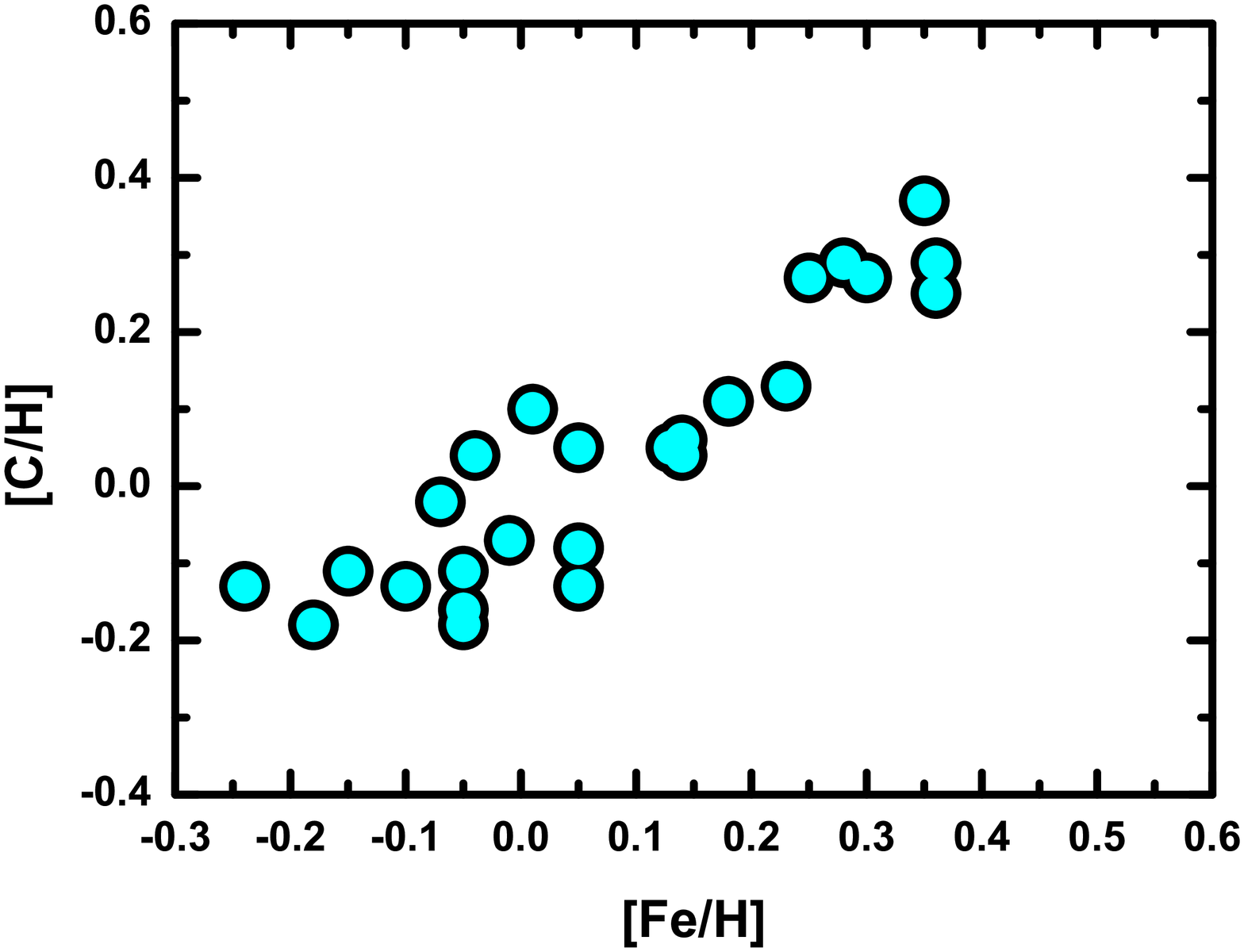}\\
\includegraphics[width=8.0cm]{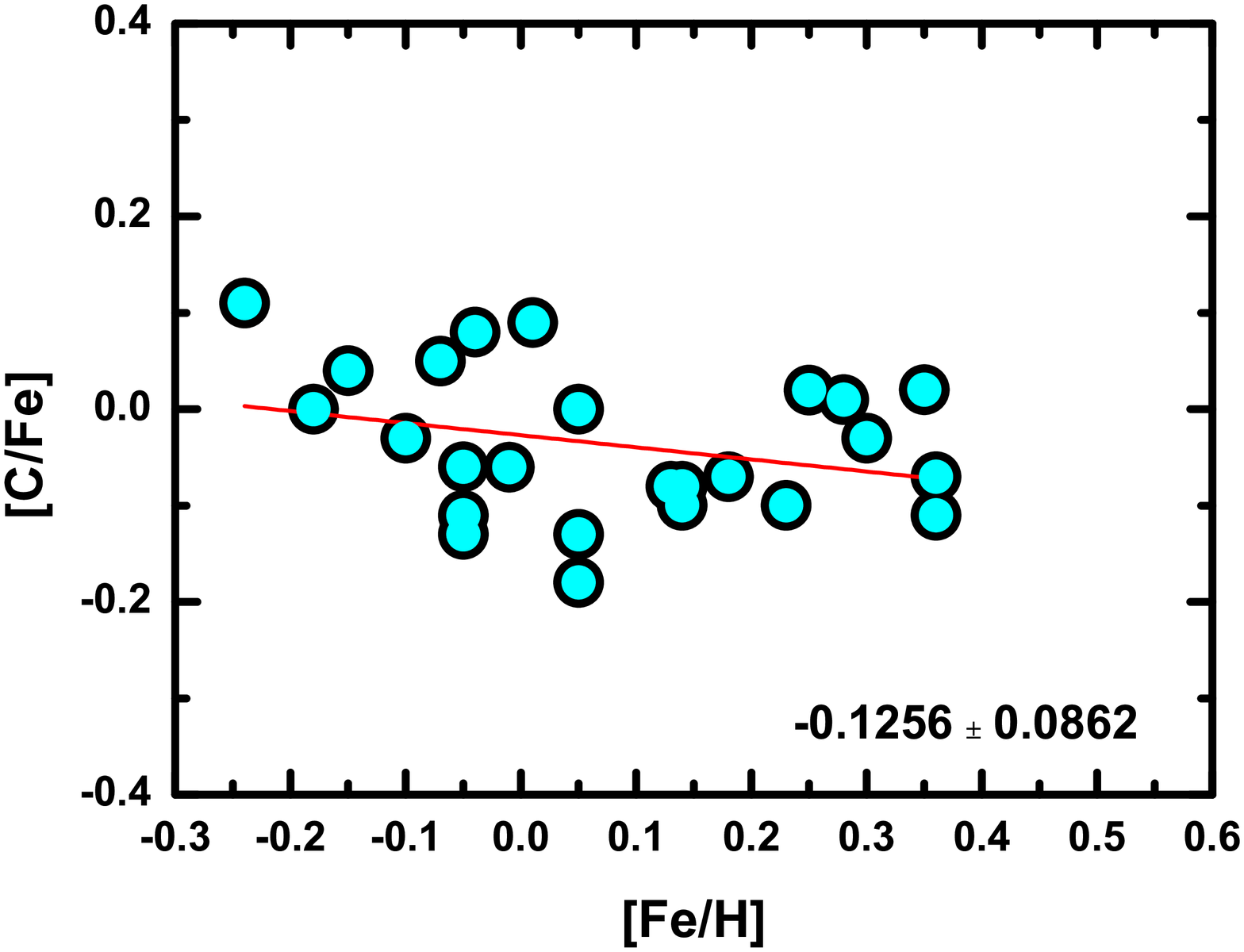}\\
\includegraphics[width=8.0cm]{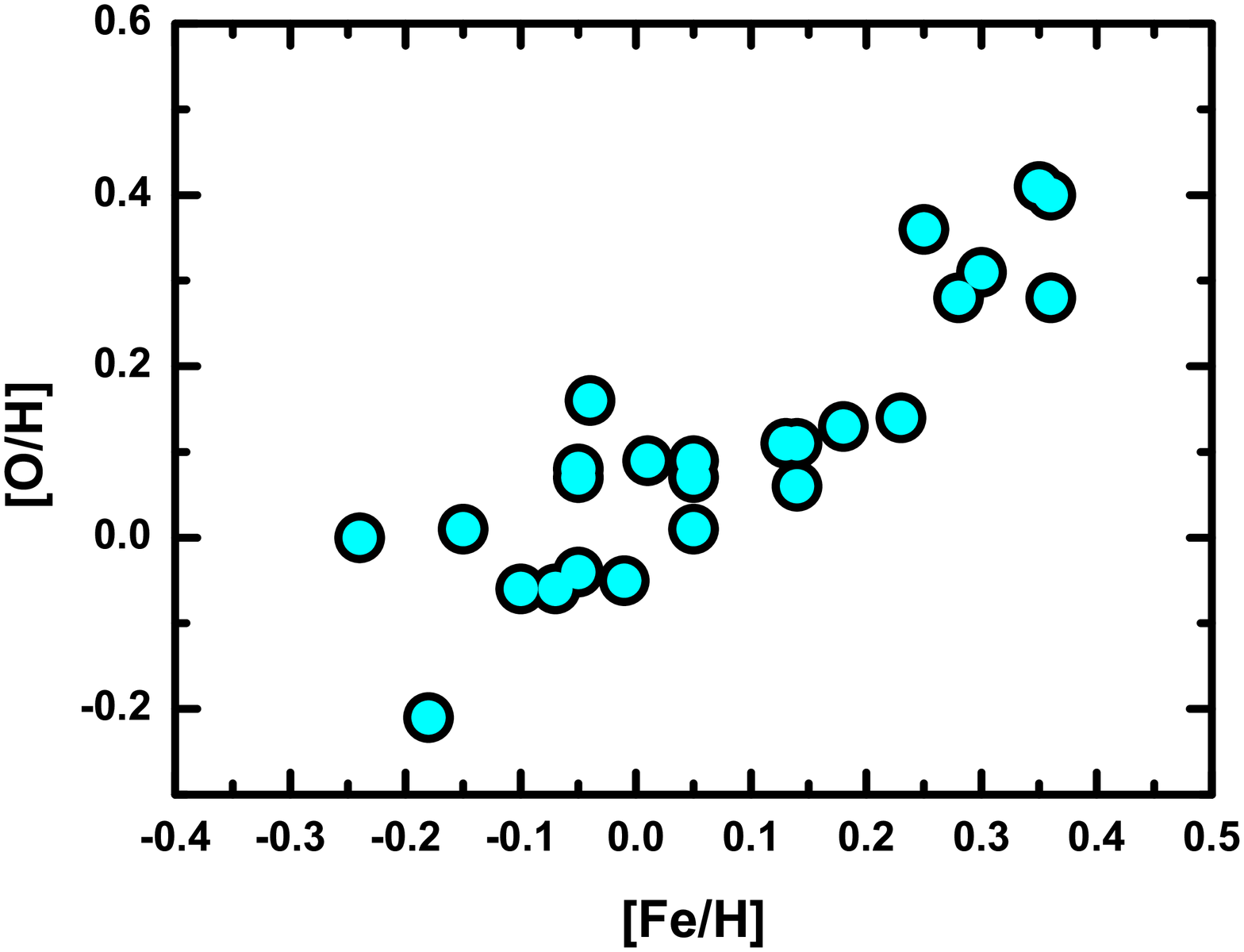}\\
\includegraphics[width=8.0cm]{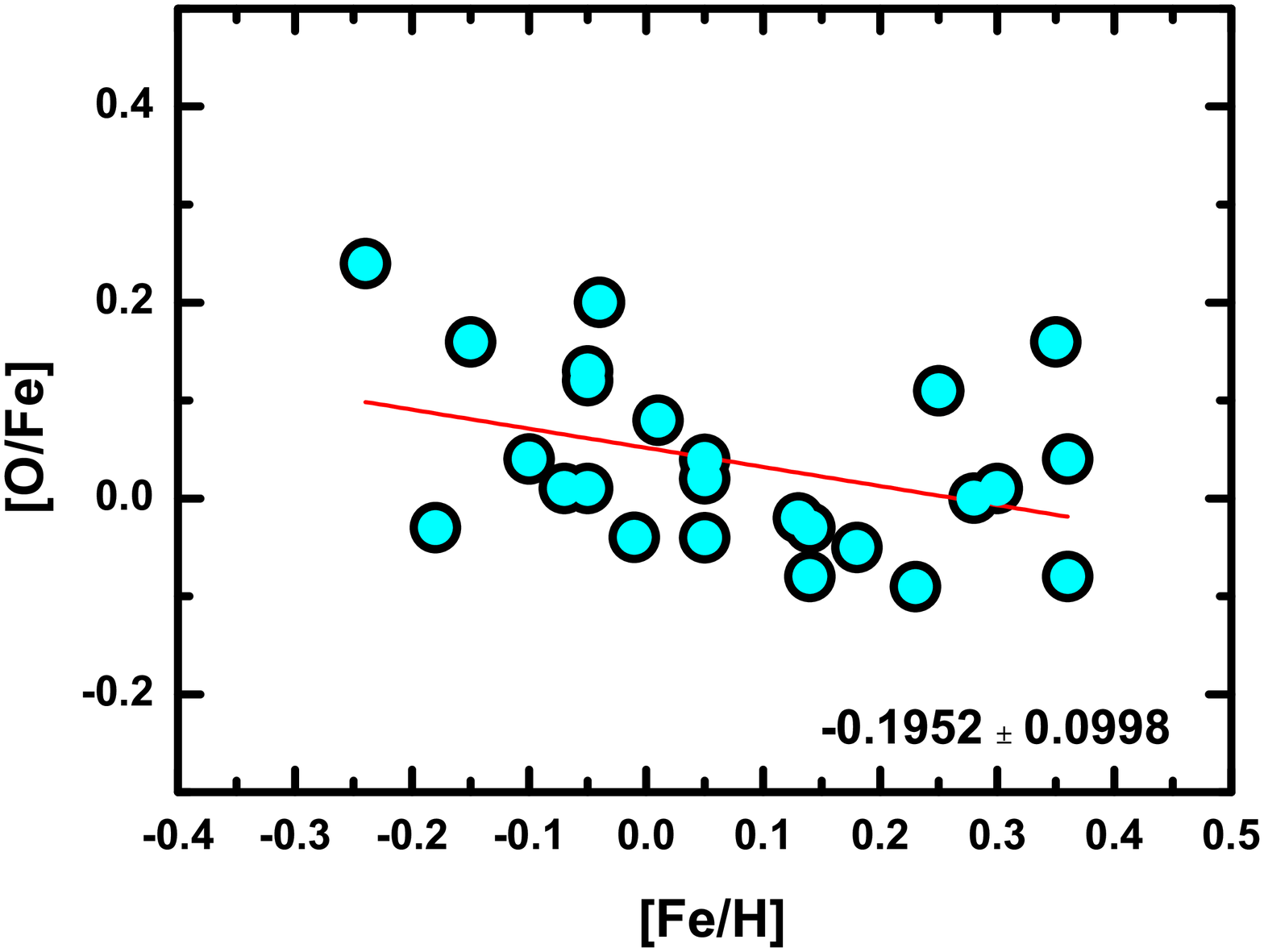}\\
\end{tabular}
\caption{ [C/H], [C/Fe], [O/H], [O/Fe] versus [Fe/H].  }
\label{cofe1}
\end{figure}

\begin{figure}
\begin{tabular}{cccc}
\includegraphics[width=8.0cm]{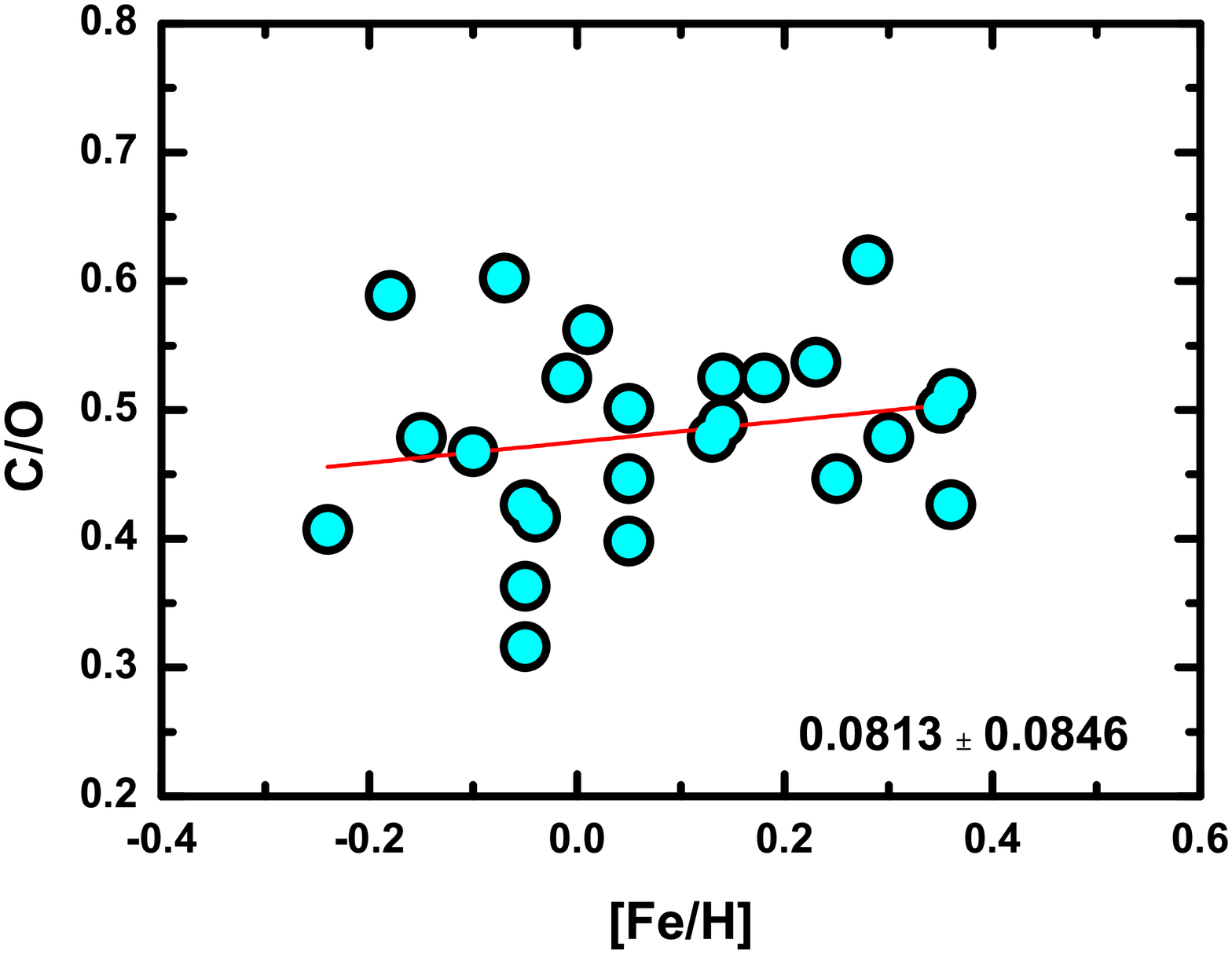}\\
\includegraphics[width=8.0cm]{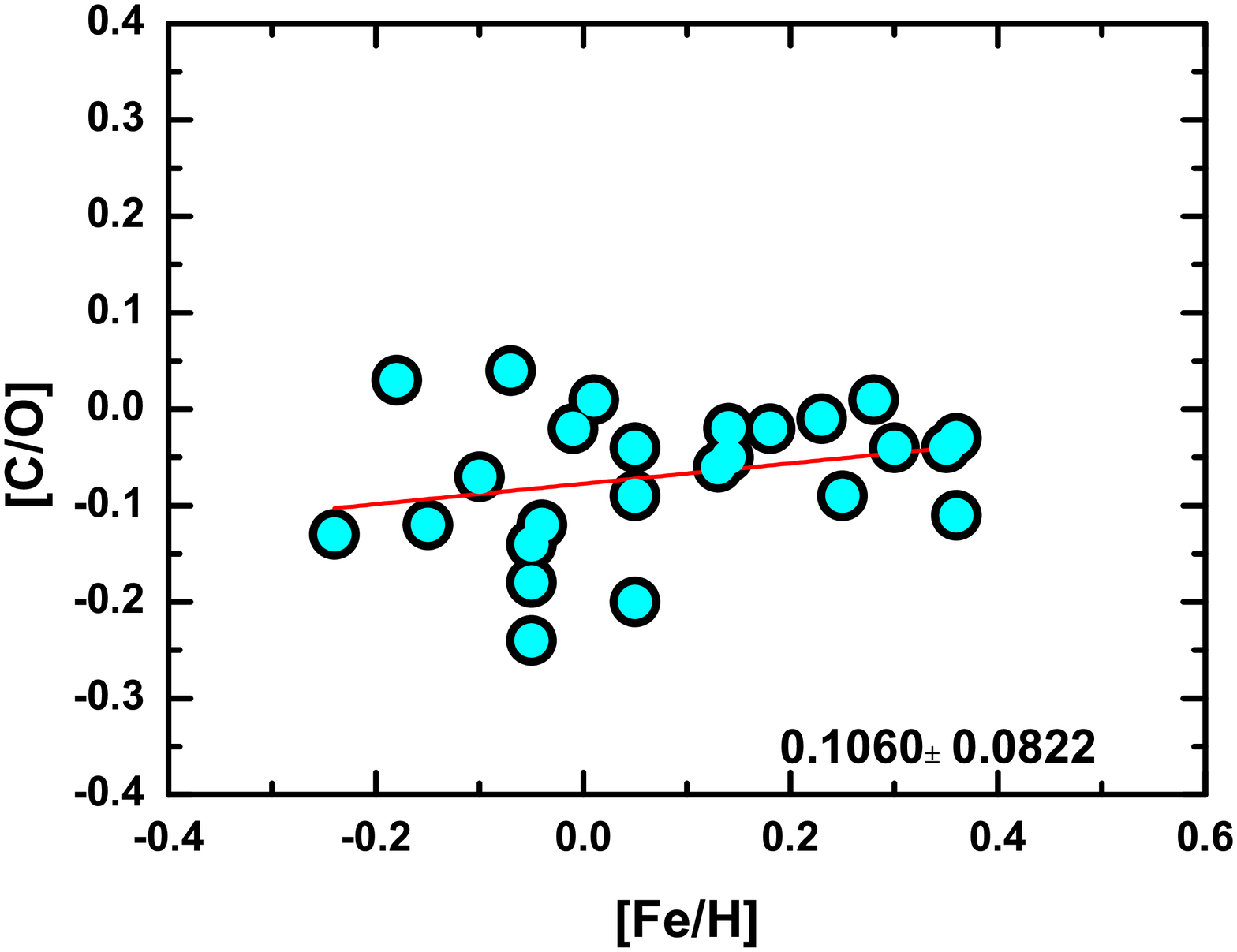}\\
\includegraphics[width=8.0cm]{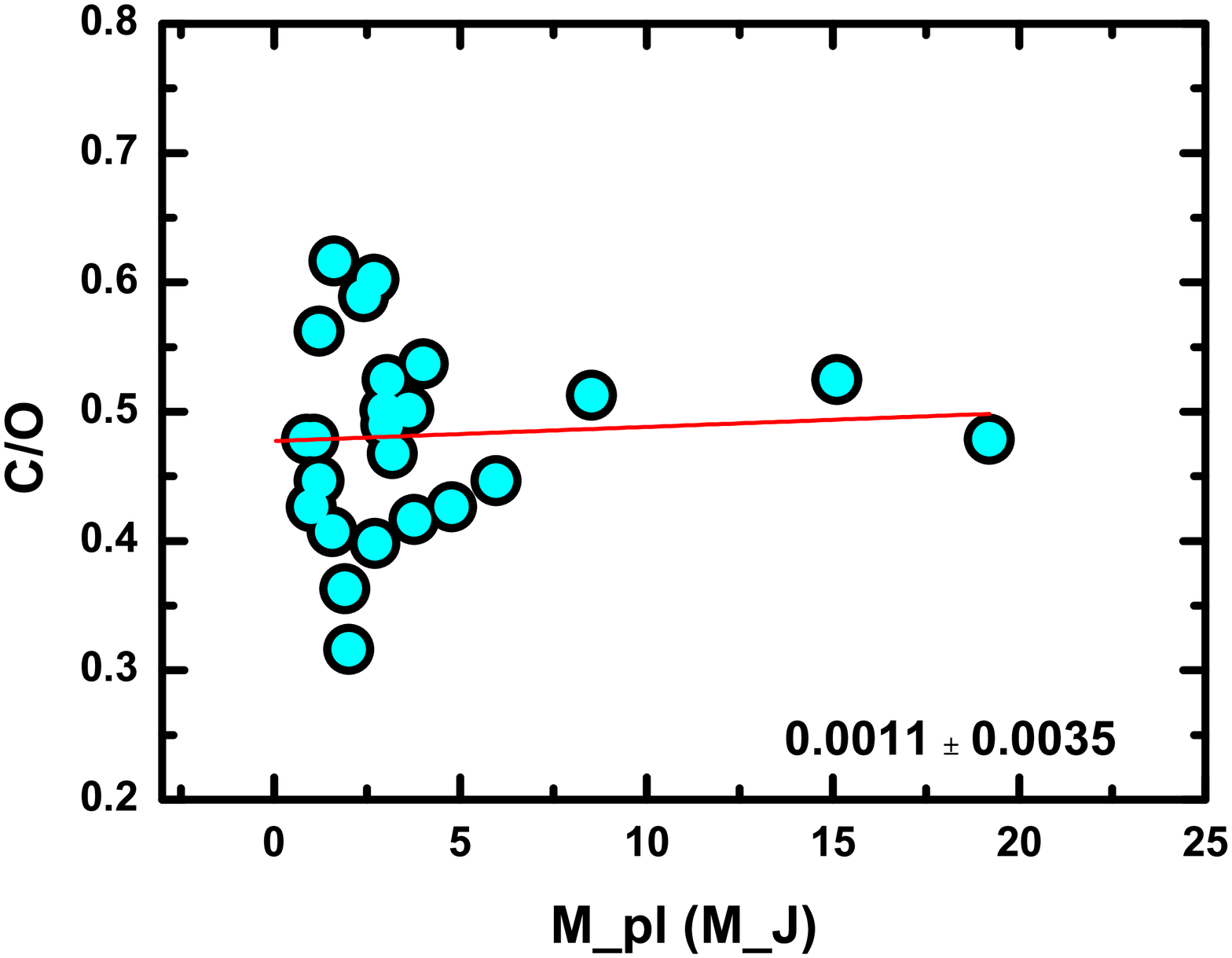}\\
\includegraphics[width=8.0cm]{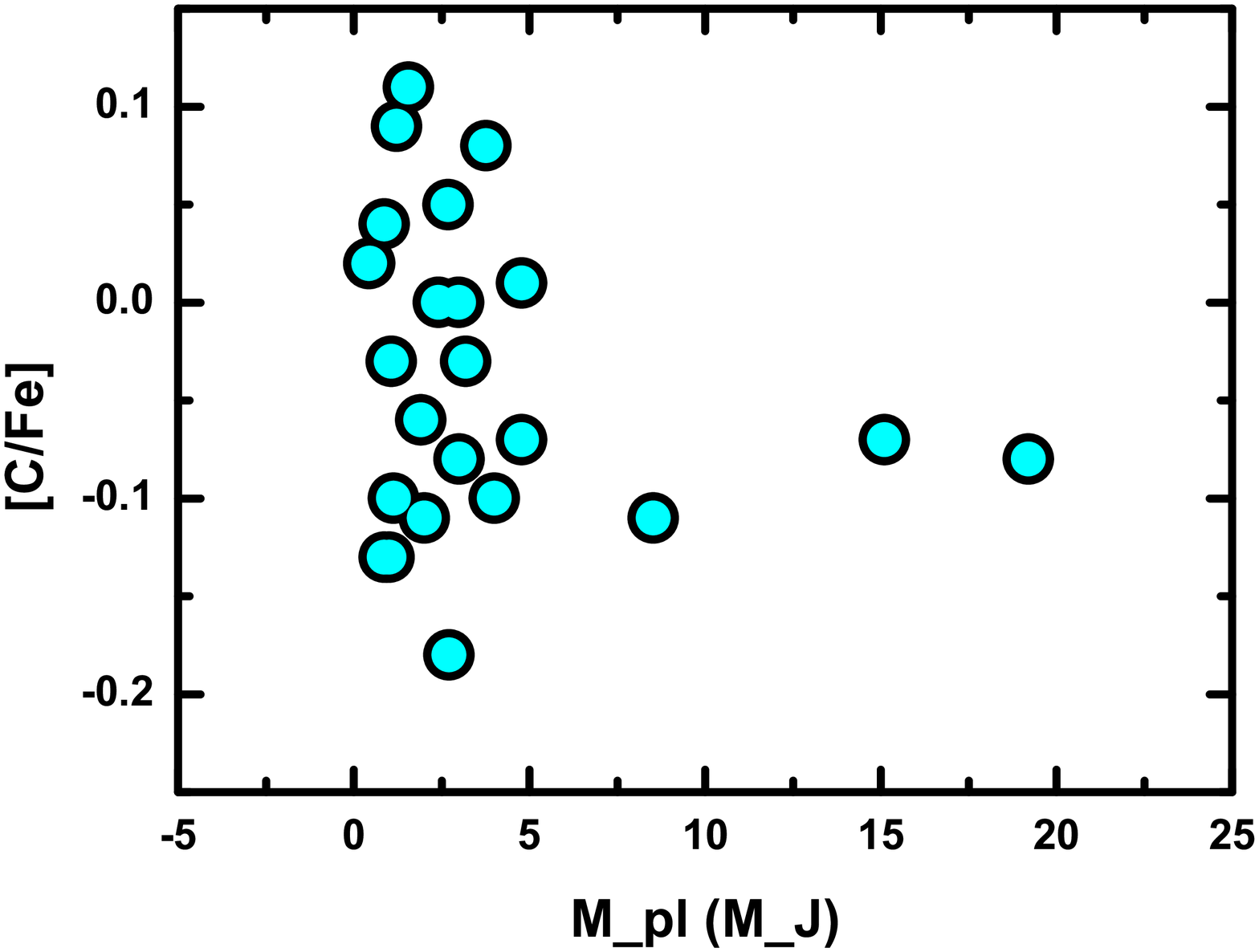}\\
\end{tabular}
\caption[]{ [C/O], C/O  versus [Fe/H]  and C/O and [C/Fe] versus planetary mass (in $M_{J}$).}
\label{cofe2}
\end{figure}

The obtained values of [C/H], [C/Fe], [O/H], [O/Fe]  vs. [Fe/H] are shown
in Fig. \ref{cofe1}.
The C/O, [C/O] vs. [Fe/H] and C/O, [C/Fe] vs. planetary masses $M_{pl}$
are depicted in Fig. \ref{cofe2}.
As can be seen in Fig. \ref{cofe1}, the trends of [C/H] and [O/H]
vs. [Fe/H] are corroborated by the data obtained earlier
\cite[see][their Figs.1,3]{delgado:10}, and \cite[][their Fig.5]{suarez:17}.
The obtained dependence of [C/H], [O/H] on the metallicity [Fe/H] reflects
the well-known fact that the production of carbon and oxygen in the Galaxy
(in absolute terms) rises with an increase of  the metallicity  due to an
additional contribution to the enrichment, besides massive collapsing
supernovae (CCSN or SN~II Type), in which oxygen is the dominant element ejected
\citep{woosley:95}, massive fast-rotating stars, \citep{maeder:12},
also from SN~Ia and  Asymptotic Giant Branch (ABS) stars, whose yields of
elements grow with increasing metallicity above [Fe/H] = --1.0
\citep[e.g.][]{timmes:95, nomoto:84, kappeler:11, prantzos:18}.

Decreasing [C/Fe] and [O/Fe] ratios with increasing stellar metallicity
(Fig.\ref{cofe1}) corroborate observations of C and O in the
Galactic disc \citep[e.g.][]{delgado:10, petigura:11, nissen:14, suarez:17}.
According to models of Galactic Chemical Evolution (GCE), 
Type la supernovae start ejecting large amounts of iron and negligible
amounts of oxygen into the interstellar medium beyond about 1 Gyr, which
causes the distinct downturn of [O/Fe] to the solar value.  
\cite{suarez:17} drew attention to two differing
patterns of behaviour of carbon at [Fe/H] below and above solar value
(see their Figure 5).
It may be associated, first of all, with different nucleosynthesis
(sources) of carbon and oxygen at such metallicities, but not with the
fact that stars with massive planets richer in metals exhibit differing C and O abundances. Reaching more reliable conclusions require the use of a larger sample of stars and involve the study of the GCE, which is beyond the scope of this paper.
Moreover, \cite{suarez:17} did not find any clear relationship between the [C/Fe] abundance ratio and planetary mass.  Fig. \ref{cofe2} (bottom panel) substantiates such a result. Therefore, the C and O abundances in our target stars demonstrate that the behaviour of carbon and oxygen in stars with planetary systems in the Galactic disc is quite typical of stars inhabiting the thin disc.

The carbon-to-oxygen ratio C/O is one of the important characteristics of the structure and chemical composition of planets.
Theoretical models of planet formation and evolution consider both refractory \citep{bond:10} and volatile elements \citep{thiabaud:15}, which can provide information about the protoplanetary disc in which the planet was formed.
Models by \cite{thiabaud:15a} show that condensation of volatiles varies with the radial distance and that the C/O ratio can be four times the solar value in certain regions of the protoplanetary disc. This may lead to the formation of planets with C/O values in their shells being more than three times the solar value.

Over several years of investigation of the chemical composition, the values of this ratio for planet hosts and non-hosts obtained in different studies have varied from C/O $>$= 1 \citep[e.g.][]{petigura:11} to near solar value (C/O$_\odot$= 0.54) \citep{teske:14}.
The differences in the obtained C abundances may be associated with a number of factors, primarily the use of various markers of the carbon abundance (atomic and molecular lines), atomic and molecular parameters (e.g. \loggf), quality of spectra, application of different model atmospheres and methods, and samples of stars, etc. In this regard, it is important to determine reliable C and O abundances and seek   for correlations between the abundances of these elements and properties of planets.

The mean value of $<$C/O$>$ for our target is $<$C/O$>$ = 0.48$\pm$0.07, which agrees well with the mean value of $<$C/O$>$  = 0.45$\pm$0.06 obtained by \cite{brewer:16}, normalized to the solar carbon abundance adopted in their study \citep{grevesse:07} for nine stars in common. At the same time, the mean C/O ratio is slightely different from the mean C/O = 0.68 $\pm$0.13 obtained by \cite{petigura:11} for four stars in common.
There is only one star in our sample in common with the study by \cite{stonkute:20}, namely HD~150706, for which the C, O abundances, as well as the C/O ratio, are reported; these values show good agreement.

 As mentioned earlier, Delgado Mena et al. (2010) and Petigura \& Marcy (2011) found  that a significant percentage of high-metallicity solar-type stars have C/O $>$ 0.8; however, it has been demonstrated in a number of later works that not so many stars actually have C/O greater than 0.8  \citep{nissen:13, brewer:16, suarez:18, stonkute:20}. \cite{nissen:20} have indicated that their 3D non-LTE C/O values support a rising trend with decreasing age for the old (red) sequence, though all the stars have C/O ratios clearly below 0.8. All nine stars with C/O = 0.7 have been confirmed to host planets and they all exhibit high metallicities ranging within 0.18 $<$ [Fe/H] $<$ 0.31. The authors have concluded that it is not feasible to deduce from their small sample of stars whether it is a high C/O ratio or a high [Fe/H] (or both) that favours the formation of planets.

The C/O ratios obtained in this study range from 0.32 to 0.62 with the mean value of $<$C/O$>$ = 0.48 $\pm$0.07, which is well consistent with the data of \cite{suarez:18}.
The stars with massive planets investigated in this study show a wide range of C/O ratios, but do not exhibit such a high C/O ratio close to 0.8.
In this study, the mean value relative to the solar one is $<$[C/O]$>$ = --0.07$\pm$0.07; it enables us to deduce that our studied stars with massive planets have [C/O] ratios slightly lower, but close to the solar one within the errors. This result does not unambiguously confirm the finding reported by \cite{pavlenko:19}, in particular that the metal-rich dwarfs with planets are overabundant in carbon with an average difference of  $<$[C/O]$>$ = 0.05 $\pm$0.05; in our case for stars with [Fe/H] $>$ 0 this value is  $<$[C/O]$>$ = --0.05 $\pm$0.05. 
In Fig.\ref{cofe2}, the C/O ratio is also shown as a function of the planetary mass, but no trend is evident. 

\subsection{Mg and Si in planet host stars}

\begin{figure}
\begin{tabular}{ccc}
\includegraphics[width=8.4cm]{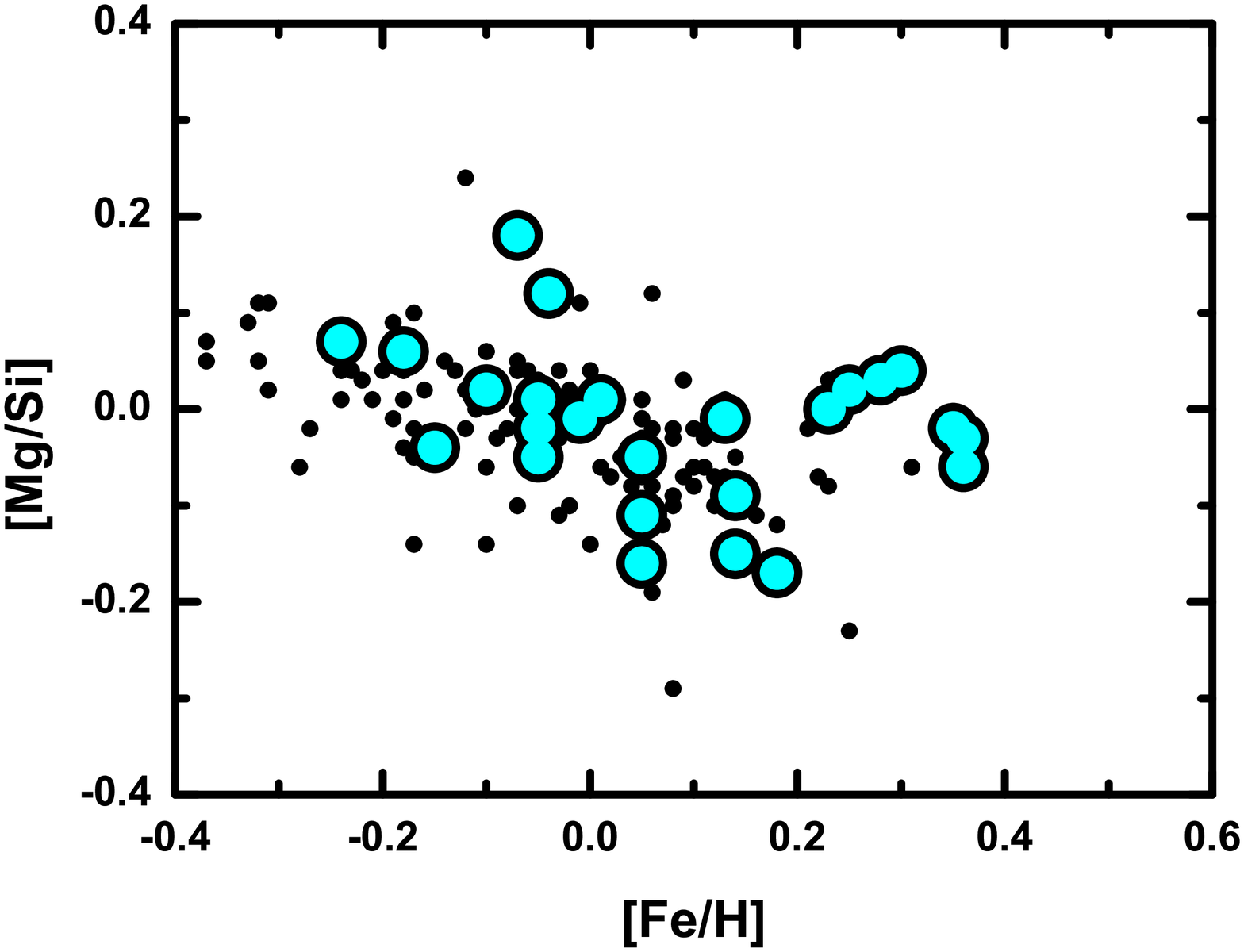}\\
\includegraphics[width=8.4cm]{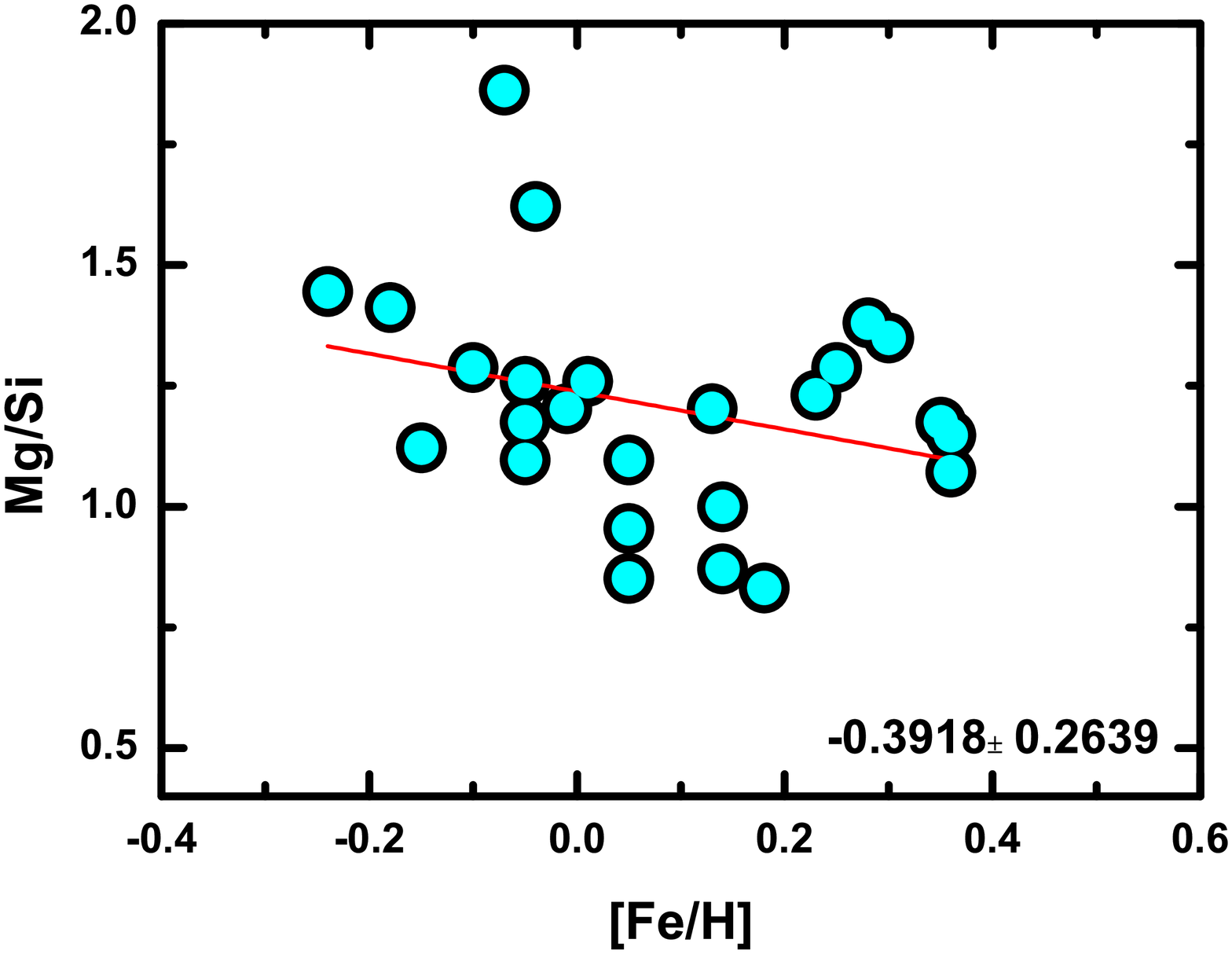}\\
\includegraphics[width=8.4cm]{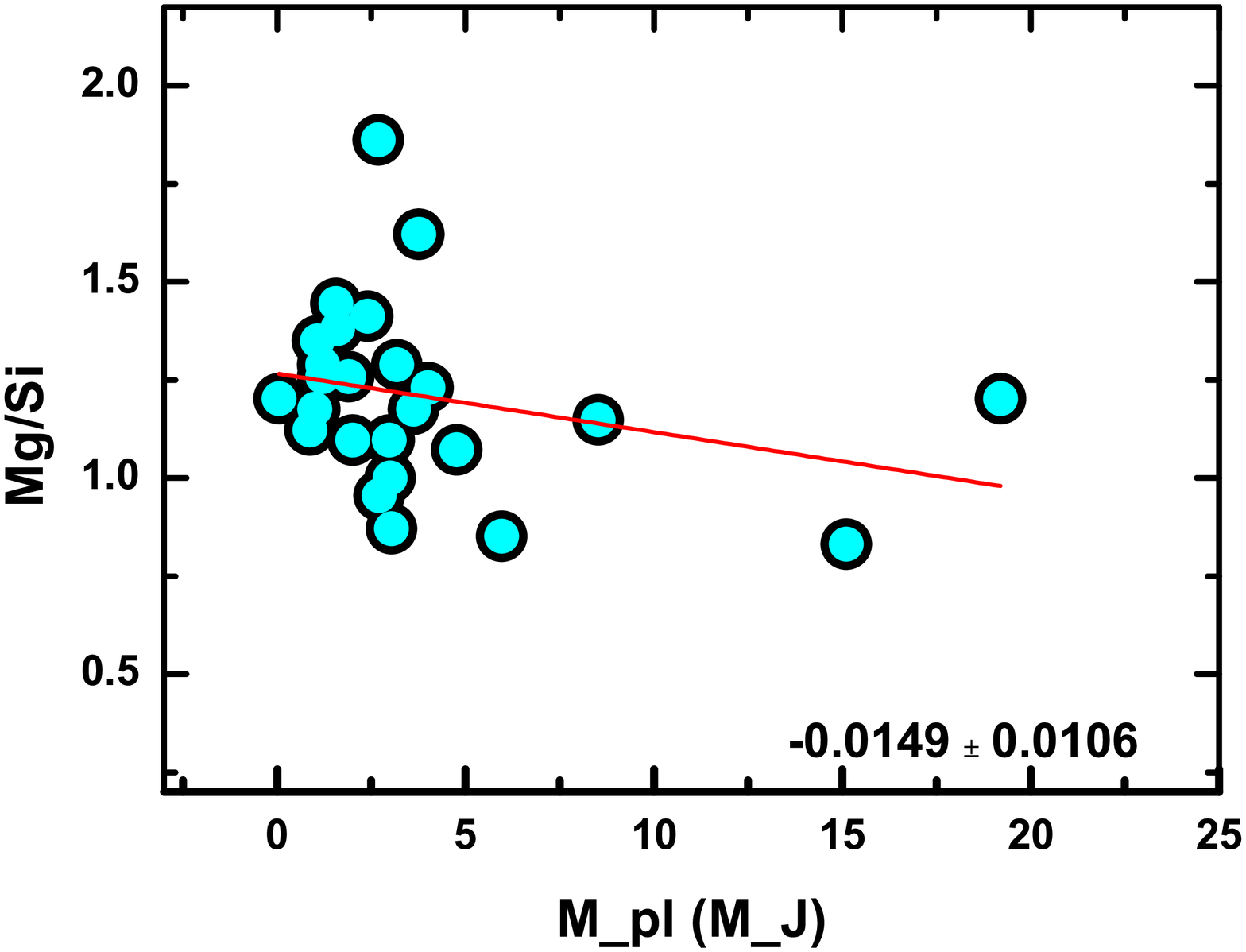}\\
\end{tabular}
\caption{ [Mg/Si] versus [Fe/H] for stars from \protect\cite{mishenina:16} (dots) and this study (open circles); Mg/Si  versus [Fe/H]  and planetary mass (in M$_{J}$). }
\label{mgsife}
\end{figure}

Magnesium and silicon are key planet-building elements and Mg/Si ratio may be used to detect low-mass planets and specify their chemical composition and mineralogy \citep{bond:10}. \cite{thiabaud:15a} have shown that the elemental ratios Mg/Si and Fe/Si in planets are essentially identical to those in the star. 
Our result for Mg and Si are listed in Table A2 in Appendix and their comparison with the values for the main sequence stars reported in \cite{mishenina:16} is depicted in Fig. \ref{mgsife} (upper panel). The stars studied in \cite{mishenina:16} have the same range of parameters as the stars of the current work and belong to the Galactic thin disc.  

As can be seen from Fig.\ref{mgsife}, the dependence of Mg/Fe on [Fe/H] is similar to the trend for dwarfs that substantiates the findings in \cite{adibekyan:15}, in particular, that the [Mg/Si] ratio depends significantly on metallicity through the GCE. Later, \cite{brewer:16} also showed that the Mg and Si behaviour could reflect the overall metallicity trend traced by [Fe/H], i.e. the GCE.

We obtained  Mg/Si ratios ranging from 0.83 to 0.96 for four stars of our sample with high-mass planets and from 1.0 to 1.8 for the remaining 21 stars.
It supports the finding in \cite{suarez:18}, that 85\%  stars with high-mass companions have the Mg/Si ratios between one and two while the other 15\% exhibit Mg/Si values below one.

Fig.\ref{mgsife} (bottom panel) illustrates the dependence of [Mg/Si] on planetary masses. Just a trace of a downward trend can be seen in this figure; however, the large scatter of the [Mg/Si] values in the region of lower masses (1-5 M$_J$), along with the small number of stars with planets of masses exceeding 10 M$_J$, do not allow us to draw a confident conclusion about the dependence of the planetary mass on the [Mg/Si] ratio in the host star.

\section{Elemental abundances versus condensation temperatures $T_{cond}$ }
\label{sec: Tcond_abun}

As reported in \cite{udry:07}, the stars known to host giant planets are expected to be more enriched in refractory elements (i.e. elements with high condensation temperature $T_{cond}$), because volatiles (having low $T_{cond}$) may evaporate from infalling  bodies before being accreted \citep[e.g.][]{gonzalez:98, smith:01, ecuvillon:06}.
As noted earlier by \cite{guillot:05}, the cores of Jupiter, Saturn, Uranus and Neptune may contain heavy metals in the form of rocks and ices, which means that these (massive) planets possess dense cores of different masses.
Based on a correlation between the difference of abundance in the Sun and mean values in non-hosts with condensation temperature $T_{cond}$, \cite{melendez:09} found that volatile elements are more abundant in the Sun relative to
the solar twins while elements that easily form dust, i.e., refractories, are underabundant.
\cite{gonzalez:13} obtained clear upward abundance trends with $T_{cond}$ for only four of eight stars hosting super-Earth-like planets. The authors came to a conclusion that there was
no clear evidence of the relationship between the volatile-to-refractory abundance ratio and the presence of rocky planets.
Later, a different assumption was made in the studies of \cite{adibekyan:14, adibekyan:15}, in particular, that the chemical peculiarities (small refractory-to-volatile ratios) of planet-hosting stars could reflect that those stars were older and indicate their inner Galaxy origin. The GCE, the place of birth in the Galaxy and, probably, stellar age play an important role in using elemental abundance estimates and clarifying their relationship with various parameters. \cite{nissen:15} concluded that while an unusually low refractory-to-volatile ratio suggests that the relationship between [El/Fe] and $T_{cond}$  may be used as a signature of the existence of terrestrial planets around stars, the dependence of [El/Fe] on the stellar age makes it difficult to employ this relationship as such an indicator.

As response to the study by  \cite{nissen:15}, \cite{spina:16} verified whether the chemical evolution of the Galactic disc could explain different observed slopes of elemental abundance with $T_{cond}$.
The authors claimed that a wide diversity of those slopes was still observed after subtracting the chemical evolution effect, which could be indicative of some other processes not related to the GCE that might have affected the element-$T_{cond}$ slopes. They supposed that such a great difference between  those slopes reflected the difference of exo-planetary systems  themselves and lies in highly dynamic stages of planetary systems during which some portion of the rocky material surrounds the star (e.g., cores of gaseous planets, rocky planets, and planetesimals) and fall onto the central star;  but other planetary systems may be subject to quieter processes. Thus, the observed variety of slopes may correspond to a great number of evolutionary paths of matter in the circumstellar discs during the formation of planetary systems.
As noted by \cite{maia:19},--terrestrial planets (or the core of giant planets) may influence the surface abundance of its host star in two ways: i) the accretion of rocky material (planetary engulfment) enriches the stellar atmosphere in refractories; ii) imprisonment of refractory rich material into rocky objects that deplete the material accreted by the star during its formation.

Since the cores of massive planets can affect the surface content of refractories,
we also attempted to compare the volatile and refractory abundance ratios with  $T_{cond}$ \citep{lodders:03} in our sample of stars hosting massive planets, adjusted for the chemical evolution effects; in other words, we subtracted the mean elemental abundances in non-hosting solar twin stars $<$[El/Fe]$>$(st) from the [El/Fe] ratios in our target stars.
The values of $<$[El/Fe]$>$(st) were taken from \cite{mishenina:16}.
In that paper, 33 stars were selected as solar twins for which the average values of the parameters were $<$\Teff$>$
= 5800 $\pm$100 K,  $<$\logg$>$ = 4.40$\pm$ 0.09,  $<$[Fe/H]$>$= 0.00$\pm$ 0.05.

\begin{figure*}
\begin{tabular}{l}
\includegraphics[width=5.8cm,height=2cm]{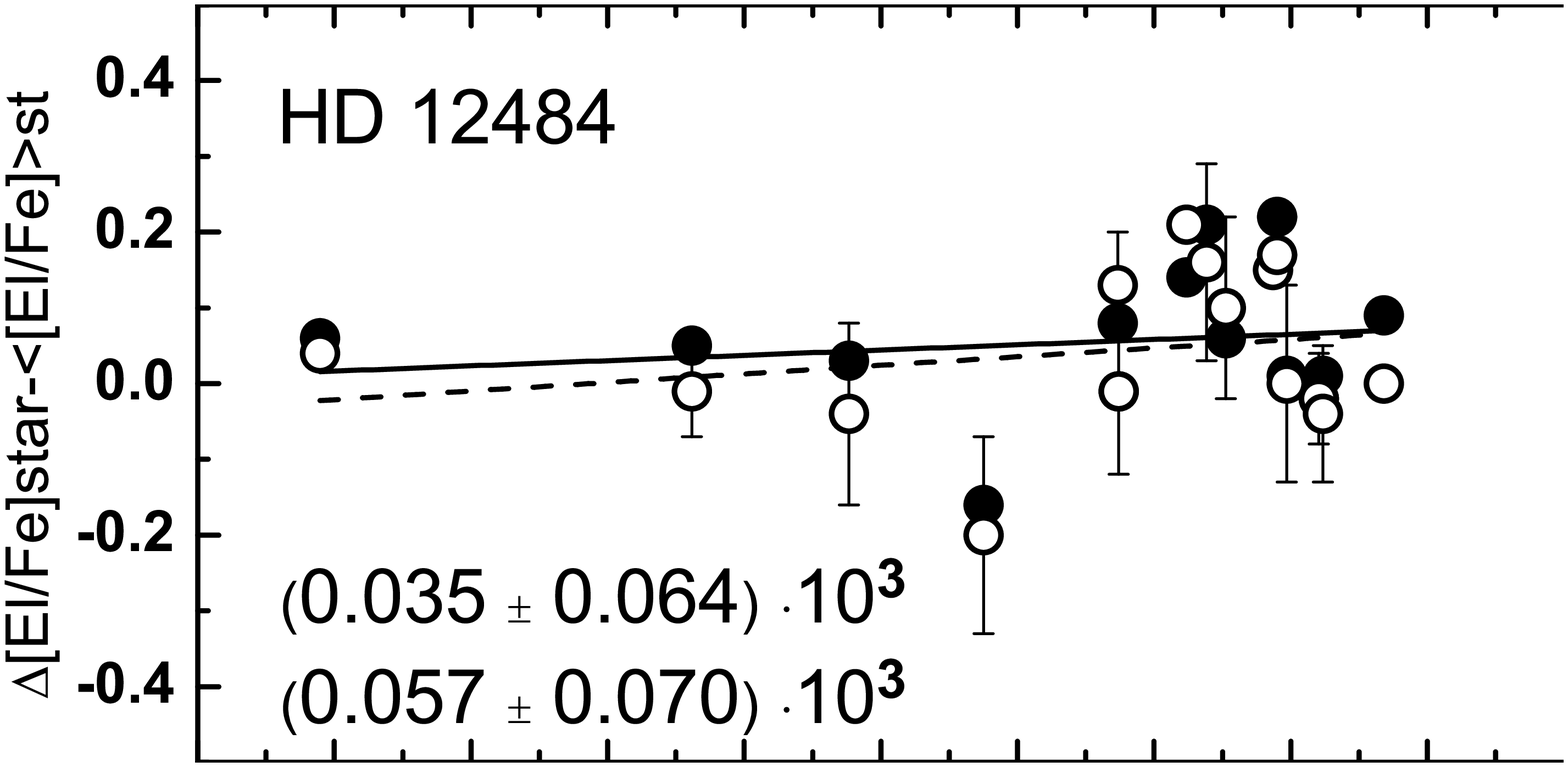}\includegraphics[width=5cm,height=2cm]{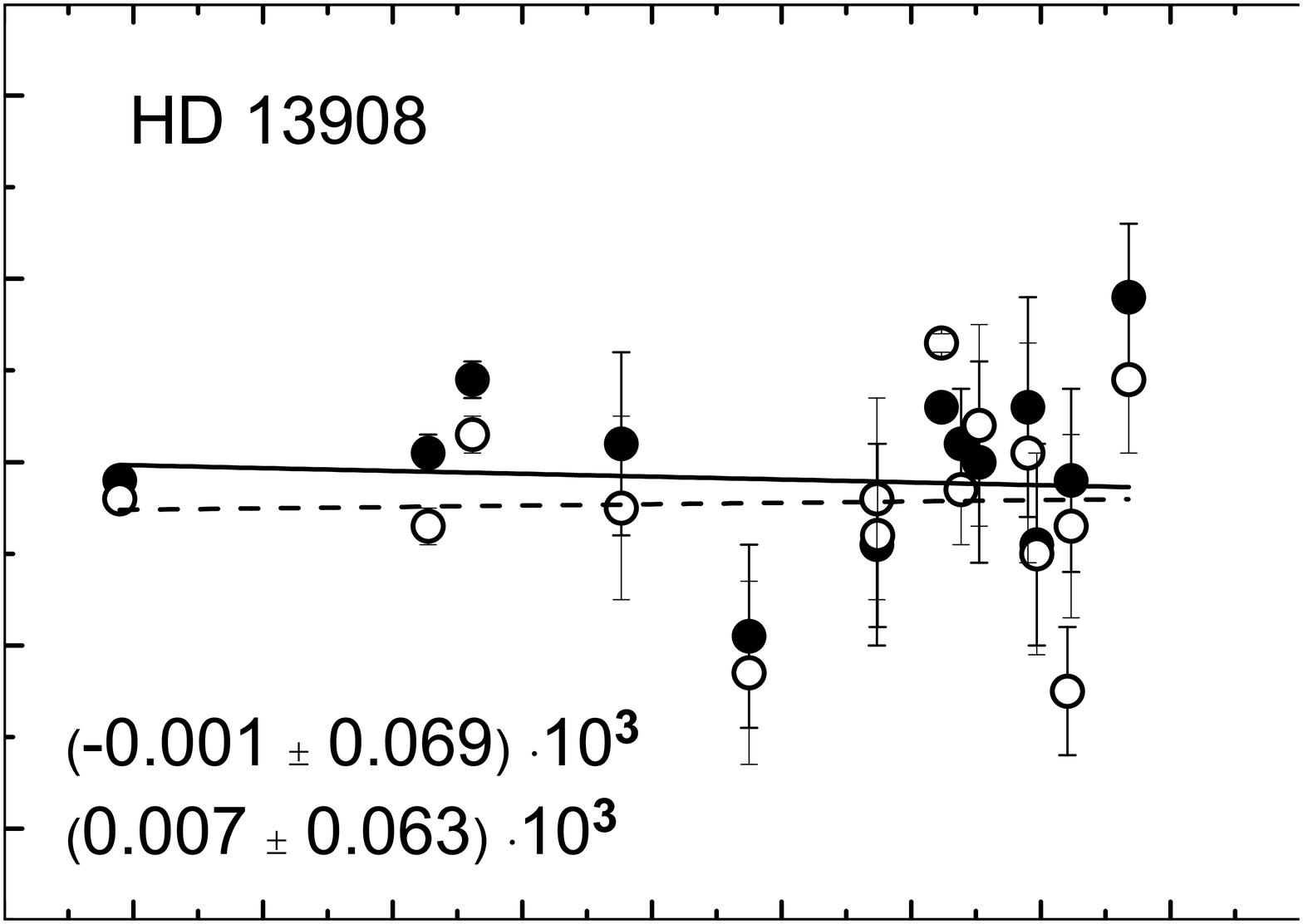}\includegraphics[width=5cm,height=2cm]{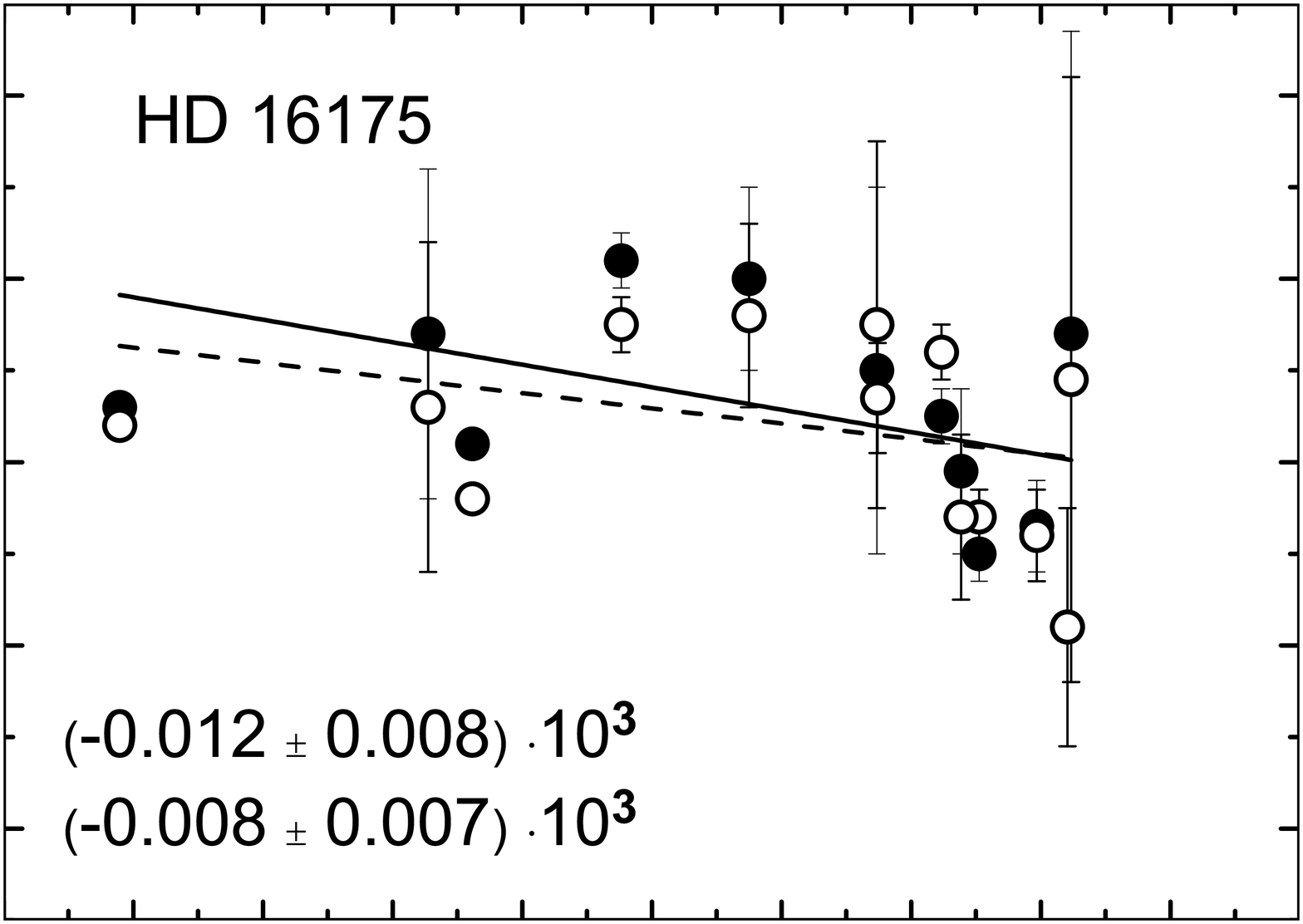}\\
\includegraphics[width=5.8cm,height=2cm]{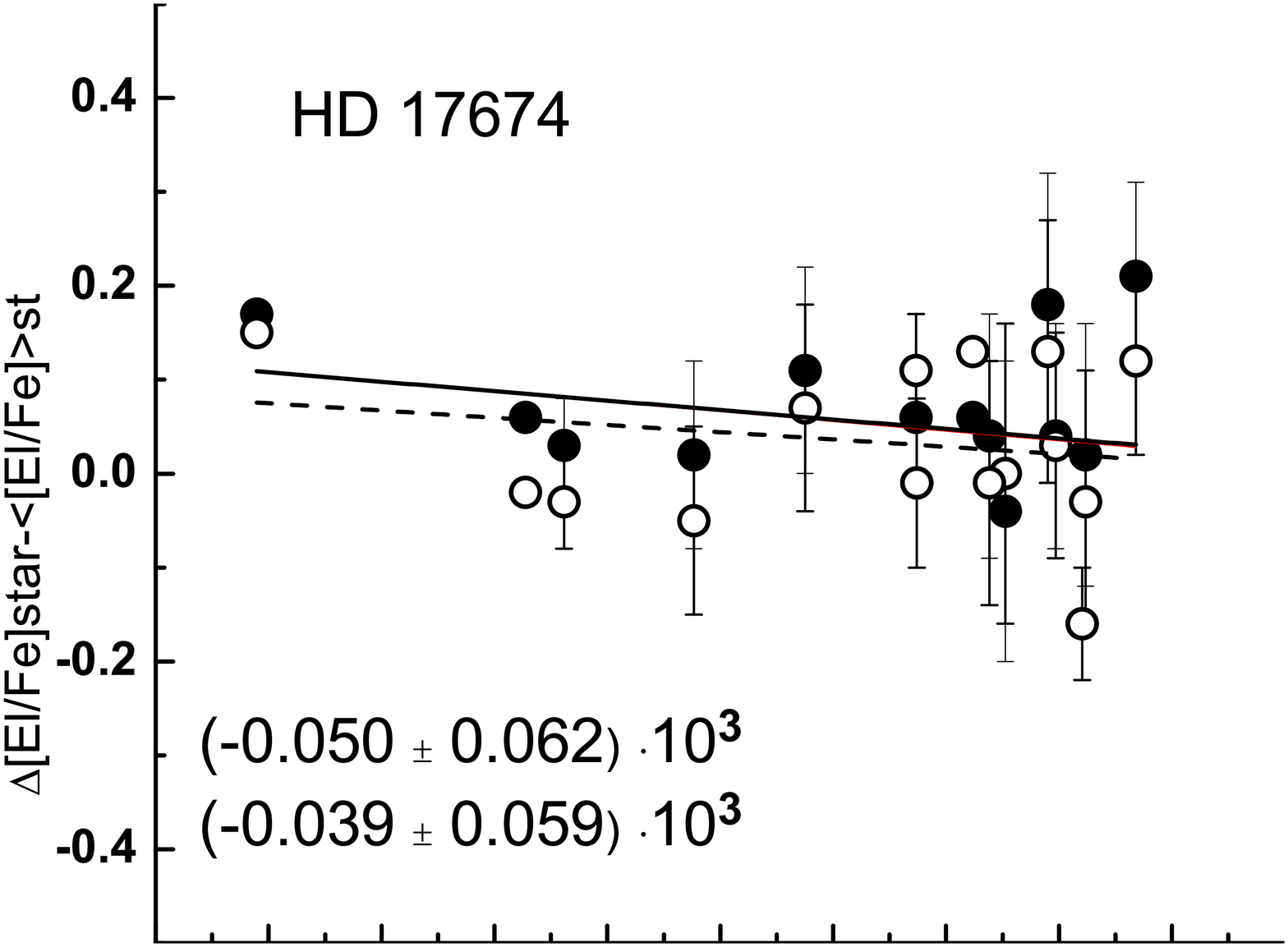}\includegraphics[width=5cm,height=2cm]{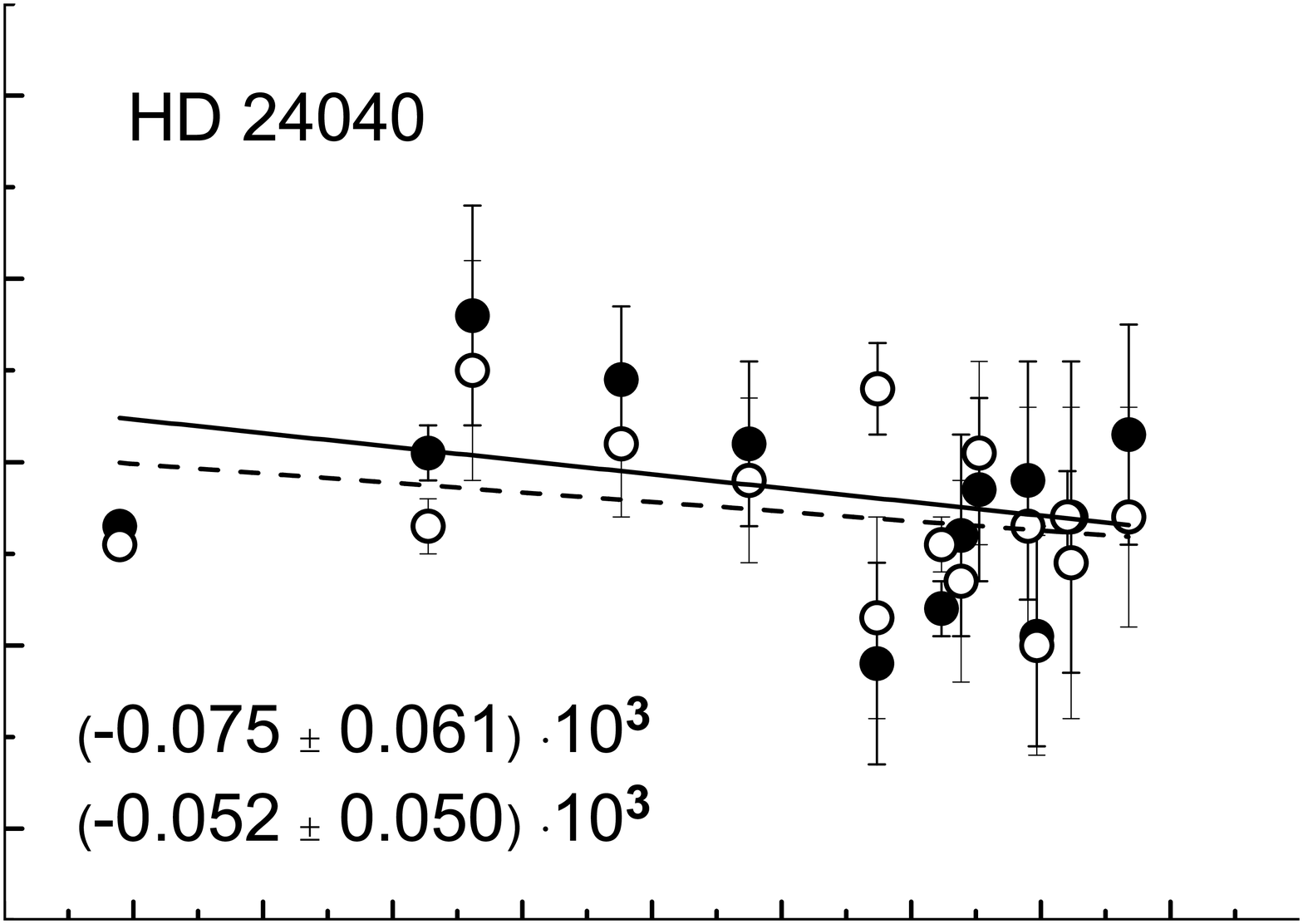}\includegraphics[width=5cm,height=2cm]{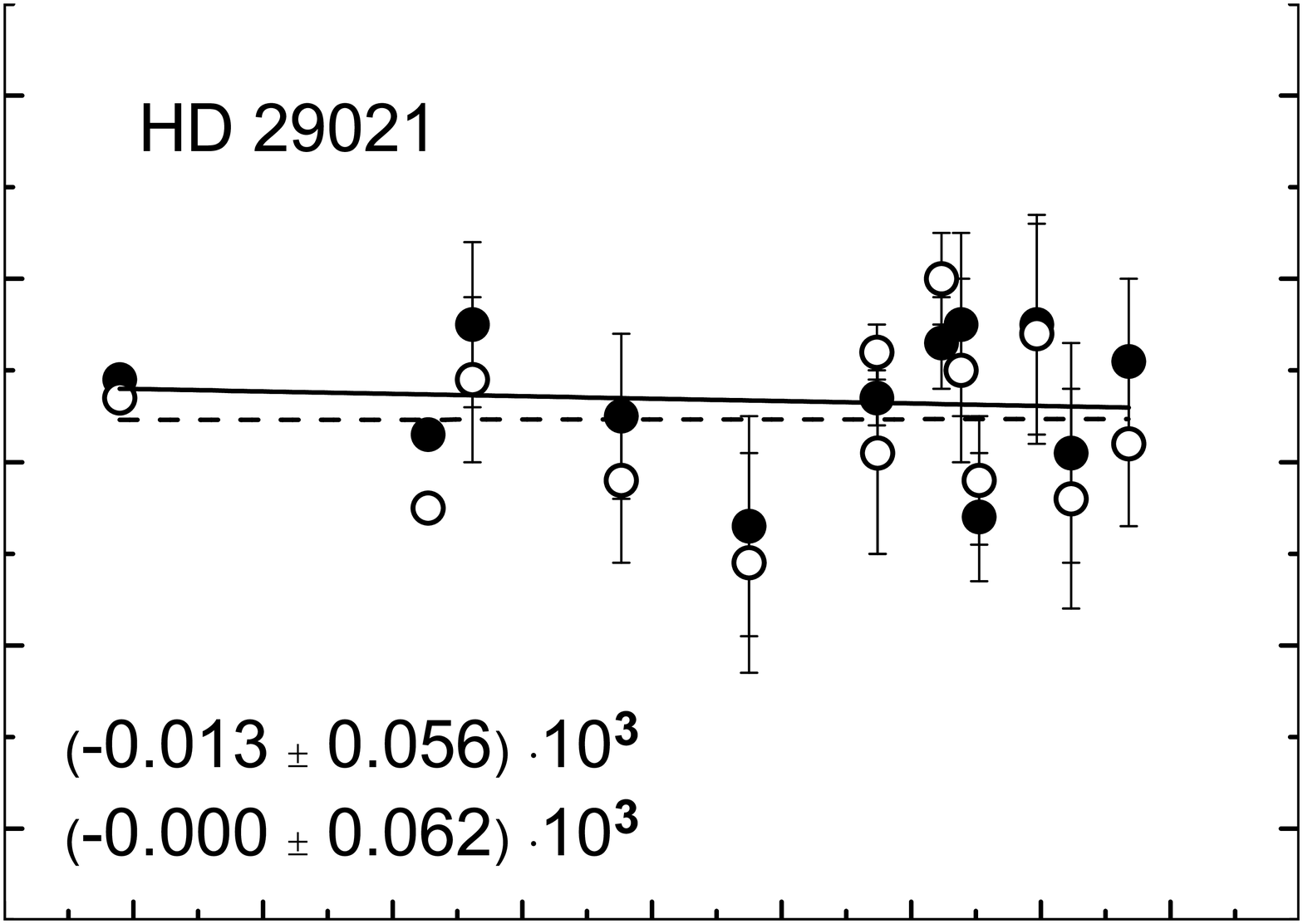}\\
\includegraphics[width=5.8cm,height=2cm]{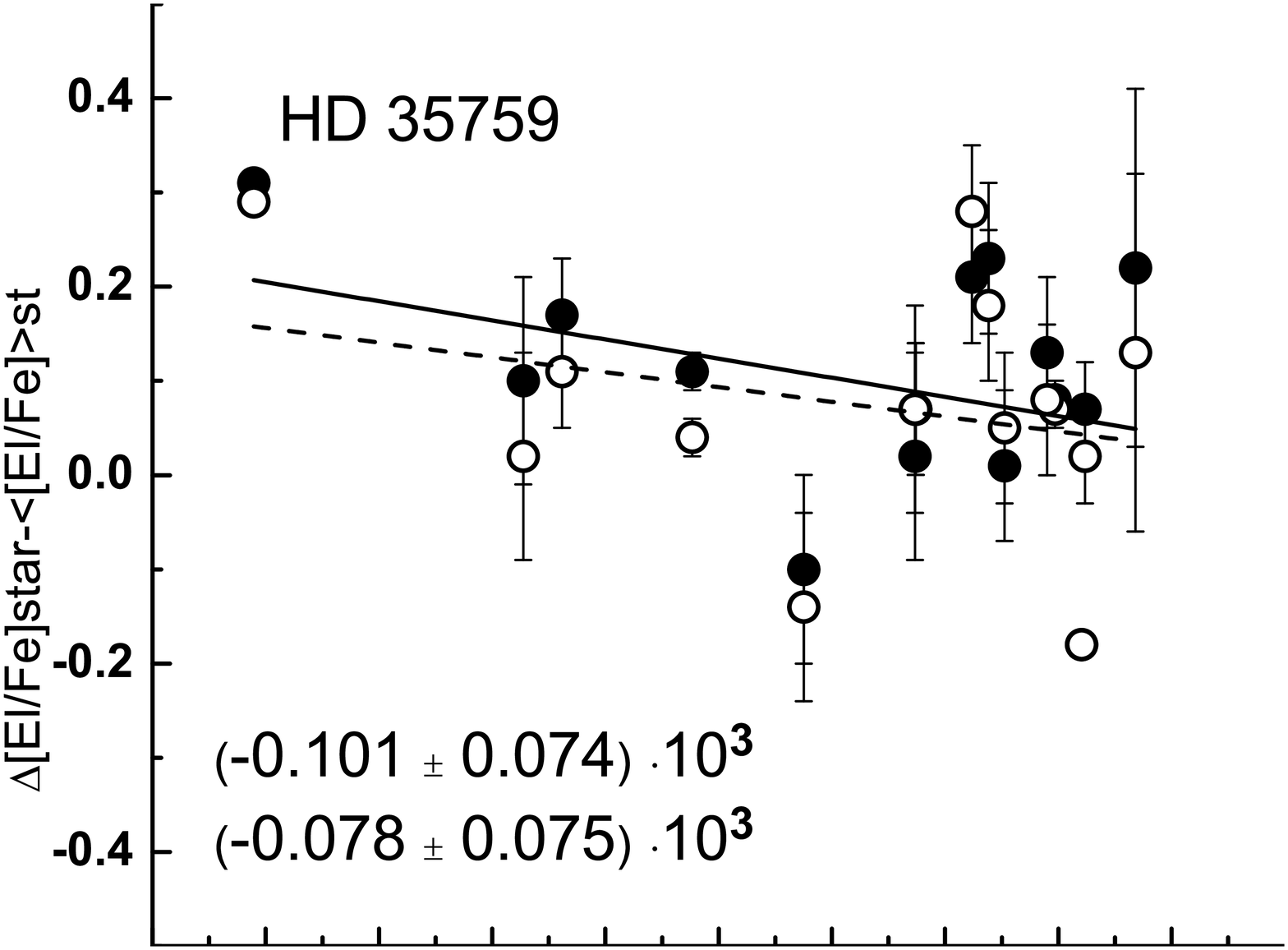}\includegraphics[width=5cm,height=2cm]{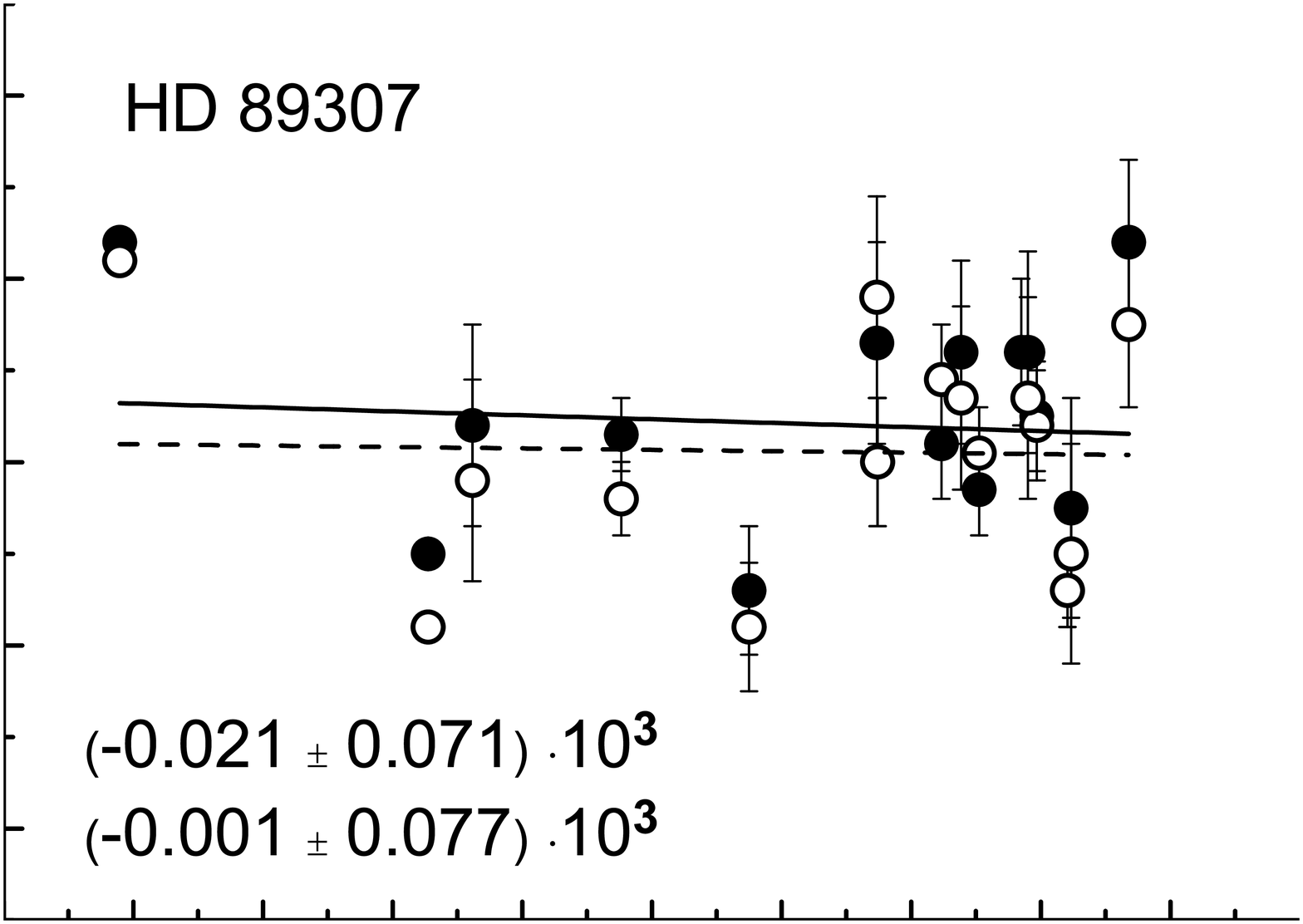}\includegraphics[width=5cm,height=2cm]{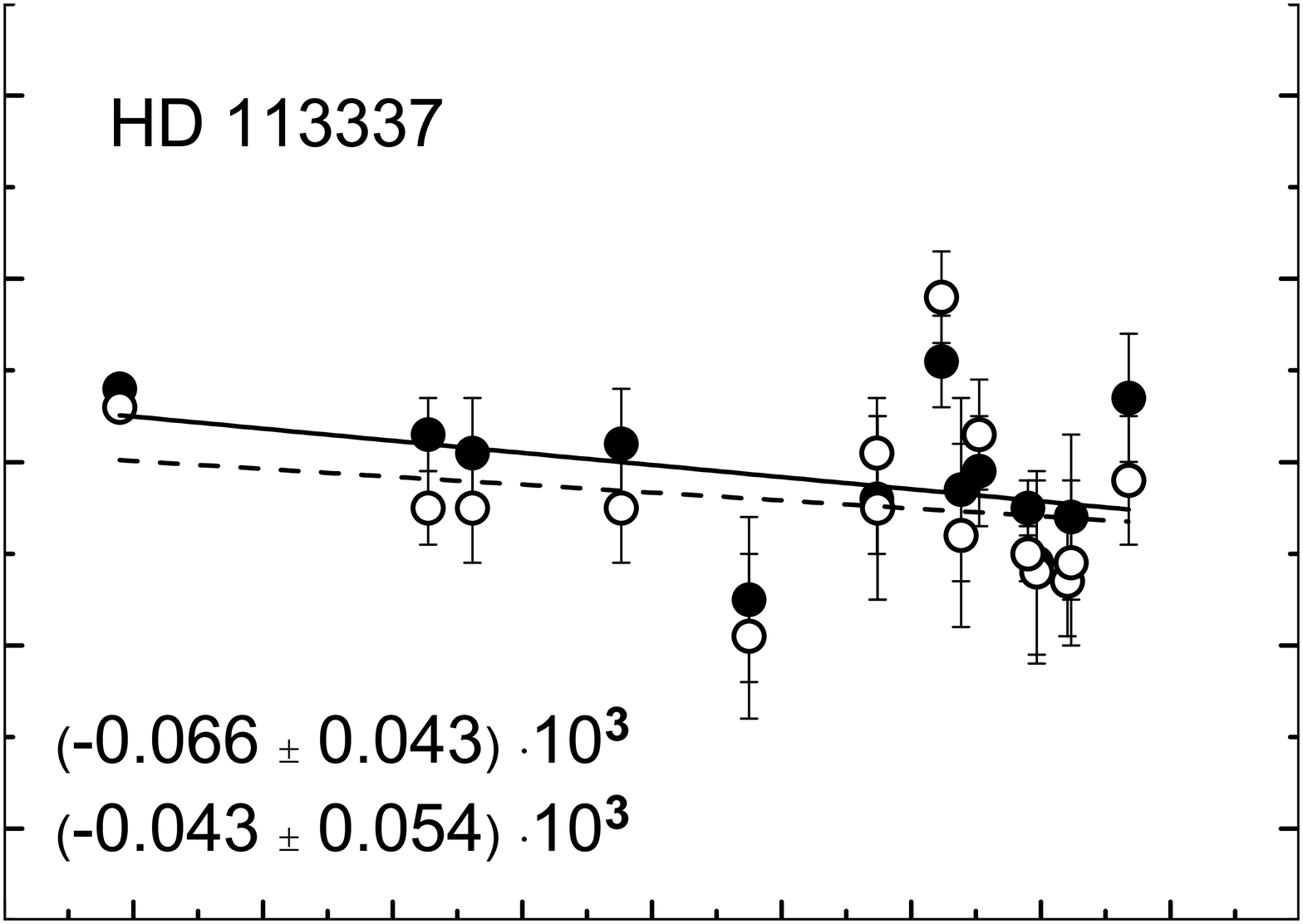}\\
\includegraphics[width=5.8cm,height=2cm]{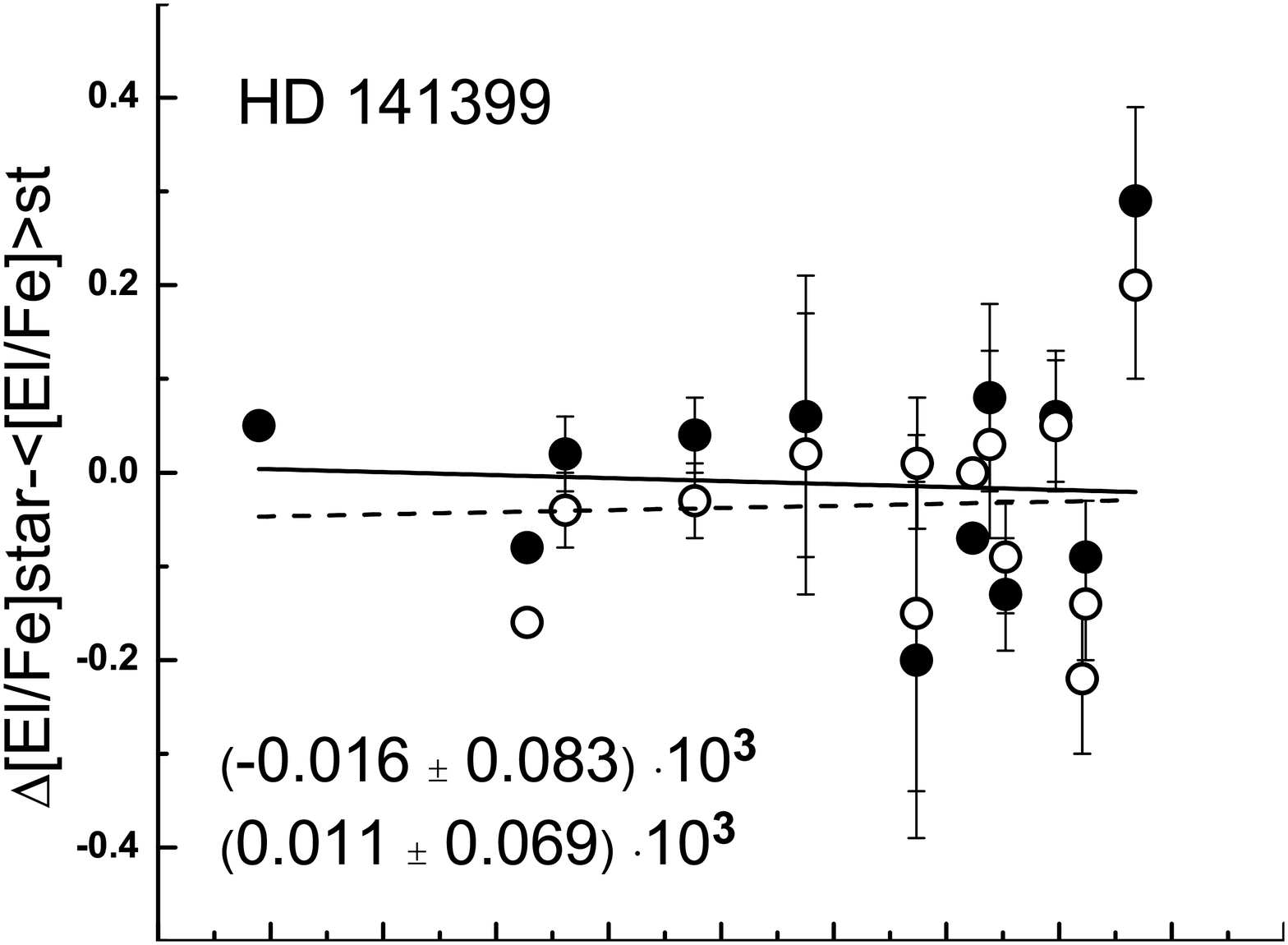}\includegraphics[width=5cm,height=2cm]{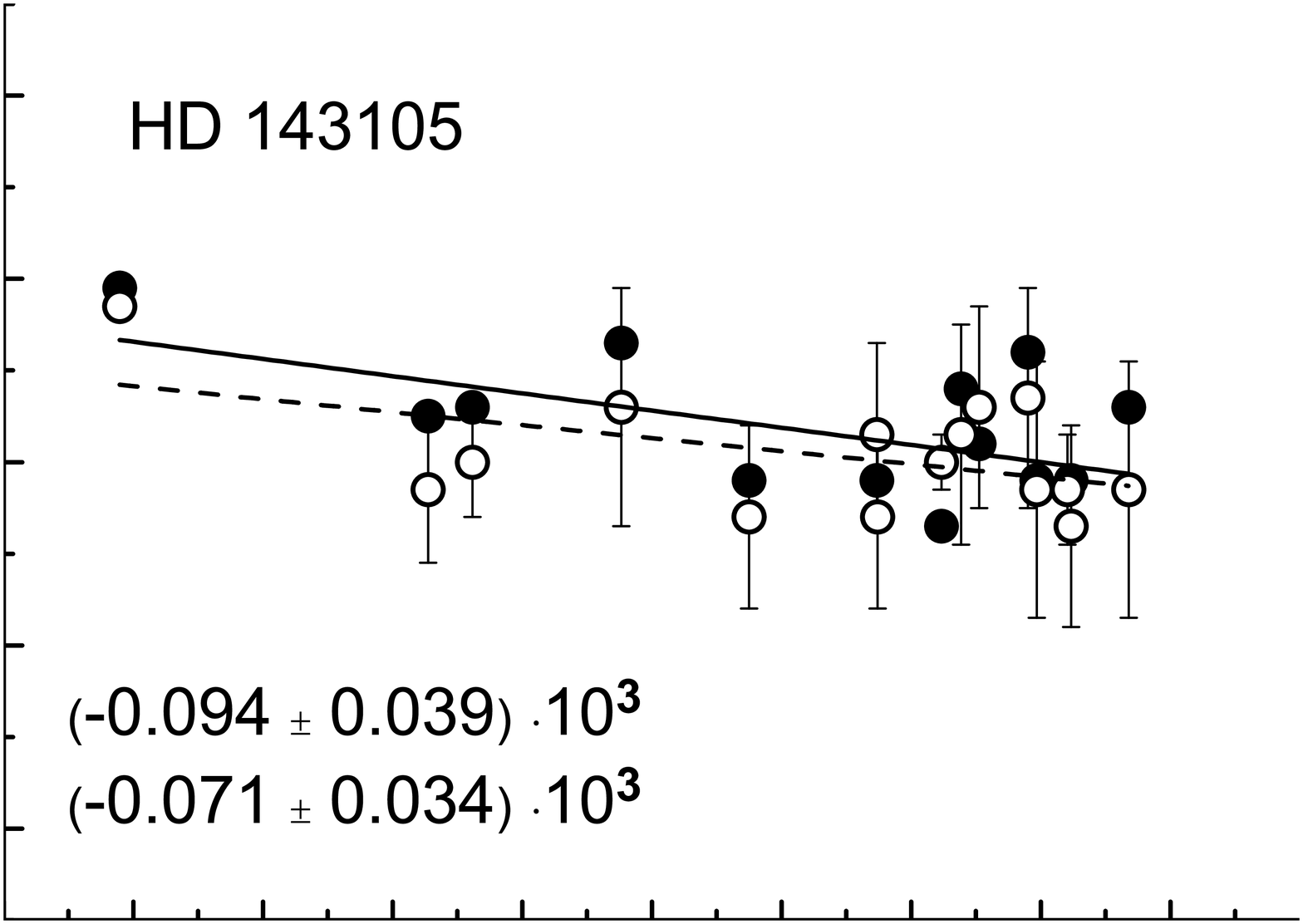}\includegraphics[width=5cm,height=2cm]{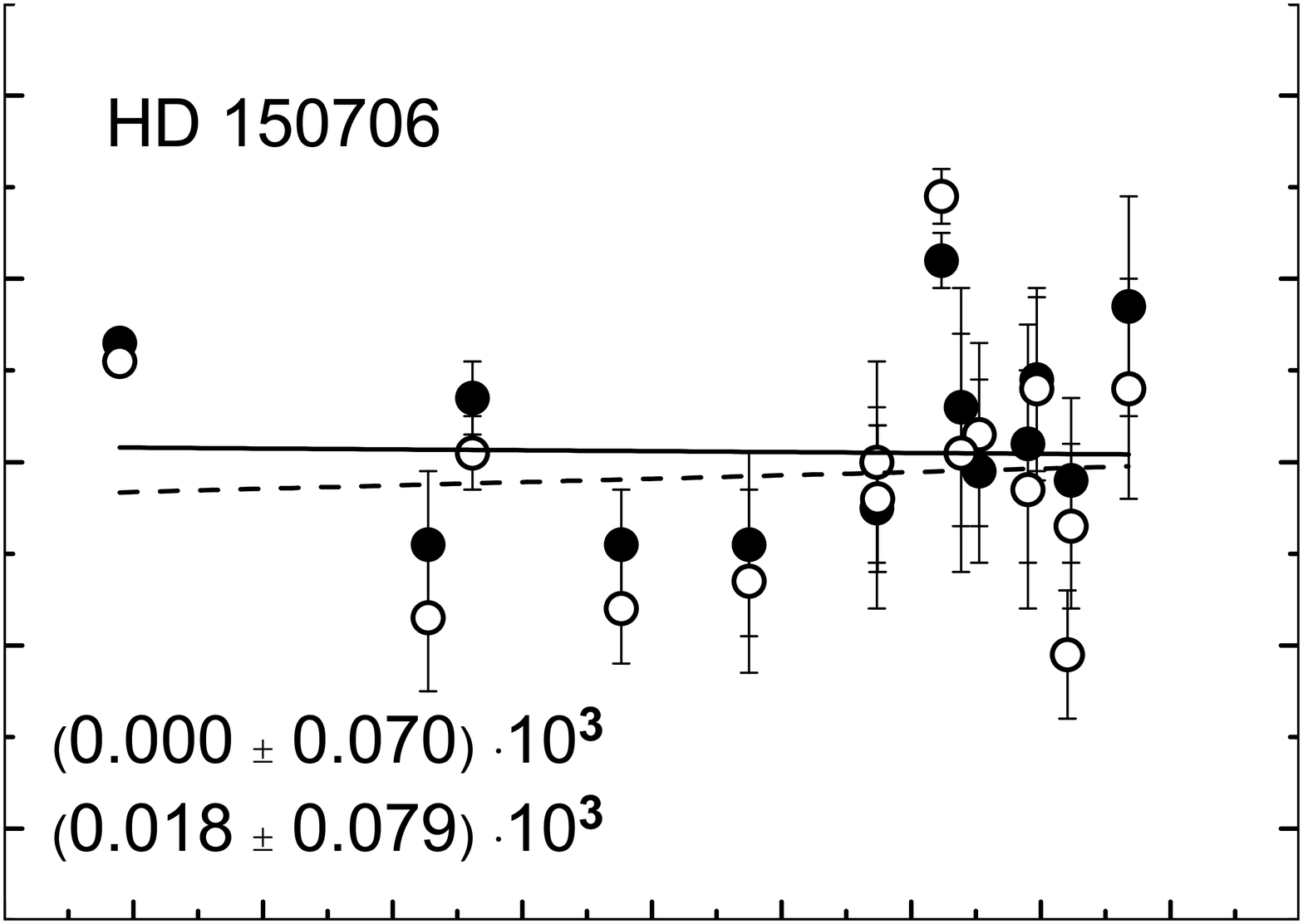}\\
\includegraphics[width=5.8cm,height=2cm]{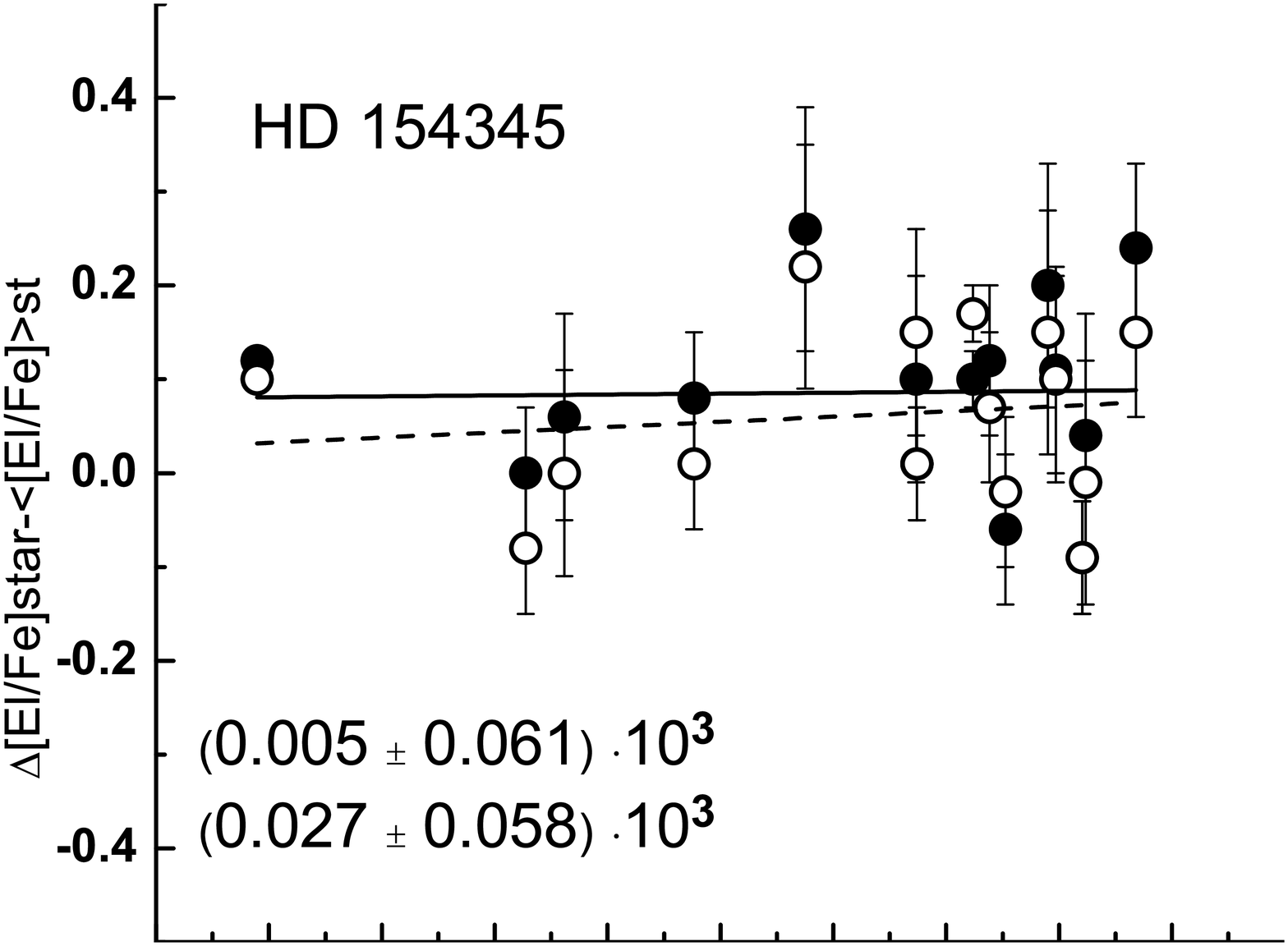}\includegraphics[width=5cm,height=2cm]{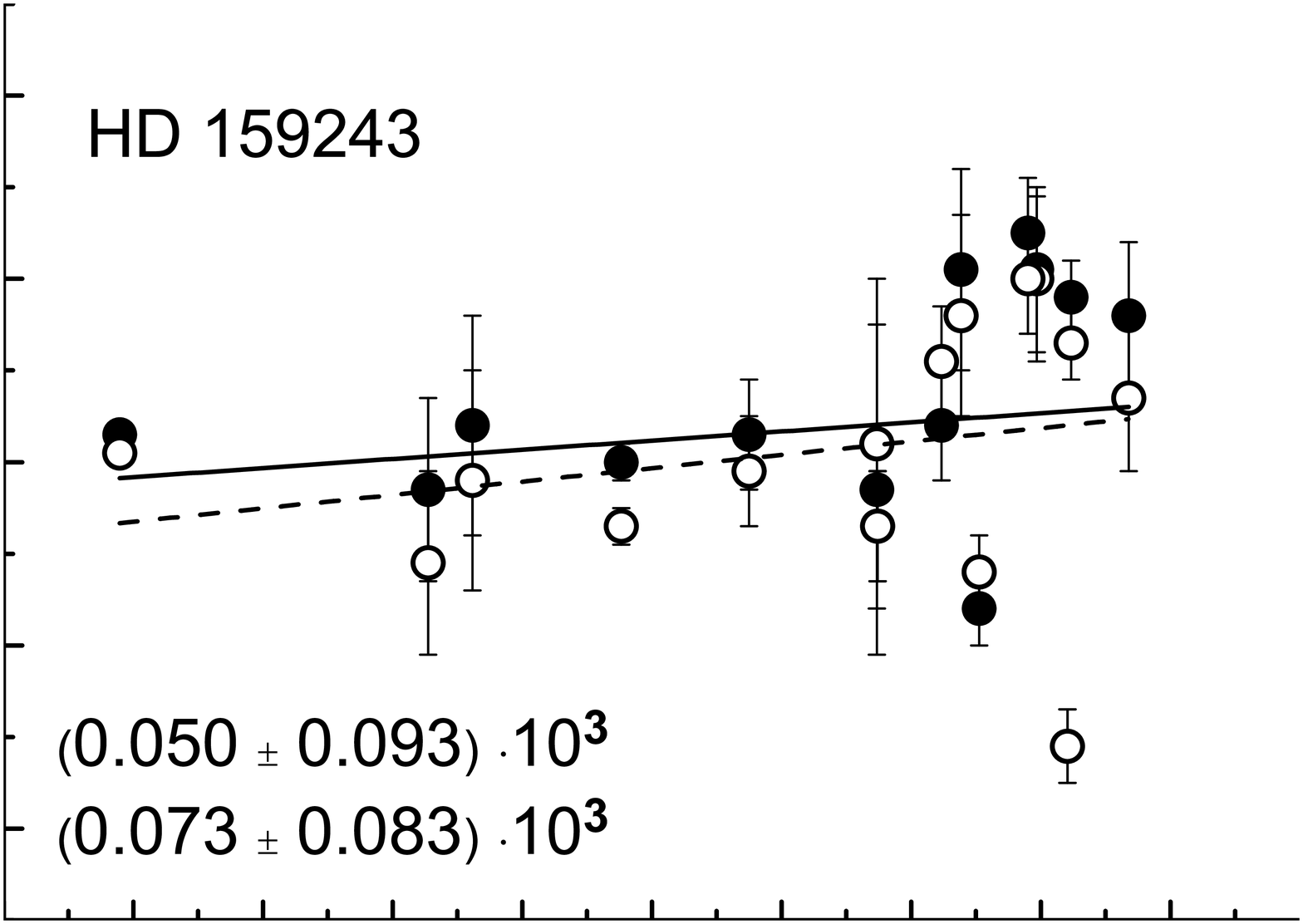}\includegraphics[width=5cm,height=2cm]{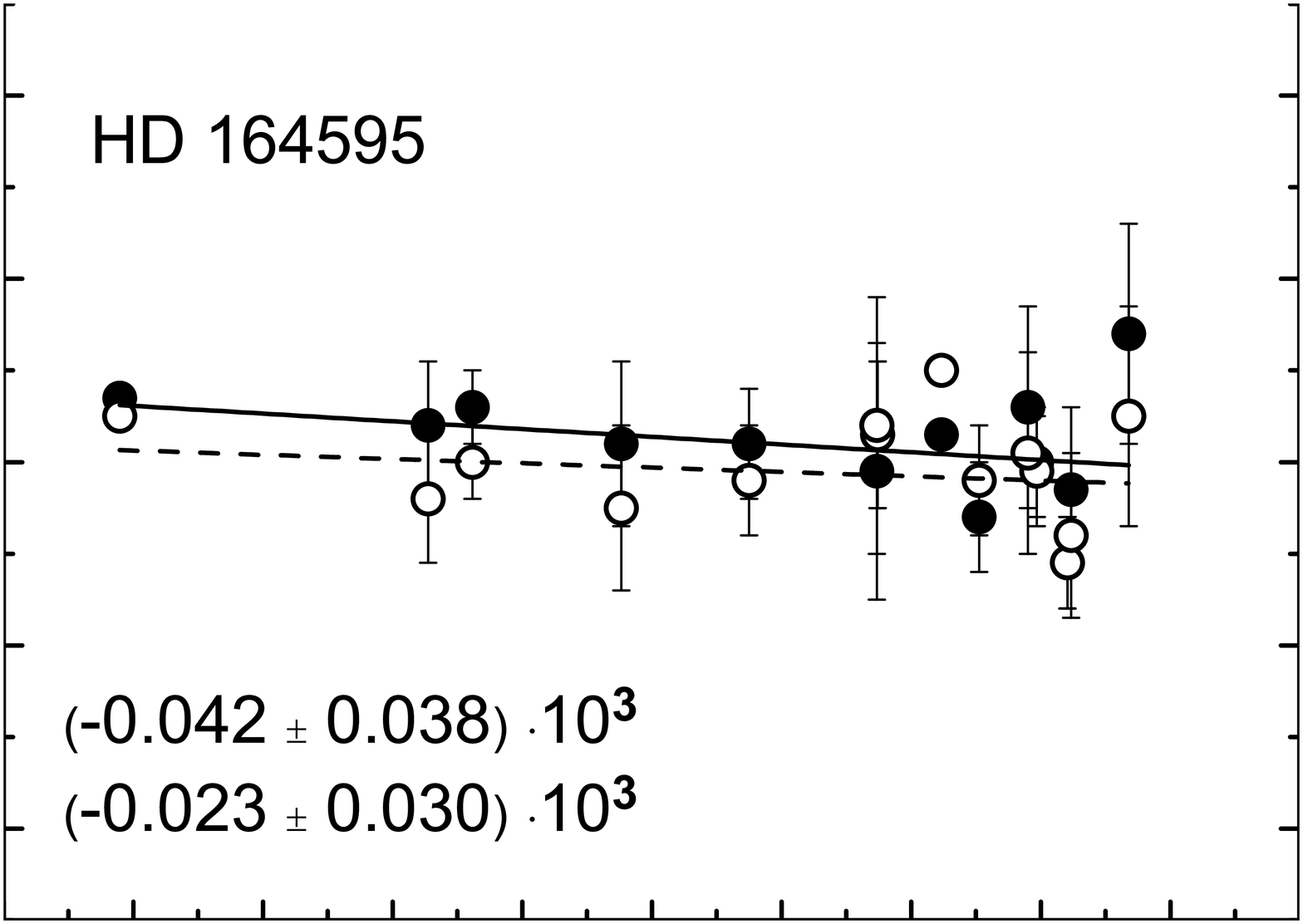}\\
\includegraphics[width=5.8cm,height=2cm]{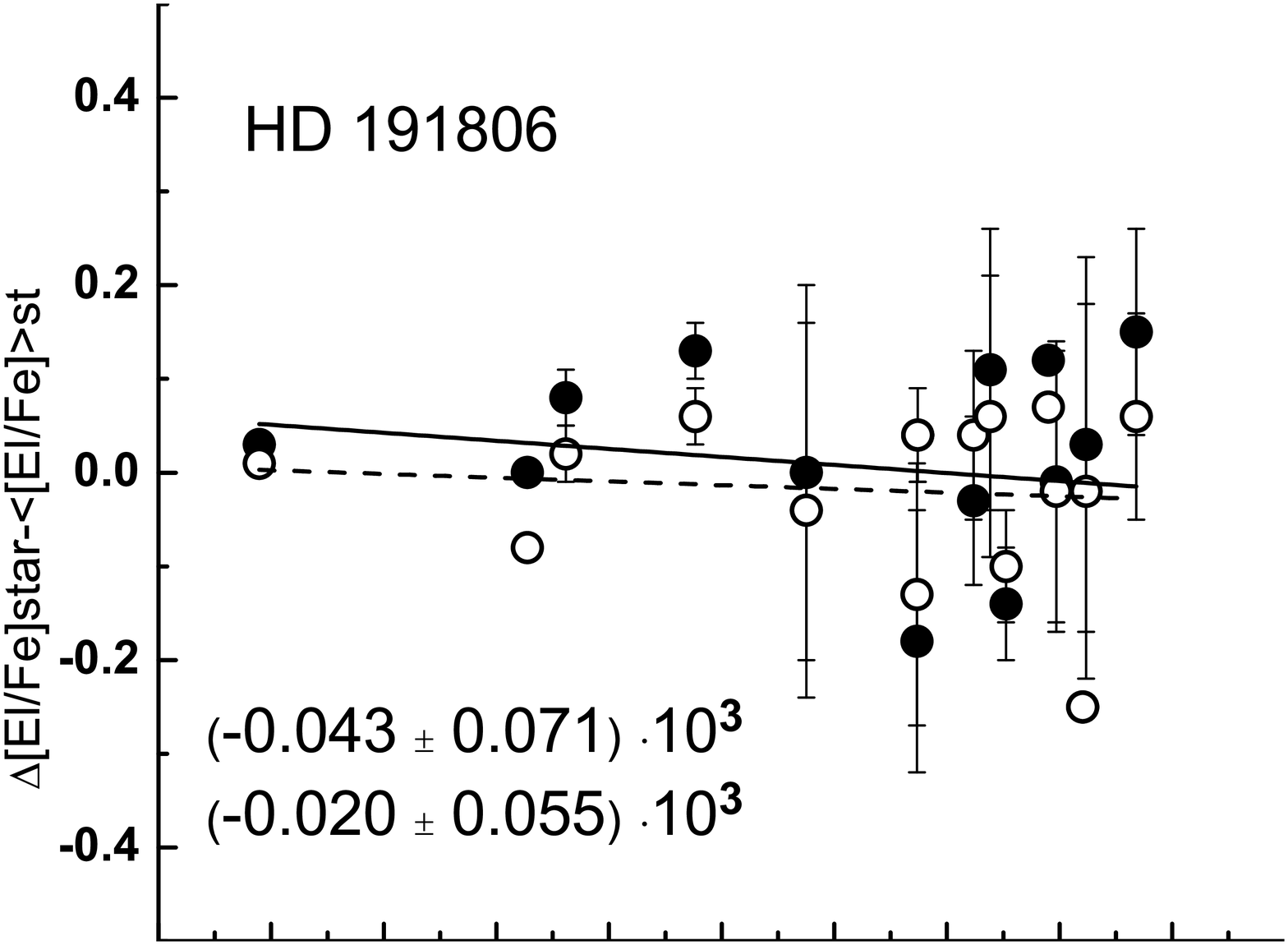}\includegraphics[width=5cm,height=2cm]{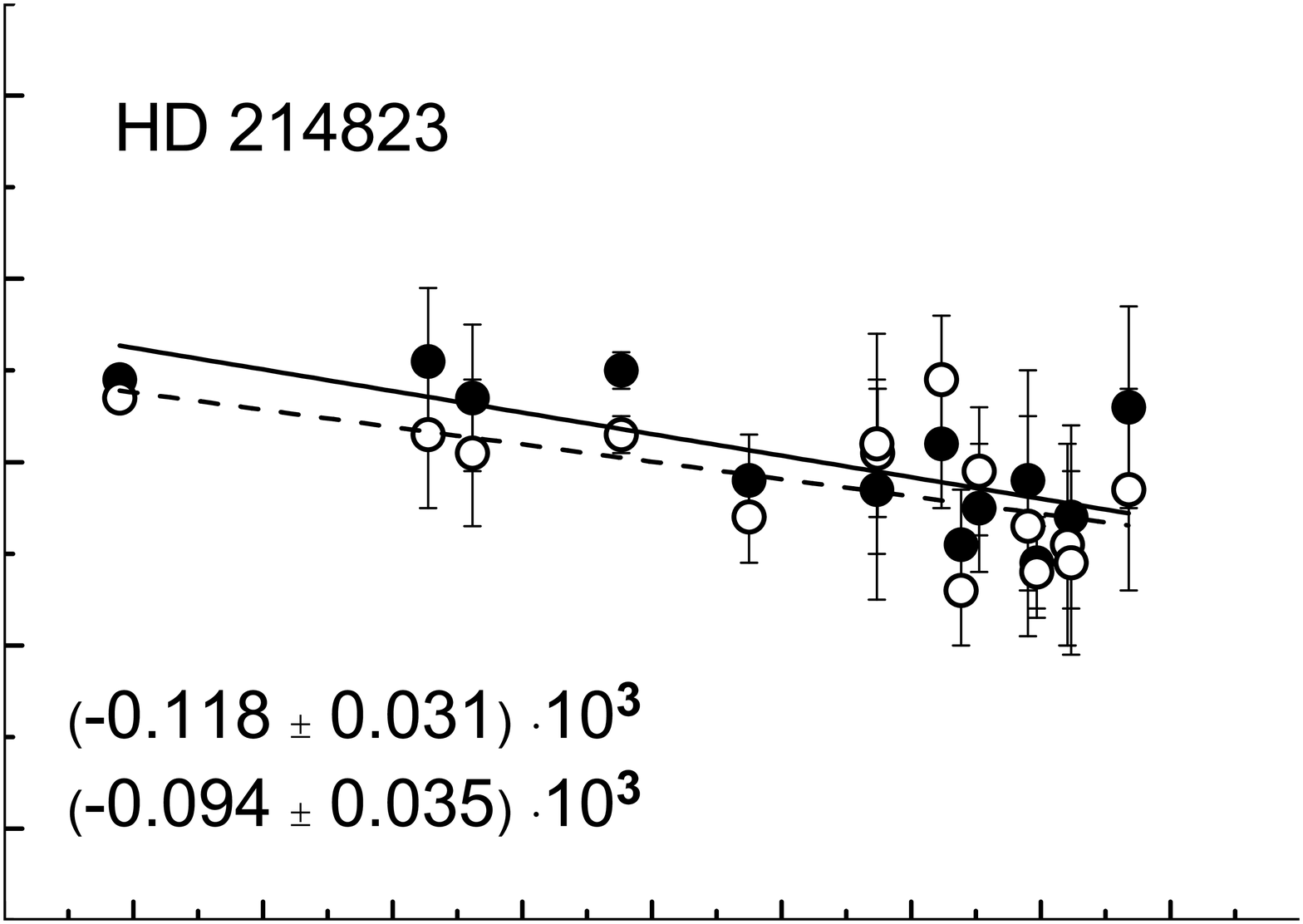}\includegraphics[width=5cm,height=2cm]{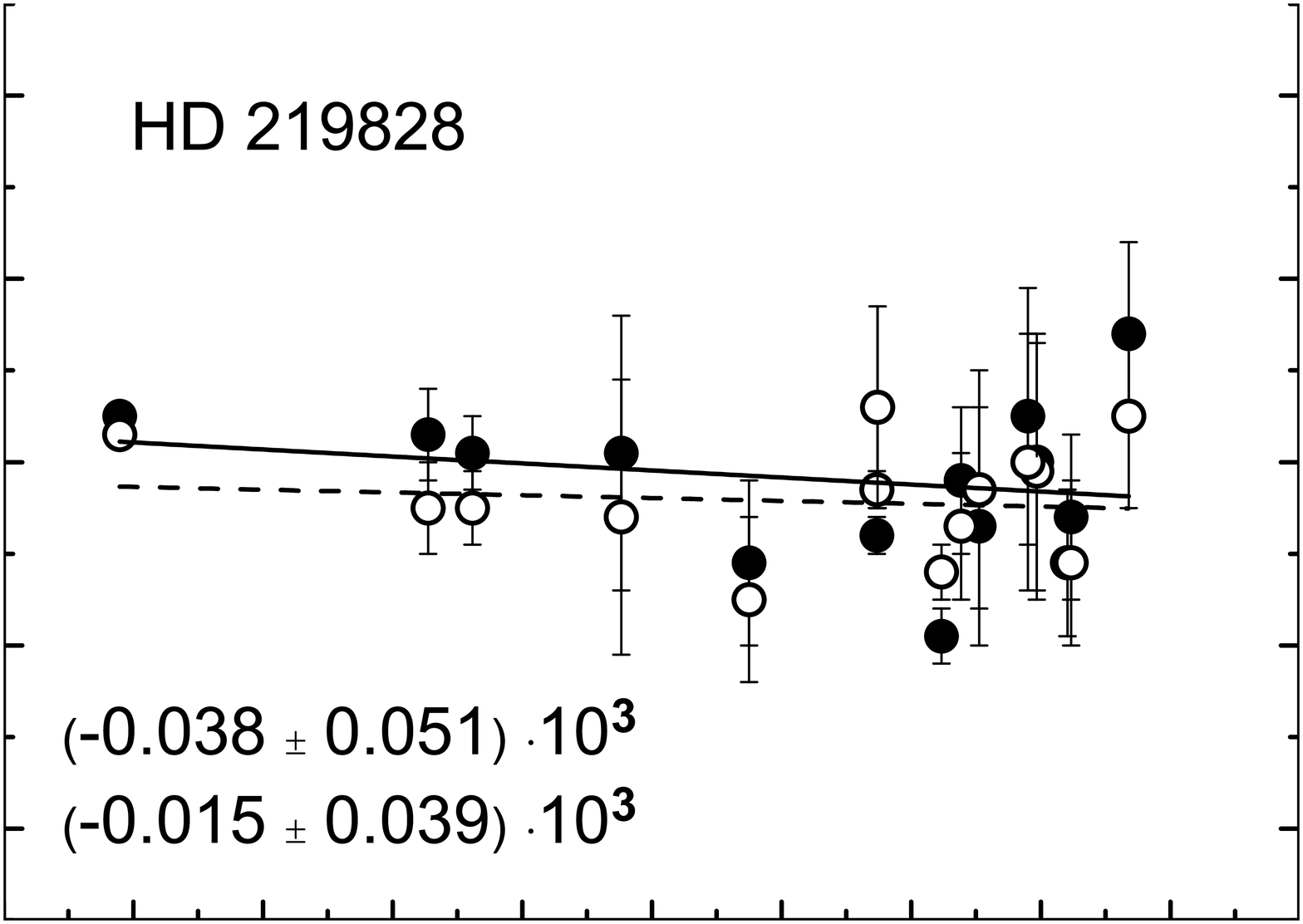}\\
\includegraphics[width=5.8cm,height=2cm]{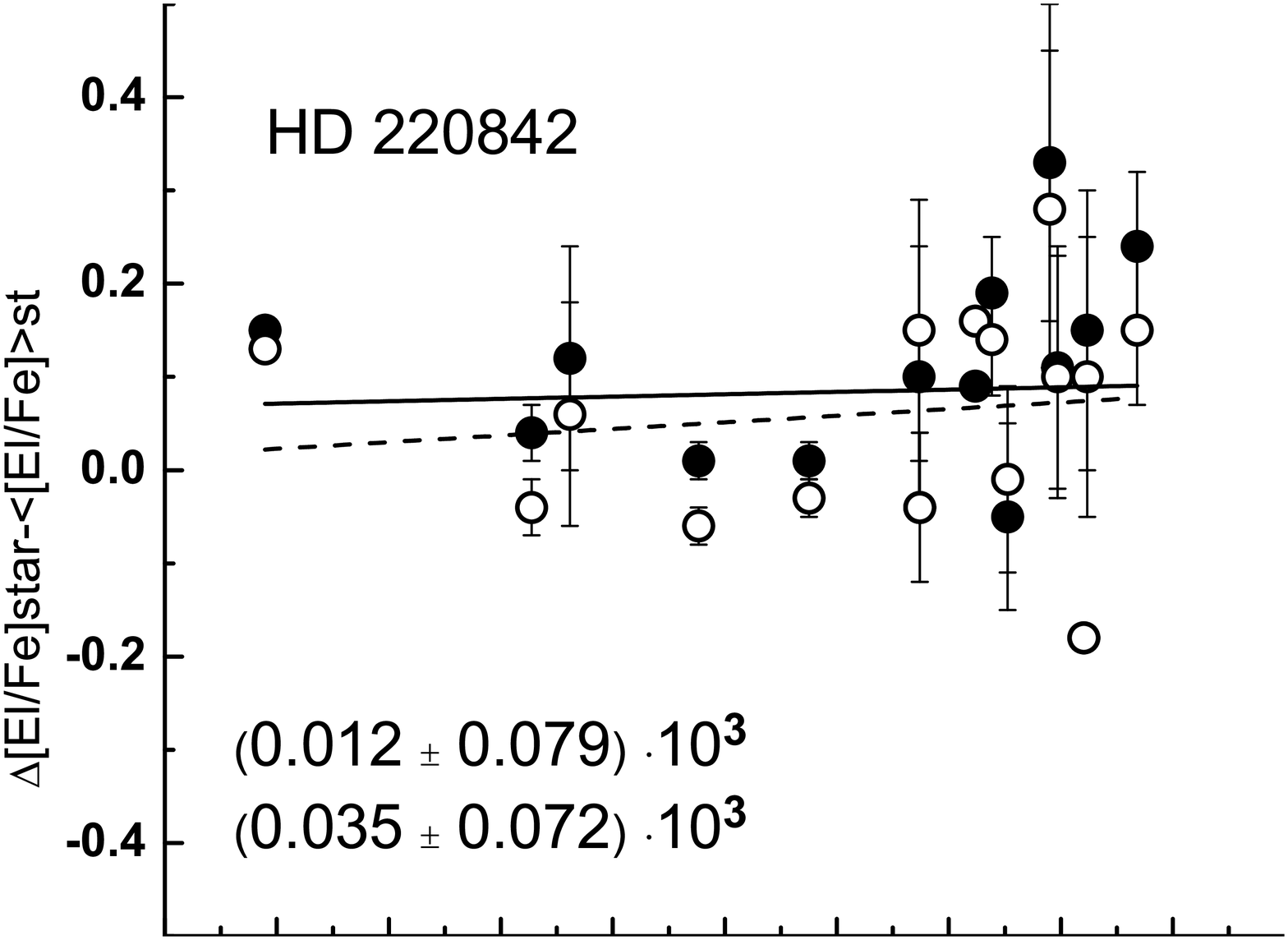}\includegraphics[width=5cm,height=2cm]{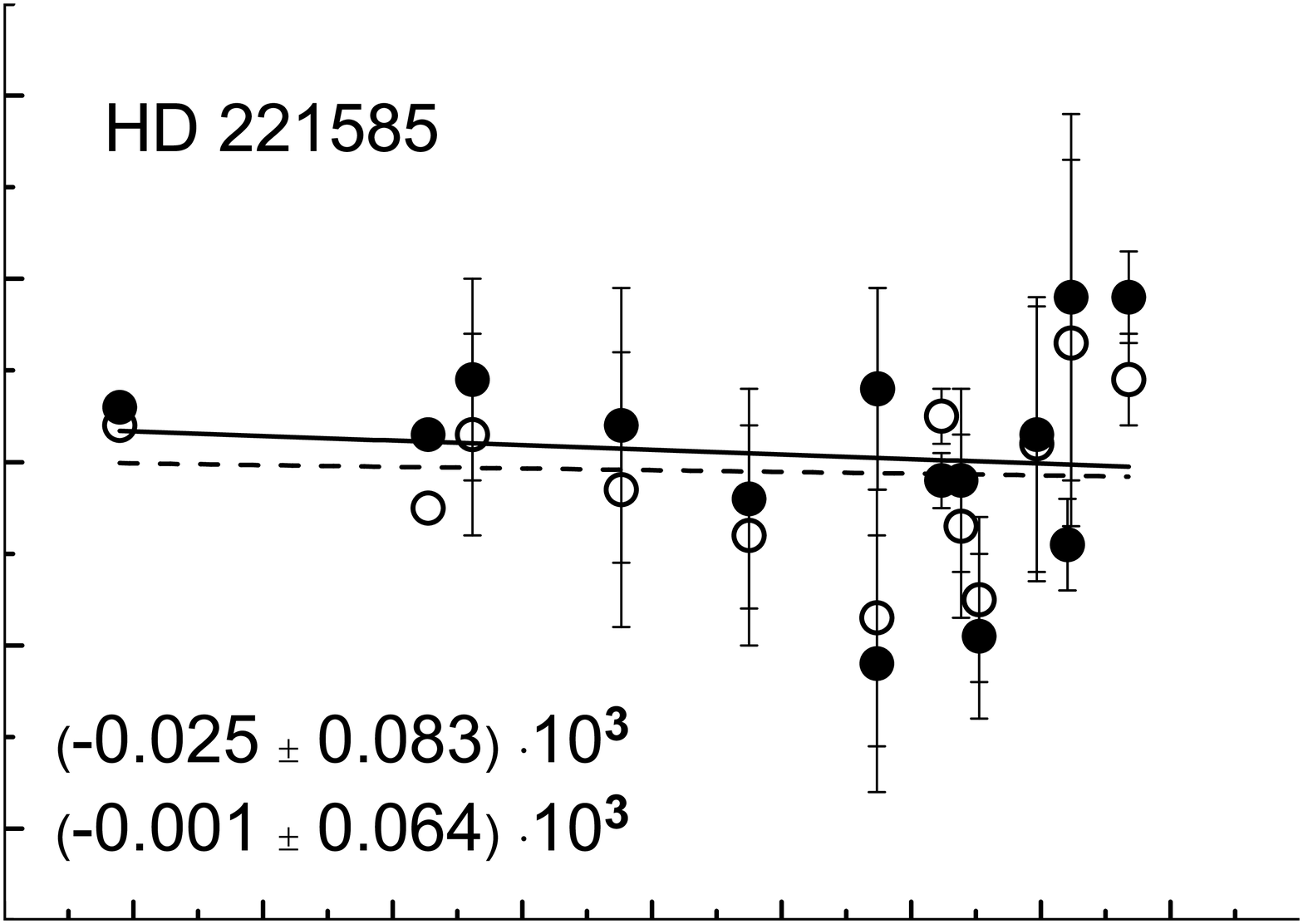}\includegraphics[width=5cm,height=2cm]{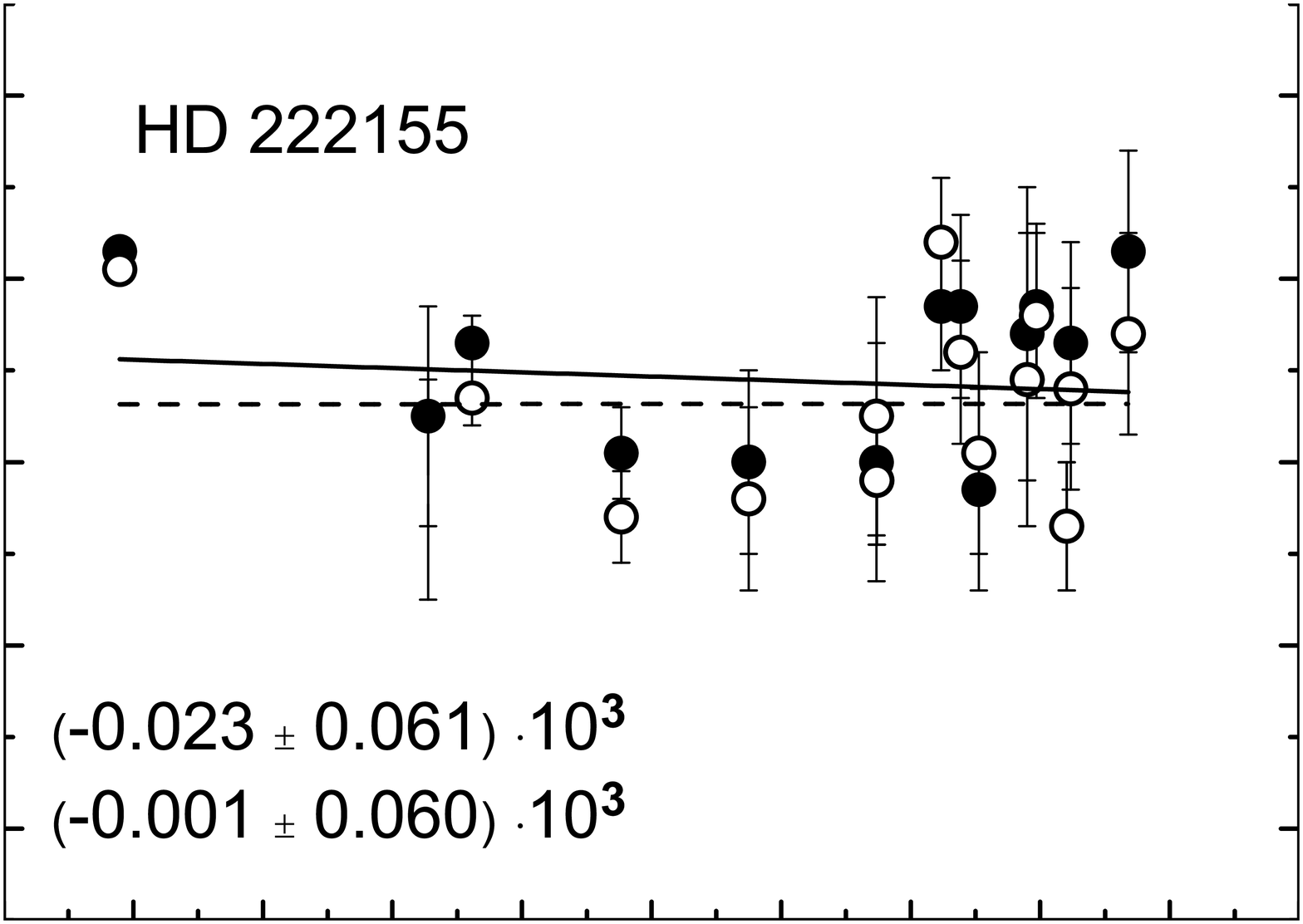}\\
\includegraphics[width=5.8cm,height=2.5cm]{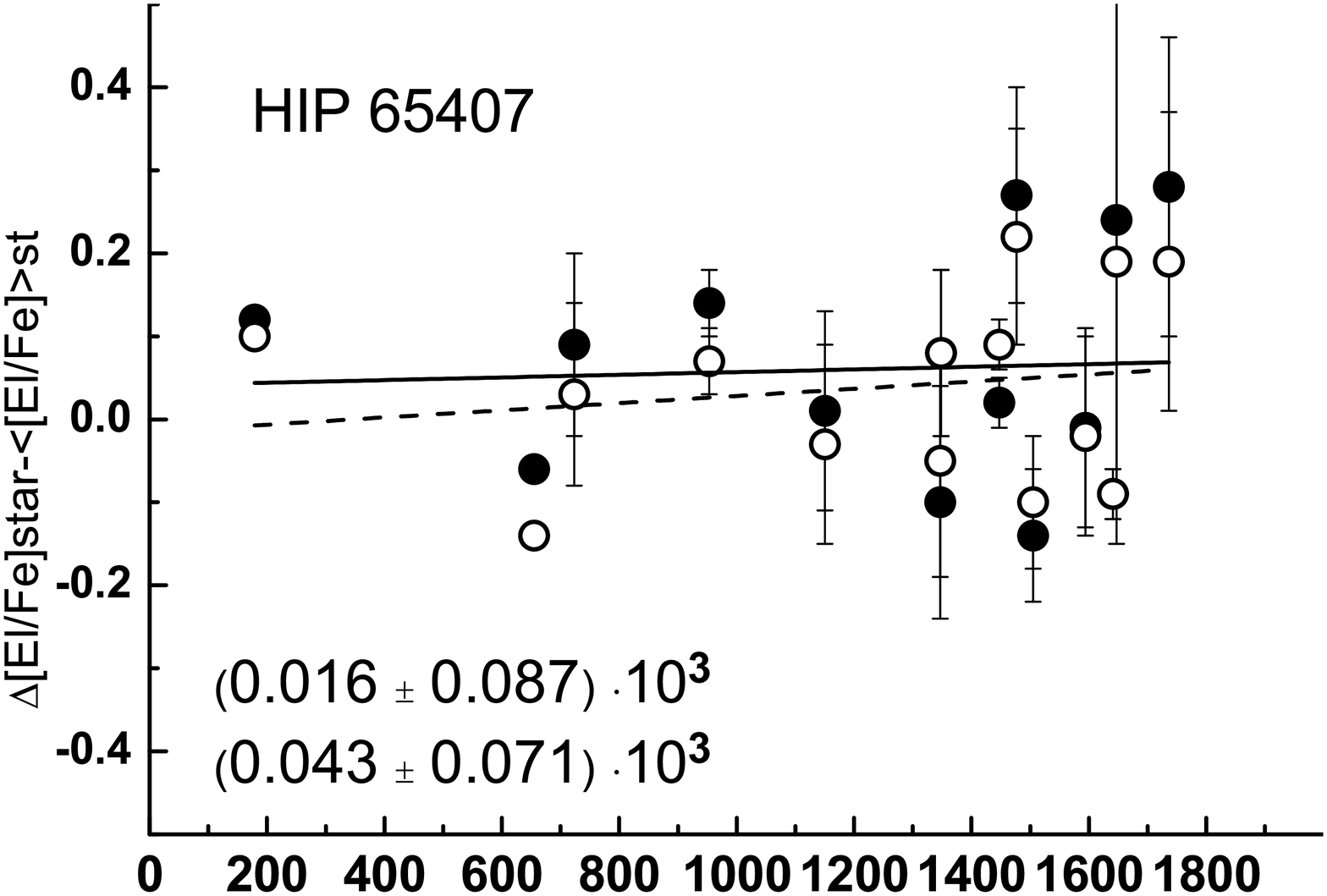}\includegraphics[width=5cm,height=2.5cm]{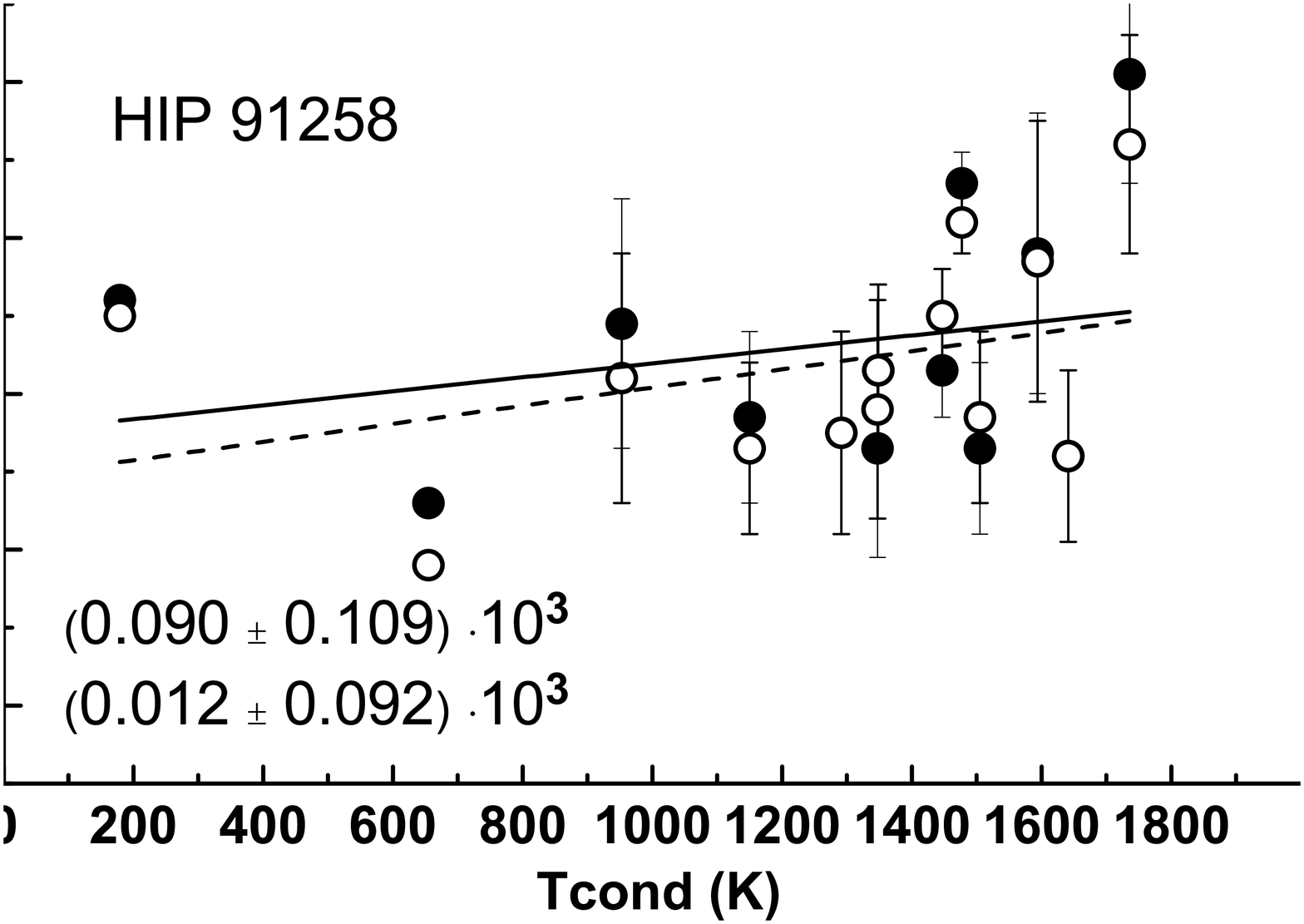}\includegraphics[width=5cm,height=2.5cm]{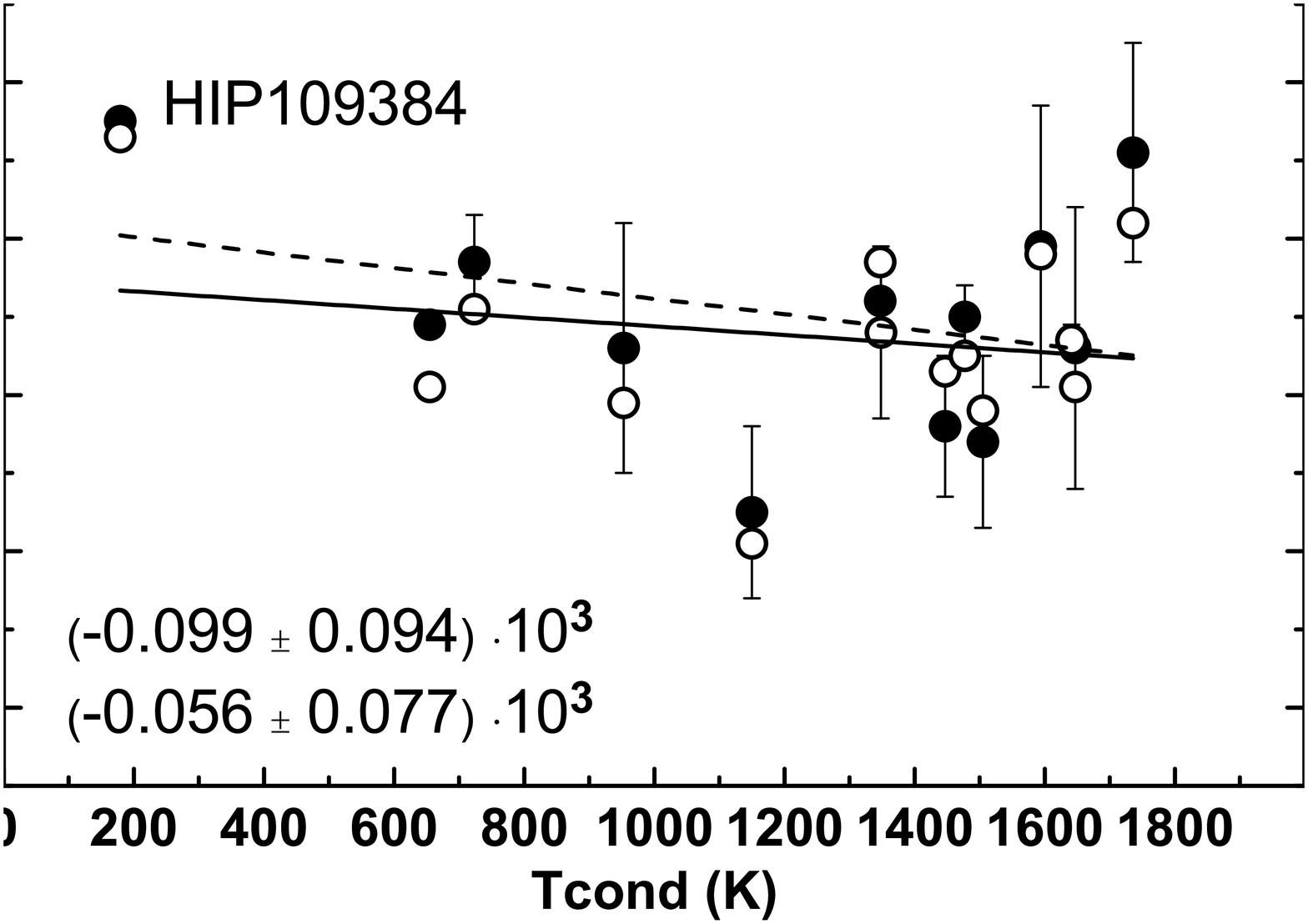}\\
\includegraphics[width=5.8cm,height=2.5cm]{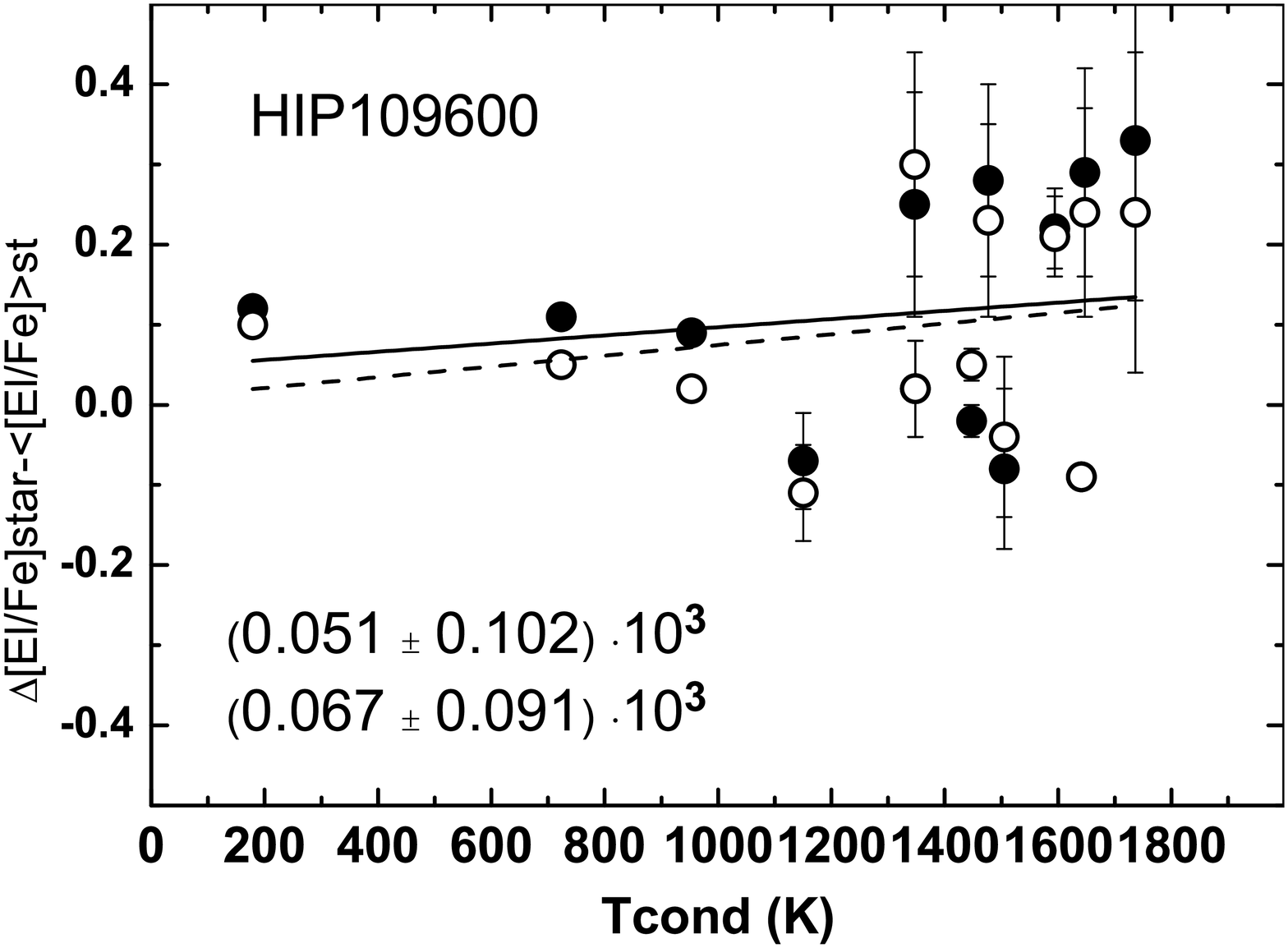}\\
\end{tabular}
\caption{$\Delta$[El/Fe] and [El/Fe] versus $T_{cond}$ for all target stars. The solid and dashed lines correspond to the abundance values 
with and without corrections, respectively, as well as the numbers indicated in each panel give (upper value with correction, lower value without correction)}
\label{el_tcond_all}
\end{figure*}

In Fig. \ref{el_tcond_all} we plot with solid line the differences between elemental abundances of our target stars and mean values of the solar twins $\Delta$[El/Fe](star)-$<$[El/Fe]$>$(st), and with dashed line [El/Fe] as a function of $T_{cond}$.

We can see that the stars exhibit both positive and negative slope in the dependencies of $\Delta$[El/Fe]-$T_{cond}$, but most of them do not show noticeable tilts to the trend line. Two stars, namely, HD143105(--0.094$\pm$0.039) and HD214823(--0.118$\pm$0.031)
show negative slopes that represent decreasing refractory-to-volatile abundance ratios, and  three stars, namely, HD24040 (--0.075$\pm$0.061), HD35759 (--0.101$\pm$0.074), HD113337 (--0.066$\pm$0.043) have only slightly decreasing trend.
For all the stars with positive slopes, the error in the slope is higher than its value. 

\begin{figure*}
\begin{tabular}{c}
\includegraphics[width=16.8cm]{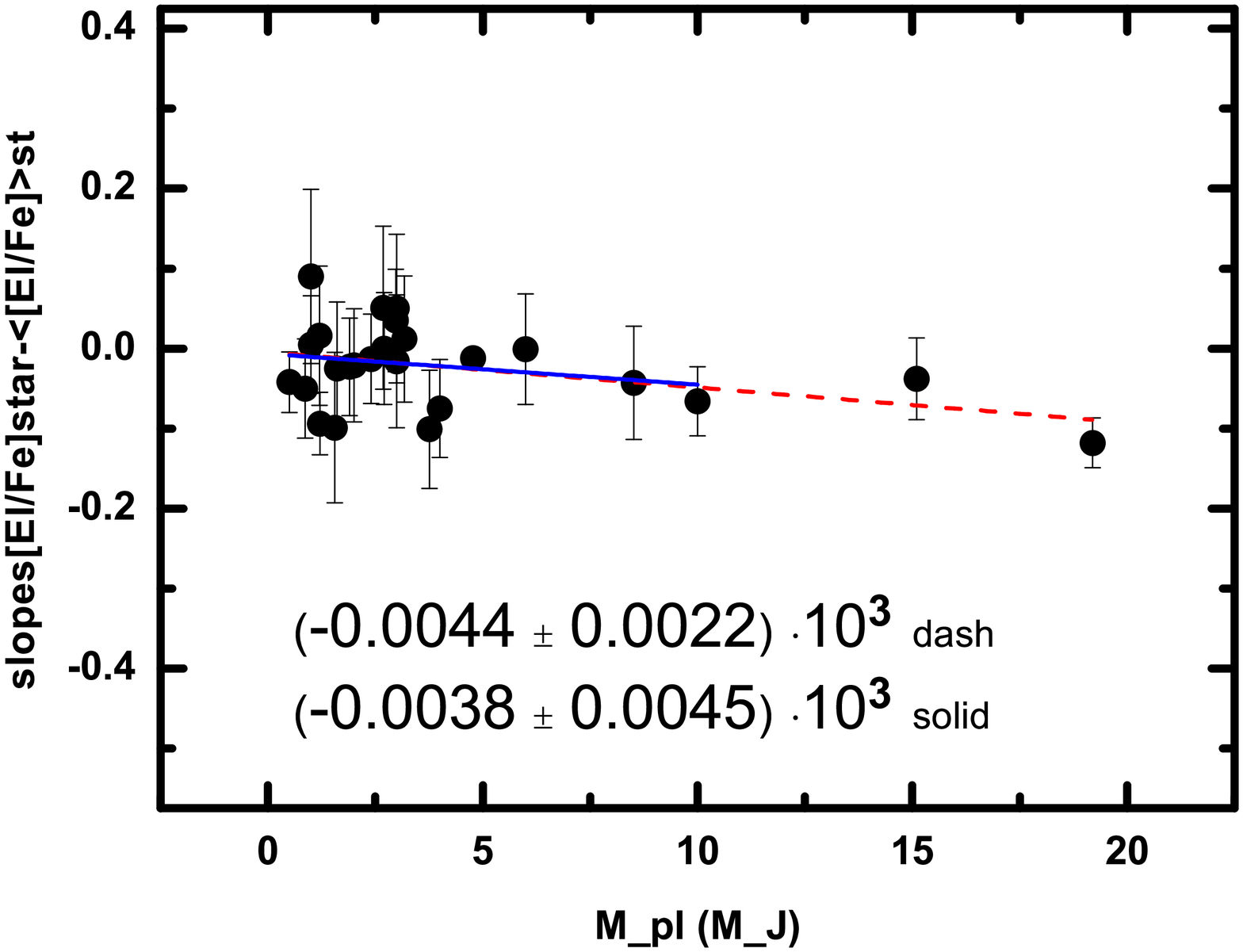}\\
\end{tabular}
\caption{Dependence of the $\Delta$[El/Fe]-$T_{cond}$ slope on the planetary mass in $M_{J}$ }
\label{slope_tcond}
\end{figure*}

What does such a result suggest?
The masses in  $M_{J}$ unit of the first two stars mentioned above are 1.21 and 19.2, respectively, and 4.00, 3.76, and 10.0 for the other three stars, respectively.
Four of five stars that exhibit negative slopes are more massive than 3 $M_{J}$.
Could this indicate some pattern in the formation of the nuclei of massive planets? It is difficult to answer to this question, since our sample also contains other that do not exhibit any trends.
Fig.\ref{slope_tcond} displays the relationship between the resulting slopes ($\Delta$ [El/Fe] versus $T_{cond}$) and planetary masses in $M_{J}$ for all stars of our target. We can see a slight trend (--0.0044$\pm$0.0022) indicating possible relationship between negative slope and planetary masses. However, when excluding the two most massive planets from the sample the slope of the linear trend changes to --0.0038$\pm$0.0045, making the dependence insignificant.

\section{Conclusions}
\label{sec: conclusions}

We have determined the abundances of 25 elements, from Li to Eu, in a sample of stars with detected massive planets with the exception of one target hosting a Neptun-size mass planetary companion. A comparison of the parameters in this study with those reported in other papers has shown a good agreement, which enables us to perform a consistent analysis of planet hosts and a sufficiently high accuracy and reliability of the derived elemental abundances. The C/O abundances for 16 of 25 sample stars were calculated for the first time; these elements play an important role in the formation of massive planets. We have analyzed the behavior of elemental abundances as a function of various parameters, such as [Fe/H] and $T_{cond}$, and planetary masses.
This study has yielded the following findings:

i) The resulting trends of [C/Fe] and [O/Fe] vs. [Fe/H] are consistent with those in the Galactic disc. Our mean values of C/O and [C/O] are $<$C/O$>$ = 0.48 $\pm$0.07 and $<$[C/O]$>$ = --0.07 $\pm$0.07, which are slightly lower than the solar ones. Different individual values of C/O ratios obtained in this study are consistent with a large variance of the C and O abundances in stars hosting massive planetary systems; it does not support the claim that stars with massive planets must be carbon rich, but reflects the fact that the stars in our sample have lower carbon abundances as compared to that in the Sun. 

ii) The trend of [Mg/Si]  vs. [Fe/H] follows the behaviour of Mg and Si in the Galactic disc. The Mg/Si ratios obtained in this study range from 0.83 to 0.96 for four stars and from 1.0 to 1.86 for the remaining 21 stars in our sample. These results agree well with the study by \cite{suarez:18}, in which it was found that 85\% stars with high-mass companions had the Mg/Si ratios ranging between one and two;

iii) A trend of $M_{pl}$ with [Fe/H] and just a trace of trend representing   decreasing Mg/Si ratios with increasing planetary mass have been detected;

iv)  The diversity of the observed abundance-$T_{cond}$ slopes found for planet-hosting stars might be associated with various processes, such as a variety of evolutionary paths of circumstellar discs, accretion or formation of the giant-planet rocky core. However, we cannot rule out that 
there is no physical relation between presence of planets and the $T_{cond}$ trend.

The data obtained in this study can impose constraints on existing models of planetary evolution and prove useful in the development and validation of theoretical studies of the formation and evolution of massive planetary systems.

\section*{Data Availability Statement}
The data that support the findings of this study are available from the corresponding author T. Mishenina upon reasonable request.

\section*{Acknowledgements}

 The authors thank the anonymous referee for a careful reading of the manuscript and suggestions
that significantly improved the manuscript. The authors are grateful to Sergey Korotin for the discussion on the barium abundance. 
V.A. was supported by FCT - Funda\c{c}\~ao para a Ci\^encia e Tecnologia (FCT) through national funds and by FEDER through COMPETE2020 - Programa Operacional Competitividade e Internacionaliza\c{c}\~ao by these grants: UID/FIS/04434/2019; UIDB/04434/2020; UIDP/04434/2020; PTDC/FIS-AST/32113/2017 \& POCI-01-0145-FEDER-032113; PTDC/FIS-AST/28953/2017 \& POCI-01-0145-FEDER-028953. V.A. also acknowledges the support from FCT through Investigador FCT contracts nr.  IF/00650/2015/CP1273/CT0001, and POPH/FSE (EC) by FEDER funding through the program ``Programa Operacional de Factores de Competitividade - COMPETE''.
This article is based upon work from the ChETEC COST Action (CA16117), supported by COST (European Cooperation in Science and Technology).

\bibliography{star_planet}

\appendix
\section{On-line material}
 
\onecolumn

\clearpage
\begin{longtable}{rrrrrrrrrrrrl}
\caption{The list of used lines}
\label{lines}\\
\hline
      Wave &   Ion  & Elow   & \loggf  &       Wave    & Ion  &  Elow  & \loggf  &    Wave   &  Ion  &    Elow& \loggf  \\
\hline
\endhead
\hline
   5688.205&  11.00 &2.104   & --0.451 &      5653.867 & 26.00&  4.387 &  --1.639&   6739.522&  26.00&   1.557&  --4.793\\
   6154.226&  11.00 &2.102   & --1.546 &      5661.346 & 26.00&  4.284 &  --1.735&   6745.101&  26.00&   4.580&  --2.159\\
   6160.747&  11.00 &2.104   & --1.245 &      5672.264 & 26.00&  4.584 &  --2.799&   6745.957&  26.00&   4.076&  --2.769\\
   4571.096&  12.00 &0.000   & --5.622 &      5678.379 & 26.00&  3.884 &  --3.019&   6746.955&  26.00&   2.609&  --4.349\\
   4730.029&  12.00 &4.346   & --2.346 &      5679.023 & 26.00&  4.652 &  --0.919&   6750.153&  26.00&   2.424&  --2.620\\
   5711.088&  12.00 &4.346   & --1.723 &      5680.240 & 26.00&  4.186 &  --2.579&   6752.707&  26.00&   4.638&  --1.203\\
   6318.717&  12.00 &5.108   & --2.102 &      5686.530 & 26.00&  4.549 &  --0.445&   6753.464&  26.00&   4.559&  --2.289\\
   6319.237&  12.00 &5.108   & --2.323 &      5691.497 & 26.00&  4.301 &  --1.519&   6761.070&  26.00&   4.584&  --2.529\\
\hline
\end{longtable}


\newpage

{\bf Appendix}
\begin{table*}
\caption[]{Elemental abundances from individual ions,  $\sigma$ is internal errors}
\begin{tabular}{lrrrrrrrrrrrrrrrr}
\hline
\hline
\noalign{\smallskip}
\multicolumn{1}{c}{}&\multicolumn{4}{c}{HD 12484}&
\multicolumn{4}{c}{HD 13908}&\multicolumn{4}{c}{HD 16175}&
\multicolumn{4}{c}{HD 17674}\\
\noalign{\smallskip}
\hline
 Ion &[M/H]
&$\sigma$&N&(M/H)&[M/H]&$\sigma$&N&(M/H)&[M/H]&$\sigma$&N&(M/H)&[M/H]&$\sigma$&N&(M/H)\\
\hline
Li~{\sc i} & --  &  --   &   1   &   2.85 &   -- &    -- &    1   &    1.38  &    -- &   --  &    1    &   3.00   &  --  &  --   &   1   &    1.80 \\
 C~{\sc i} & 0.05  &  0.10 &   3   &    8.48 & --0.08 &  0.08 &    3   &    8.35  &    0.29 &   0.10&    3    &   8.72   &   --0.11 &  0.06 &   3  &    8.32 \\
 O~{\sc i} & 0.09  &  --    &   1  &    8.78 &   0.01 &   --  &    1  &    8.70  &    0.40 &   --   &    1    &   9.09   &    0.01  &       &   1  &    8.70 \\
Na~{\sc i} & 0.01  &  0.12 &   3   &    6.25 &   0.00 &   0.10&    3   &    6.24  &    0.43 &   0.03&    2    &   6.75   &   --0.19 &  0.10 &   3   &    6.05 \\
Mg~{\sc i} & 0.07  &  0.12 &   4   &    7.67 & --0.06 &   0.04&    6   &    7.54  &    0.28 &   0.15&    4    &   7.88   &   --0.11 &  0.08 &   4   &    7.49 \\
Al~{\sc i} & 0.03  &  0.06 &  2    &    6.48 &  -0.20 &   0.07&    2   &    6.25  &    0.18 &   0.13&    2    &   6.63   &   --0.30 &  0.06 &   2   &    6.15 \\
\hline
\end{tabular}
\begin{list}{}{}
\item[] N -- is the number of lines used in analysis
\end{list}
\end{table*}

\begin{table*}
{\bf Table A2  (continued)}\\
\begin{tabular}{lrrrrrrrrrrrrrrrr}
\hline
\hline
\noalign{\smallskip}
\multicolumn{1}{c}{}&\multicolumn{4}{c}{HD 113337}&
\multicolumn{4}{c}{HD 141399}&\multicolumn{4}{c}{HD 143105}&
\multicolumn{4}{c}{HD 150706}\\
\noalign{\smallskip}
\hline
 Ion &[M/H]
&$\sigma$&N&(M/H)&[M/H]&$\sigma$&N&(M/H)&[M/H]&$\sigma$&N&(M/H)&[M/H]&$\sigma$&N&(M/H)\\
\hline
Li~{\sc i}  & -- &   --   &   1   &   -- & -- &  --   &   1   &    0.71 &  -- &   --   &   1   &    2.30 &    --  &  --   &   1    &   2.60 \\
 C~{\sc i}  &  0.06 &   0.05 &   3   &    8.49 &  0.14  &  0.07 &   3   &    8.70 &  0.10 &   0.09 &   3   &    8.53 &   --0.13 &  0.09 &   3    &   8.30 \\
 O~{\sc i}  &  0.11 &   --  &   1   &    8.8  &  0.41  &  --   &   1   &    9.10 &  0.09 &   --   &   1   &    8.78 &    0.07  &  --   &   1    &   8.76 \\
Na~{\sc i}  &  0.09 &   0.06 &   3   &    6.33 &  0.36  &  0.04 &   2   &    6.68 &  0.07 &   0.13 &   3   &    6.31 &   --0.11 &  0.06 &   3    &   6.13 \\
Mg~{\sc i}  &  0.12 &   0.08 &   6   &    7.72 &  0.41  &  0.1  &   4   &    8.01 &  0.11 &   0.10 &   5   &    7.71 &   --0.03 &  0.07 &   6    &   7.57 \\
Al~{\sc i}  &  0.01 &   0.06 &   2   &    6.46 &  0.40  &  0.08 &   2   &    6.70 & --0.01&   0.06 &   2   &    6.43 &   --0.16 &  0.07 &   2    &   6.29 \\
\hline
\end{tabular}
\begin{list}{}{}
\item[] N -- is the number of lines used in analysis
\end{list}
\end{table*}

\begin{table*}
{\bf Table A2  (continued)}\\
\begin{tabular}{lrrrrrrrrrrrrrrrr}
\hline
\hline
\noalign{\smallskip}
\multicolumn{1}{c}{}&\multicolumn{4}{c}{HD 214823}&
\multicolumn{4}{c}{HD 219828}&\multicolumn{4}{c}{HD 220842}&
\multicolumn{4}{c}{HD 221585}\\
\noalign{\smallskip}
\hline
 Ion &[M/H]
&$\sigma$&N&(M/H)&[M/H]&$\sigma$&N&(M/H)&[M/H]&$\sigma$&N&(M/H)&[M/H]&$\sigma$&N&(M/H)\\
\hline
Li~{\sc i}   &  --  &  --   &   1   &    1.50 &  --  &  --   &   --  &    2.30 &  --  &  --   &   1     &  1.16 &   -- &   --   &   --  &    1.62 \\
 C~{\sc i}   &  0.06  &  0.10 &   3   &    8.49 &  0.11  &  0.06 &   3   &    8.54 & --0.13 &  0.05 &   3     &  8.30 &   0.29 &   0.06 &   3   &    8.72 \\
 O~{\sc i}   &  0.11  &  --   &   1   &    8.80 &  0.13  &  --   &   1   &    8.82 & --0.06 &  --   &   1     &  8.63 &   0.28 &   --   &   1  &    8.97 \\
Na~{\sc i}   &  0.16  &  0.02 &   3   &    6.40 &  0.13  &  0.15 &   3   &    6.37 & --0.16 &  0.02 &   3     &  6.08 &   0.30 &   0.15 &   3   &    6.54 \\
Mg~{\sc i}   &  0.15  &  0.06 &   6   &    7.75 &  0.16  &  0.05 &   4   &    7.76 & --0.07 &  0.09 &   6     &  7.53 &   0.30 &   0.15 &   4   &    7.90 \\
Al~{\sc i}   &  0.04  &  0.11 &   2   &    6.49 &  0.08  &  0.08 &   2   &    6.53 & --0.28 &  0.01 &   2     &  6.17 &   0.24 &   0.05 &   2   &    6.69 \\
\hline
\end{tabular}
\begin{list}{}{}
\item[] N -- is the number of lines used in analysis
\end{list}
\end{table*}

\begin{table*}
{\bf Table A2  (continued)}\\
\begin{tabular}{lrrrr}
\hline
\hline
\noalign{\smallskip}
\multicolumn{1}{c}{}&\multicolumn{4}{c}{HIP 109600}\\
\noalign{\smallskip}
\hline
 Ion &[M/H]
&$\sigma$&N&(M/H)\\
\hline
Li~{\sc i}  &  -- &  --   &   1    &   1.00  \\
 C~{\sc i}  &  --0.02 &  0.04 &   2    &   8.41  \\
 O~{\sc i}  &  --0.06 &  --  &   1    &   8.63  \\
Na~{\sc i}  &  --0.13 &  0.01 &   2    &   6.19  \\
Mg~{\sc i}  &   0.11  &  0.07 &   3    &   7.70  \\
Al~{\sc i}  &  --0.01 &  0.01 &   2    &   6.29  \\
\hline
\end{tabular}
\begin{list}{}{}
\item[] N -- is the number of the lines used in analysis.
\end{list}
\end{table*}

\end{document}